\documentclass[11pt,preprint]{aastex}
\usepackage{apjfonts}

\def\gax{\mathrel{\raise.3ex\hbox{$>$}\mkern-14mu\lower0.6ex\hbox{$\sim$}}}
\def\lax{\mathrel{\raise.3ex\hbox{$<$}\mkern-14mu\lower0.6ex\hbox{$\sim$}}}
\def\gtorder{\mathrel{\raise.3ex\hbox{$>$}\mkern-14mu
             \lower0.6ex\hbox{$\sim$}}}
\def\ltorder{\mathrel{\raise.3ex\hbox{$<$}\mkern-14mu
             \lower0.6ex\hbox{$\sim$}}}

\begin{document}

\title{Unmasking the Supernova Impostors}

\author{
   C.~S. Kochanek$^{1,2}$, 
   D.~M. Szczygie\l$^{1,2}$,
  K.~Z. Stanek$^{1,2}$
  }

\altaffiltext{1}{Department of Astronomy, The Ohio State University, 140 West 18th Avenue, Columbus OH 43210}
\altaffiltext{2}{Center for Cosmology and AstroParticle Physics, The Ohio State University, 191 W. Woodruff Avenue, Columbus OH 43210}

\begin{abstract}
The canonical picture of a supernova impostor is a $-11 \ltorder M_V \ltorder -14$
optical transient from a massive ($M_* \gtorder 40M_\odot$) star during which the star ejects a dense 
shell of material.  Dust formed in the ejecta then obscures the star.  
In this picture, the geometric expansion of the shell leads to clear
predictions for the evolution of the optical depths and hence the
evolution of the optical through mid-IR emissions.  Here we
review the theory of this standard model and then
examine the impostors SN~1954J, SN~1997bs, SN~1999bw, 
SN~2000ch, SN~2001ac, SN~2002bu, SN~2002kg and SN~2003gm, as well
as the potential archetype $\eta$ Carinae.  SN~1999bw,
SN~2000ch, SN~2001ac, SN~2002bu and SN~2003gm all show mid-IR
emission indicative of dust, and the luminosities of SN~1999bw, SN~2001ac, 
SN~2002bu and SN~2003gm are dominated by dust emission.  We
find only find upper limits on dust emission from SN~1954J, SN~1997bs and 
SN~2002kg.   { \it We find, however, that the properties of these sources are broadly 
inconsistent with the predictions of the canonical model.}
Ignoring SN~2000ch and SN~2002kg, which are simply variable stars
with little or no dust, there appear to be two classes of objects.
In one class ($\eta$ Carinae, SN~1954J, SN~1997bs, and (maybe) SN~2003gm), the 
optical transient is a signal that the star is entering a phase with
very high mass loss rates that must last far longer than the visual 
transient.  The second class (SN~1999bw, SN~2001ac, SN~2002bu and (maybe) SN~2003gm)
has the different physics of SN~2008S and the 2008 NGC~300 transient,
where they are obscured by dust re-forming in a pre-existing wind 
after it was destroyed by an explosive transient.  There are no
cases where the source at late times is significantly fainter than
the progenitor star.  All these dusty 
transients are occurring in relatively low mass ($M_*\ltorder 25M_\odot$)
stars rather than high mass ($M_*\gtorder 40M_\odot$) stars radiating near 
the Eddington limit like $\eta$ Carinae. The durations
and energetics of these transients cannot be properly characterized 
without near/mid-IR observations, and the fragmentary nature of 
the available data leads to considerable uncertainties in our understanding
of the individual sources.  Continued monitoring 
of the sources at both optical and near/mid-IR wavelengths should
resolve these ambiguities. 
\end{abstract}


\section{Introduction}
\label{sec:introduction}

The eruptions of luminous blue variables (LBV) sometimes mimic the 
spectroscopic properties of Type~IIn core collapse supernovae (SNe)
and approach their luminosities (see, e.g., \citealt{Smith2011}).  Unfortunately, the Type~IIn 
(narrow $\ltorder 2000$~km/s line widths) spectral type (\citealt{Schlegel1990},
\citealt{Filippenko1997})
seems to be a property
of almost any  stellar eruption, explosion or evolutionary process with a strong interaction 
between the ejected material and the circumstellar medium, ranging
from proto-planetary nebulae to some of the most luminous SNe (see, e.g., \citealt{Prieto2009}).   
\cite{Vandyk2002} introduced the label of ``supernova impostors'' for
LBV transients with Type~IIn spectra that approach the luminosity 
of Type~II SN but are believed to be somewhat fainter 
 than the lowest luminosity SNe  ($M\ltorder -14$ rather than
$ -16 \ltorder M \ltorder -18$).  The ``prototypical'' luminous LBV transient
is the eruption of $\eta$ Carinae in the mid-1800s (see the reviews by
\citealt{Humphreys1994}, \citealt{Humphreys1999}, \citealt{Smith2009b}, \citealt{Vink2009}). 
We should note, however, that a typical SN impostor like SN~1997bs (see \citealt{Smith2011}) has completely different
velocities, ejected masses and durations from the typical Galactic analogue
even if many of these differences can be explained as a selection effect
(see \citealt{Kochanek2011b}).

The need for clearer distinctions between genuine SNe and stellar eruptions is illustrated
by the ongoing debate over the nature of SN~2008S and the transient later
that year in NGC~300.  The dust enshrouded progenitors (\citealt{Prieto2008},
\citealt{Prieto2008a}) are astonishingly rare, only a few per galaxy, and
appear to be extreme AGB stars (\citealt{Thompson2009}, \citealt{Khan2010})
with graphitic dust (\citealt{Prieto2009}, \citealt{Wesson2010}) rather
than the silicate dust typically seen in massive stars. A broad range
of physical mechanisms have been proposed for these events:
an extension of the LBV
phenomenon (\citealt{Smith2009a}, \citealt{Berger2009}), some other
type of massive star transient (\citealt{Bond2009}, \citealt{Humphreys2011}), electron capture
SN or some other form of low luminosity SN (\citealt{Thompson2009}, 
\citealt{Botticella2009}, \citealt{Pumo2009}), a binary merger
(\citealt{Kashi2010}), or a mass ejection associated with forming
a white dwarf (\citealt{Thompson2009}).  At this point, several years
after the optical peaks, both events
are still bright mid-IR sources (\citealt{Ohsawa2010}, \citealt{Prieto2010a},
\citealt{Szczygiel2011}) that are well-explained by an explosive transient
(\citealt{Kochanek2011}).  A further consideration is that there
appears to be a mismatch between massive star formation rates
and supernova rates that could be solved if significant numbers of these fainter
transients were SN rather than impostors  
(see \citealt{Lien2010}, \citealt{Horiuchi2010}, \citealt{Botticella2011}).
The alternatives are
either that massive star formation rate estimates are significantly
in error (by a factor of two) or that the failed SN rate is
comparable to the normal SN rate (see the discussion in \citealt{Kochanek2008}). 

The obvious difference between eruptions and supernovae is that stars survive
eruptions but not core collapse.  Candidate surviving stars have been
found for SN~1954J (\citealt{Smith2001}, \citealt{Vandyk2005}), 
SN~1961V (\citealt{Filippenko1995}, \citealt{Vandyk2002}, \citealt{Chu2004}), 
SN~1997bs (\citealt{Vandyk1999}, \citealt{Li2002}), 
SN~2000ch (\citealt{Wagner2004}, \citealt{Pastorello2010}),
and SN~2002kg/V37 (\citealt{Weis2005}, \citealt{Maund2006}).
With the exceptions of SN~2000ch and SN~2002kg/V37, which have varied repeatedly, 
it is probably safe to say that none of these identifications are
absolutely certain and that some are likely wrong.  Even with near
perfect astrometry one can still be fooled by binary companions
(see \citealt{Kochanek2009}) or chance projections.  It would be
best to have some additional tests.

The standard picture of the impostors is that the visual transient is
associated with the radiatively driven ejection of a significant amount
of mass (see \citealt{Humphreys1994}), although \cite{Dessart2010} discuss
shock driven models in some detail.   The high density of the ejected material leads to dust formation, 
and the expanding shell of material then obscures the surviving star for 
an extended period.  In \cite{Kochanek2011b}
we examined the formation of dust in the transients of hot stars, and
found that dust formation and (great) eruptions should be essentially synonymous, 
as the conditions seen in eruptions are also necessary and sufficient
conditions for forming dust in the ejecta.   
The most frequently proposed test of this picture is 
to search for the mid-IR emission produced by reprocessing the emissions
of the surviving star in the dusty ejecta.  For
the spectacular Galactic example of $\eta$ Carinae, $\sim 90\%$ of
the emission is absorbed and reradiated in the mid-IR, leading to
a spectral energy distribution (SED) that peaks near $20\mu$m  
(see \citealt{Humphreys1994}).  Indeed, most of the candidate survivors
are usually optically fainter than their progenitors, and this is generally explained as being
due to absorption in the ejecta. However, there have been attempts to find 
the predicted mid-IR emission only for SN~1954J (\citealt{Smith2001})
and SN~1961V (\citealt{Kochanek2010}, \citealt{Vandyk2011}), although
\cite{Vandyk2011} show some results from a broader survey.  Finding the dust
emission then allows us to characterize the luminosity of the
star and the mass of the ejecta.  This test also largely 
eliminates false detections of non-binary survivors due to chance superpositions
because of the small angular extent of the shell and the
relative rarity of other luminous dusty stars (see, e.g., \citealt{Thompson2009}, 
\citealt{Khan2010}).  

There is a second test, namely that the surviving star should tend
to become systematically brighter in the optical.  As the shell expands, its optical
depth drops, and the survivor should reappear.
This steady brightening is observed for
$\eta$ Carinae (see, e.g., \citealt{Humphreys1994}).  The rate of brightening
further constrains the optical depth of the ejected material.
Neither the mid-IR emission nor the optical brightening eliminates the possibility
that the ``surviving'' star is really a binary companion to the 
progenitor (see \citealt{Kochanek2009}), but superposed on the 
steady trend, there should also be the general low level variability 
of LBVs, which should help to eliminate this possibility.  
\cite{Chu2004} use the absence of
any LBV-like variability of the candidate surviving stars for
SN~1961V as one of their arguments in favor of a SN rather than
an LBV explanation for this transient.

% other SN in Thompson
% SN1994N - nodata
% SN1997D - nodata
% SN1999eu - NGC1097 17 Mpc nothing
% SN1999br - NGC4900 26 Mpc nothing
% SN2001dc - NGC5777 30 Mpc nothing
% SN2003Z -- NGC2742 23 Mpc nothing
% SN2005cs -- NGC5195 8 Mpc nothing
% other impostor candidates in Thompson/Smith
% NGC2366 V1 -- 4 Mpc nothing
% SN2006bv UGC07848 51 Mpc nothing
% SN2006fp UGC12182 25 Mpc nothing
% SN2007sv UGC05979 28 Mpc maybe
% NGC4656 OT -- 28 Mpc nothing
% SN2009ip NGC72549 - just missed
% UGC2773 OT no data
% SN2010da NGC 300 -- progenitor found
% SN2010dn NGC3184 -- no progenitor obvious, post transient observations 

Here we carry out these tests, as possible,  for SN~1954J/V12,
SN~1997bs, SN~1999bw, SN~2000ch, SN~2001ac, SN~2002bu,
SN~2002kg/V37 and SN~2003gm.  We also, for comparison, examine
the properties of $\eta$ Carinae.  We do not consider  
SN~2008S or the 2008 NGC~300-OT 
because their evolution cannot be described by the ``standard'' 
LBV eruption picture -- they require  explosive transients and dust 
re-forming in a pre-existing wind 
(\citealt{Kochanek2011}).  While we still view SN~1961V as a supernova
(\citealt{Kochanek2010}, \citealt{Smith2011}), we include a
short discussion in response to the rebuttal case by \cite{Vandyk2011}.
We do not consider many of the candidate
impostors in \cite{Smith2011} because they are either too 
distant (e.g. SN~2006bv) or lack (adequate) mid-IR observations 
(e.g. SN~2009ip).  Table~\ref{tab:objects} summarizes the distances and Galactic 
extinctions we use for each source, the transient properties
from \cite{Smith2011}, and the elapsed time since the transient
of the primary data we examine.  The radiated energy of the
transient is estimated as $E = \nu L_\nu \Delta t_{1.5}$ where
$\nu L_\nu$ is the luminosity corresponding to the peak absolute
magnitude and $\Delta t_{1.5}$ is the time scale for the transient
to fade by $1.5$~mag.  We have included SN~1961V,
SN~2008S and the NGC~300-OT for comparison even though we will not model their
properties in detail because we have done so elsewhere (\citealt{Kochanek2010},
\citealt{Kochanek2011}).

We start in \S\ref{sec:motivation} with an overview of the 
evolution of dusty shells.  We first outline the requirements
for dust formation and then make several estimates of expected
ejecta mass for either radiatively-driven or explosive transients.
Then we describe the optical depth and mid-IR evolution of 
expanding dusty shells.  In \S\ref{sec:data}, we summarize the available
data and how it will be analyzed.  In \S\ref{sec:objects}, we discuss
each source individually in terms of how well its properties can
be explained by an expanding dusty shell.
Finally, in \S\ref{sec:discussion}
we summarize the results and outline the need for future observations.

\section{The Physics of Dust in Stellar Transient Ejecta}
\label{sec:motivation}

The scenario we consider consists of a massive luminous star that for
a brief period of time (months to decades) undergoes a period of 
dramatically higher  mass loss.  Our general focus is not the initial 
transient but its possible mechanisms, the long term evolution of the ejected material
and its observational consequences.  
In \S\ref{sec:form} we outline the conditions for forming dust, 
and then in \S\ref{sec:mloss} we set limits on the mass 
loss rates.  Next, in \S\ref{sec:opdepth} we consider
the optical depth of the dust, the resulting optical depths, their
evolution and the observational consequences.  
In \S\ref{sec:dusty} we describe the DUSTY dust radiation transfer models 
we use for detailed models of SEDs.  Finally, in \S\ref{sec:caveats}
we discuss the effects of more complicated geometries and inhomogeneity.

\subsection{Dust Formation}
\label{sec:form}

The general assumption about the impostor transients is that they are cool 
($T_{peak} \simeq 7000$~K), luminous ($L_{peak}\simeq 10^7L_\odot$)
eruptions from otherwise massive ($M_* \gtorder 40M_\odot$),
hot ($T_* > 10^4$~K), luminous ($L_* > 10^{5.8} L_\odot$) stars (e.g. \citealt{Humphreys1994},
\citealt{Vink2009}).  In \cite{Kochanek2011b} we explored the physics
of dust formation around hot stars, finding that a necessary condition
for dust formation was that the ejecta wind becomes non-dust optically
thick and establishes a cool, external ``pseudo-photosphere'' with $T_{peak} \simeq 7000$~K.  
Hotter transients produce so many soft UV ($<13.6$~eV) photons that 
small dust grains photo-evaporate faster than they can grow.  
Such temperatures are typical of all these eruptions, and indeed of real SNe as well,
and are largely a consequence of the rapid rise in non-dust opacities
with temperature in this regime (\citealt{Davidson1987}).   As we discuss
below, such dense winds must also be optically thick in order to be 
radiatively accelerated.

We can view the ejecta as a time varying wind surrounding the star characterized
by mass loss rate $\dot{M}$, wind velocity $v_w$, and the mass fraction of 
condensible species $X_g$.  Not all the condensible mass need condense onto
grains, so the ultimate mass fraction of dust $X_d \leq X_g$.  The wind is produced by
a star/transient of luminosity $L_*$, photospheric radius $R_*$ and effective
temperature $T_*$ where $L_* = 4 \pi R_*^2 \sigma T_*^4$.  In \cite{Kochanek2011b}
we found that the requirement for dust formation is roughly 
$\dot{M} \gtorder 10^{-2.5} M_\odot$/year.  This limit is relatively
insensitive to other parameters because it is associated with the
``exponential'' drop in the production of photons towards short wavelengths.
  
Once the wind conditions are such that the dust formation region is shielded from
the hot stellar photosphere, dust formation can commence once the equilibrium
temperature of small grains is low enough (e.g. \citealt{Draine2011}).  This sets the formation radius at
\begin{eqnarray} 
      R_f &= &\left( { L Q_P(T,a_{min}) \over 16 \pi \sigma T_d^4 Q_P(T_d,a_{min}) }\right)^{1/2} \nonumber \\
               &= &5.2 \times 10^{14} \left( { L \over 10^6 L_\odot } \right)^{1/2}
                         \left( { 1500~\hbox{K} \over T_d } \right)^2 
                         \left( { Q_P(T_*,a_{min}) \over Q_P(T_d,a_{min}) } \right)^{1/2}~\hbox{cm} 
             \label{eqn:rform}
\end{eqnarray}
where $T_d$ is the dust destruction temperature, $Q_P(T,a_{min})$ is the Planck-averaged 
absorption efficiency for the smallest grains, and $L$ and $T$ are the transient luminosity
and temperature.  For small grains, $R_f$ is independent
of $a_{min}$ because $Q_P \propto a$, and a reasonable approximation for both the graphitic and
silicate dust models of \cite{Draine1984} is that   
\begin{equation} 
      R_f \simeq 1.0 \times 10^{15}  \left( { L \over 10^6 L_\odot } \right)^{1/2}
                   \left( { 1500~\hbox{K} \over T_d } \right)^2~\hbox{cm}
\end{equation}
under the assumption that the transient temperature is $T_* \simeq 7000$~K (see \citealt{Kochanek2011b}).  
Including the Planck factors makes the formation radius roughly three times larger than if
they are ignored.  The growth rate of the dust is proportional to the density, so dust
growth is essentially complete by the time $R \simeq 2 R_f$ because of the 
rapidly dropping density of the expanding ejecta.  Thus, the time scale on which
the dust begins to form is
\begin{equation}
      t_f = { R_f \over v_w } \simeq 0.6 \left( { L \over 10^6 L_\odot } \right)^{1/2}
                \left( { 1500~\hbox{K} \over T_d } \right)^2 \left( { 500~\hbox{km/s} \over v_w} \right)~\hbox{years}
             \label{eqn:tform}
\end{equation}  
for expansion velocity $v_w$.
In most cases, this is longer than the duration of the optical transient peak
(see Table~\ref{tab:objects}), suggesting that dust formation is limited by the 
fainter post-peak luminosity rather than the luminosity at peak and that the
initial decline in the luminosity is usually not due to dust formation.  Dust will not form if the transient temperature rises
significantly above $7000$~K because the increasing numbers of soft UV photons photo-evaporate
grains before they can grow (see \citealt{Kochanek2011b}).  
While the objects we consider generally lack the detailed
light curves needed to probe this phase, the signature of dust formation should be an abrupt
dimming in the optical combined with a sudden rise in the near/mid-IR due to the initially 
high dust temperatures ($T \simeq T_d \simeq 1500$~K).

The opacity of the grains depends somewhat on the maximum size to which they grow, $a_{max}$, 
(see \citealt{Draine2011}) and $a_{max} \propto \dot{M}$ depends on the density of the wind (see \citealt{Kochanek2011b}).   
Thus, the dust opacity at any wavelength depends on the grain composition and growth history,
but not by enormous factors given plausible distributions.  Hence, for our present study
we all simply adopt the standard opacities used by DUSTY 
(\citealt{Ivezic1997}, \citealt{Ivezic1999}, \citealt{Elitzur2001}).\footnote{These opacities can be derived
from the parameters DUSTY supplies for its dusty wind solutions, although it would be 
helpful if DUSTY simply reported the values.}  
We provide
the detailed parameters below in \S\ref{sec:dusty} and scale our analytic results
to an opacity of $\kappa_V = 100$~cm$^2$/g.   Many uncertainties in the opacities
can also simply be absorbed into the uncertainties in the dust mass fraction $X_d$.

\subsection{The Amount of Ejecta}
\label{sec:mloss}

The total mass of the ejecta plays a key role for any expectation about the subsequent
evolution.  The peak mass loss rate inferred for the Great Eruption of $\eta$ Carinae
is roughly $\dot{M}=1 M_\odot$/year for a period of order $\Delta t =10$~years, leading
to a total ejecta mass of $M \simeq 10 M_\odot$ (\citealt{Smith2003}).  Other shells
observed around hot stars in the Galaxy are inferred to have masses from $0.1M_\odot \ltorder M \ltorder 10M_\odot$
(e.g. \citealt{Clark2009}).  Unfortunately, we have little knowledge of their mass loss
rates.  Typical estimates assume that some estimate of the radial spread in the dusty
ejecta can be used to estimate the lifetime of the transient, and the generic estimates
are $\sim 10^4$~years and thus $\dot{M} \sim 10^{-4} M_\odot$/year.  As we argue in
\cite{Kochanek2011b}, this argument for the lifetimes is almost certainly wrong based on
both the data (it would mis-predict the statistical properties of the shell geometries
and the rate of observable transients) and our theoretical models (no dust would form
at such low mass loss rates).  Rather, the radial spread must be dominated by the 
spread in ejecta velocities and the true ages are much shorter and the mass loss rates
much higher.  The lower bound based on the need to form dust is $\dot{M} \gtorder 10^{-2.5}M_\odot/\hbox{year}$
(\citealt{Kochanek2011b}).  Very crudely, this suggests bounds of 
$0.1M_\odot \ltorder M \ltorder 10M_\odot$ and $10^{-2.5}M_\odot/\hbox{year} \ltorder \dot{M} \ltorder M_\odot/\hbox{year}$
for mechanisms having anything in common with the Galactic systems.    

If the ejection is radiatively driven, the radiation has to provide
the energy and momentum needed to generate the outflow, with negligible contributions
from the initial thermal energy of the material.  The momentum imparted to 
the ejecta satisfies $\dot{M} v_\infty = \tau L_{peak}/c$ where $v_\infty$ is the
asymptotic velocity and the optical depth $\tau$
determines the number of times a photon is scattered before escaping
(see the reviews of hot stellar winds by \citealt{Kudritzki2000} or
\citealt{Puls2008}).  For the 
high mass loss rates of interest, the wind has to be optically thick, with  
\begin{equation}
       \tau \simeq 2 \times 10^3 \left( { \dot{M} \over M_\odot/\hbox{year} }\right)
                   \left( { 10^7 L_\odot \over L_{peak} } \right)
                   \left( { M_* \over 40 M_\odot }    \right)^{1/2} 
                   \left( { 10^6 L_\odot \over L_* }  \right)^{1/4}
                   \left( { T_* \over 2 \times 10^4~\hbox{K} } \right)
                    { v_\infty \over v_e }~\hbox{km/s},
             \label{eqn:accel}
\end{equation}
where 
\begin{equation}
          v_e= \left( { 2GM_* \over R_* }\right)^{1/2}
             = 430 \left( { M_* \over 40 M_\odot }    \right)^{1/2} 
                   \left( { 10^6 L_\odot \over L_* }  \right)^{1/4}
                   \left( { T_* \over 2\times 10^4~\hbox{K} } \right)~\hbox{km/s},
         \label{eqn:vesc}
\end{equation}
is the escape velocity from the stellar surface set by the quiescent
luminosity $L_*$ and temperature $T_*$. For a line-driven wind,
$v_\infty = v_e |1-L_*/L_E|^{1/2}$ where $L_E$ is the (continuum opacity-dependent)
Eddington luminosity (again, see \citealt{Kudritzki2000} or \citealt{Puls2008}).  
Thus, the optical depth requirement
for accelerating the wind (Eqn.~\ref{eqn:accel}) generally results in the formation of a ``pseudo-photosphere''
in the wind (i.e. $\tau>1$) that will shield the dust formation region from soft ultraviolet
radiation and allow the formation of dust (see \citealt{Davidson1987}, \citealt{Kochanek2011b}).  
There is also an upper bound
on the mass loss rate set by using all available energy to lift material
out of the potential well of the star.  This is referred to as the 
``photon tiring limit'' by \cite{Owocki2004}. Given an observed   
transient luminosity $L_{peak}$ and temperature $T_{peak}$, the upper 
limit on the mass loss is
\begin{equation}
      \dot{M}_{max} \ltorder { 2 L_{peak} R_* \over G M_* } = 
             1.3 \left( { L_{peak} \over 10^7 L_\odot } \right) \left( { L_* \over 10^6 L_\odot }\right)^{1/2}
             \left( { 40 M_\odot \over M_* } \right) \left( { 2\times 10^4 \over T_* }\right)^2 
            \left( { T_0 \over T_{peak} } \right)^4~M_\odot/\hbox{year}.
            \label{eqn:mdotmax}
\end{equation}
Here we have modeled the losses due to accelerating the wind as a redshifting 
of the transient spectrum, as might be appropriate for a continuum opacity.  The
true transient luminosity is $L_{peak}(T_0/T_{peak})^4$ where $T_0>T_{peak}$ is the
unobserved temperature of the transient at the stellar surface.  The radiative
efficiency of this process is of order
\begin{equation}
       f = { 2 L_{peak} \over \dot{M} v^2 } 
      \simeq \left( { \dot{M}_{max} \over \dot{M} }\right) \left( { T_{peak} \over T_0 } \right)^4
\end{equation} 
and $f \gtorder 1$ unless the transient is in the extreme limit where most of the
underlying luminosity is absorbed into expelling the ejecta ($\dot{M} \simeq \dot{M}_{max}$
with $T_0> T_{peak}$).   The ``tiring limit'' on the mass loss rate (Eqn.~\ref{eqn:mdotmax})
combined with the (somewhat arbitrary) event duration $\Delta t_{1.5}$ from Table~\ref{tab:objects} leads to
a first rough estimate $M_E = \dot{M}_{max} \Delta t_{1.5}$ of the ejected mass.  As reported
in Table~\ref{tab:objects2}, most of the sources can have ejected only $0.1$-$1.0M_\odot$
of material even at the ``tiring limit,'' albeit with significant uncertainties from the 
sensitivity of the mass loss rate to the stellar escape velocity ($\propto L_*^{1/2} T_*^{-2}$).
Only SN~1961V, due to its high peak luminosity, and $\eta$~Carinae, due to its long
transient duration, can have ejected more than $10 M_\odot$ as radiatively driven transients.

In an explosion, the ejecta are shock heated and accelerated to temperatures and velocities
exceeding the escape velocity and then expand.  The observed luminosity is a balance between
the diffusion of radiation out of the ejecta and losses due to the expansion.  If the 
expansion time scale is $t_{exp}=R/v_w$ and the diffusion time scale for electron scattering
is $t_{diff} \simeq 9M_E\sigma_T /4\pi R m_p c$, then the transient duration is approximately the geometric mean  
\begin{equation}
    t_{rad} \equiv \left( t_{exp} t_{diff} \right)^{1/2}
         = \left( { 9 M_E \sigma_T \over 4 \pi m_p c v_w } \right)^{1/2}
         = 0.6 \left( { M_E \over M_\odot } { 500~\hbox{km/s} \over v_w } \right)^{1/2}~\hbox{years},
\end{equation}
if the expansion time is shorter than the diffusion time (e.g. \citealt{Falk1977}).  If we identify $t_{rad}$ with 
the event time scale $\Delta t_{1.5}$, then we obtain an estimate of the ejected mass
\begin{equation}
      M_{diff} = { 4 \pi m_p c v_w t_{rad}^2 \over 9 \sigma_T }
              = 0.018 \left( { t_{rad} \over 30~\hbox{days} }\right)^2 \left( { v_w \over 500~\hbox{km/s} }\right)~M_\odot.
      \label{eqn:mdiff}
\end{equation}
Table~\ref{tab:objects2} reports $M_{diff}$ for each source.  These mass estimates tend to
be slightly smaller than the estimated upper limits $\dot{M}_{max}\Delta t_{1.5}$ for radiatively driven 
transients except when the transient durations are long.  SN~1961V and $\eta$~Carinae
again stand out as having the largest mass loss estimates.  The radiative efficiency
of an explosive transient is roughly determined by
\begin{equation}
    { t_{exp} \over t_{diff} } = { c m_p L_{peak} \over 9 M_E \sigma_T v_w \sigma T_{peak}^4 }
        \simeq 0.02 \left( { L_{peak} \over 10^7 L_\odot } \right)
                    \left( { 7000~\hbox{K} \over T_{peak} } \right)^4
                    \left( { M_\odot \over M } \right)
                    \left( { 500~\hbox{km/s} \over v_w } \right),
\end{equation}
which means that explosive transients are generally radiatively inefficient because they lose
energy due to expansion faster than it can be diffusively radiated.

This suggests that uncertainties about the mechanism can be described by the radiative efficiency $f$,
where the ejected mass is
\begin{eqnarray}
      M_E = { 2 E_{rad} \over f v^2 } 
          = 0.04 f^{-1} \left( { L_{peak} \over 10^7 L_\odot} \right) \left( { \Delta t_{1.5} \over 30~\hbox{days} }\right)
                  \left( { 500~\hbox{km/s} \over v_w } \right)^2~M_\odot. 
            \label{eqn:me}
\end{eqnarray}
Radiatively driven transients tend to have $f>1$ and explosive transients tend to have $f<1$, while
radiatively driven transients close to the tiring limit and weak explosions have $f \simeq 1$.
This leads to the final mass estimate in Table~\ref{tab:objects2}.  

\begin{figure}[p]
\centerline{\includegraphics[width=5.5in]{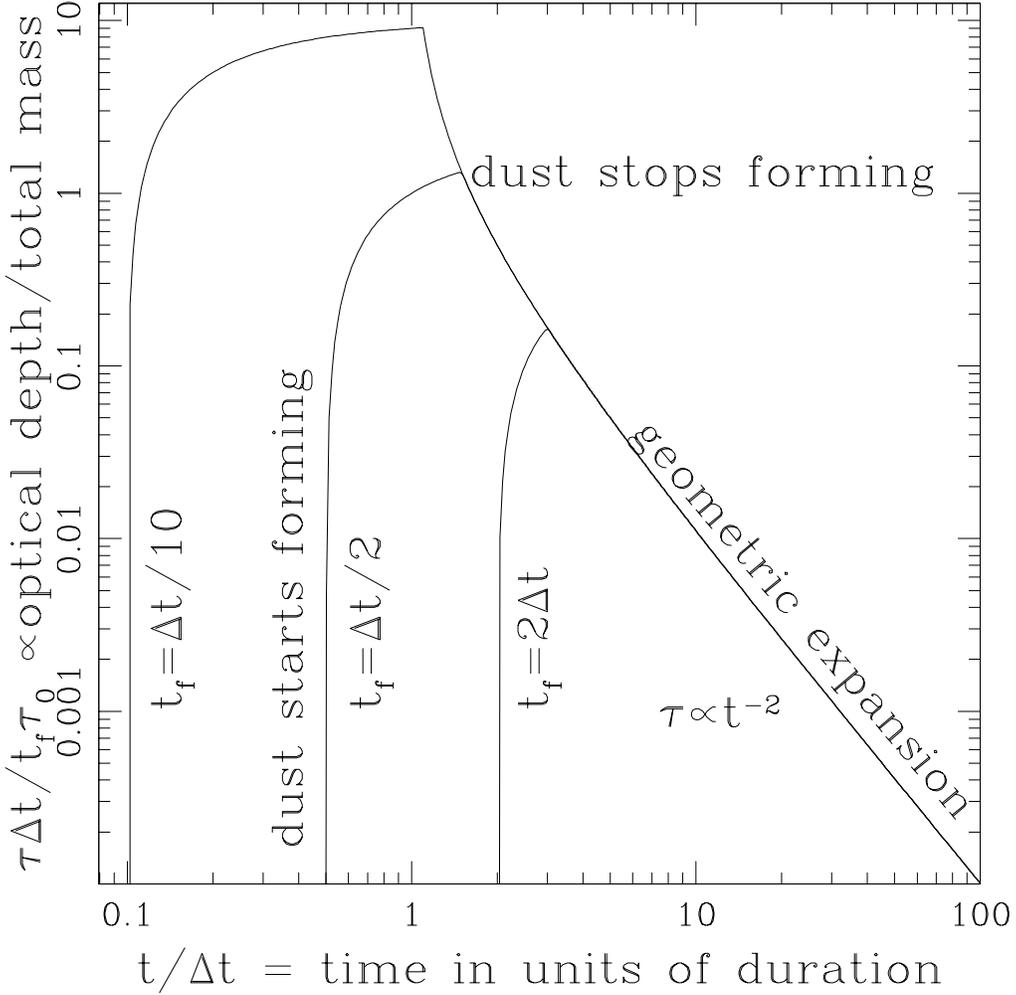}}
\caption{ The optical depth evolution of a transient (Eqn.~\ref{eqn:twind}).
  The time is in units of the transient duration, $t/\Delta t$.  The optical
  depth is normalized as $\tau\Delta t/(t_f\tau_0)$, which corresponds to 
  the optical depth per unit ejected mass ($\propto \tau/M_E$).  The scalings
  are shown for dust formation time scales that are short ($t_f=\Delta t/10$),
  comparable ($t_f=\Delta t/2$) or long ($t_f=2\Delta t$) compared to the  
  transient duration.  The optical depth rises from $t_f$ until $t_{peak}=\Delta t+t_f$,
  but it has a higher peak and an increasingly long plateau for a given 
  amount of ejected mass as the formation time scale becomes shorter 
  compared to the transient duration.  The maximum possible optical
  depth corresponds to a fully formed wind with the mass loss rate
  of the transient.  The optical depth then drops on time scale 
  $t_f$ and finally asymptotes to the $\tau \propto t^{-2}$ of an 
  expanding shell.   
  }
\label{fig:schem}
\end{figure}

\subsection{Optical Depth and Temperature Evolution}
\label{sec:opdepth}

Where dust formation and growth depends on density through $\dot{M}$ and $v_w$, the
optical depth also depends on the duration of the transient $\Delta t$ and the time of observation $t$. 
These are related to the mass of the ejecta $M_E=\dot{M} \Delta t$, the radius of the
shell $R_{out} = v_w t$  and the thickness of the shell $R_{out}-R_{in}=\Delta R = v_w \Delta t$.
These quantities likely vary with azimuth, leading to some smearing of the dust properties,
temperatures and opacities with viewing angle.  The outermost ejecta is at  
\begin{equation}
    R_{out} =  1.6 \times 10^{16} \left( { v_w \over 500 \hbox{km/s} } \right)   
              \left( { t \over \hbox{decade} } \right)~\hbox{cm},
      \label{eqn:radius}
\end{equation}
and the optical depth of the shell at wavelength $\lambda$ is
\begin{equation}
      \tau_\lambda = { \dot{M} \kappa_\lambda \over 4 \pi v_w } \left( { 1 \over R_{in} } - {1\over R_{out}}\right)
       \label{eqn:opdepth2}
\end{equation}
where $\kappa_\lambda$ is the dust opacity.  If we are focused on the apparent extinction
of a central source, the appropriate optical depth (or opacity)
for these analytic approximations is the effective absorption optical depth
$\tau_e = \left(\tau_{abs}(\tau_{abs}+\tau_{scat})\right)^{1/2}$ rather than the
total $\tau_{tot} = \tau_{abs} + \tau_{scat}$, absorption $\tau_{abs}$ or 
scattering $\tau_{scat}$ optical depths.  These distinctions are important,
particularly for silicate dusts with their larger scattering opacities
(e.g. \citealt{Draine1984}).

It is useful to recast this in terms of the time evolution of the optical depth under the assumption
of a constant expansion velocity,
\begin{equation}
    \tau(t) = \tau_0 t_f \left[ { 1\over \hbox{max}(t_f,t-\Delta t) } - { 1 \over \hbox{max}(t_f,t) }\right]
      \quad\hbox{where}\quad \tau_0 = { \dot{M} \kappa_\lambda \over 4 \pi v_w R_f }.
   \label{eqn:twind}
\end{equation}
Here $t_f = R_f/v_w$ is the time scale on which dust begins to form (Eqn.~\ref{eqn:rform}), $\Delta t$ is the duration of the
transient, and $\tau_0$, the optical depth for a wind extending from $R_f$ to infinity,
 sets the peak optical depth.  As illustrated in Fig.~\ref{fig:schem}, the optical depth begins to rise at $t=t_f$ when
material first reaches the dust formation radius. It peaks with optical depth $\tau_0 \Delta t/(t_f+\Delta t)$
at time $t_{peak}=t_f +\Delta t$ when all the material has been ejected and the last material to be 
ejected lies at the dust formation radius.  If the duration of the transient $\Delta t$ is long compared 
to the flight time to the dust formation radius, $t_f$, the optical depth approaches the value for a
fully formed wind, $\tau_0$, and is then almost constant for the remaining duration of the transient.  
The optical depth then drops as 
\begin{equation} 
       \tau(t=t_{peak}+\delta)= { \tau_0 t_f \Delta t \over (t_{peak}+\delta)(t_f + \delta) } 
        \label{eqn:tevol}
\end{equation}
where $\delta$ is the time elapsed since the optical depth peak.  The initial drop of $\tau \propto 1/(t_f+\delta)$
on timescales of $t_f=R_f/v_w$ is rapid, and then once $\delta > t_{peak}$ it asymptotes to the quadratic decline of a thin shell, 
$\tau(t)=\tau_0 t_f \Delta t/t^2$.  Taking the limit of a thin shell at late times, the optical depth is
\begin{equation}
     \tau_\lambda =  { \dot{M} \kappa_\lambda \over 4 \pi v_w } { \Delta R \over R_{out}^2 }
               = { M_E \over 4 \pi } { \kappa_\lambda \over R_{out}^2 }
               = 64 \left( { M_E \over M_\odot } \right) \left( { \kappa_\lambda \over 100~\hbox{cm}^2/\hbox{g} }\right)
                    \left( { 500~\hbox{km/s} \over v_w } \right)^2 \left( { \hbox{decade} \over t }\right)^2.
    \label{eqn:taukap}
\end{equation}
Table~\ref{tab:objects2} gives the estimated visual optical depth $\tau_V$ for the sources we consider at
the time of the primary observations assuming an opacity
of $\kappa_V=100$~cm$^2$/g and scaled by $M_E$ (Eqn.~\ref{eqn:me}).  

A transient can maintain high optical depth only for the duration of the transient, after
which the optical depth begins to decline very rapidly and the central source should become
steadily brighter.  
The rate of brightening for $t>t_{peak}=\Delta t + t_f$ is 
\begin{equation}
      { d m \over dt } = - { 2.5 \over \ln 10 } { \tau \over t } \left( 1 + { t \over t - \Delta t }\right)
               \rightarrow - { 5 \over \ln 10 } { \tau \over t },       
\end{equation}
where the limit is for very late times when $t \gg t_{peak}$.  Close to the end of the transient,
the rate of change can be very much faster because the term in parenthesis approaches 
$2+\Delta t/t_f$ as $t \rightarrow t_{peak}$ (see Fig.~\ref{fig:schem}).   Even at late times, a source detected   
with a significant optical depth at time $t$ doubles in flux
on the fast time scale of $t/(3\tau)$.
This evolution could be balanced by changes
in the star, and modeling the star as a black body, the luminosity at wavelength $\lambda$ evolves as
\begin{equation}
     \lambda L_\lambda = L(t) G(x) \exp(-\tau_{e,\lambda}(t))
  \quad\hbox{where}\quad G(x) = { 15 \over \pi^4 } { x^4 \over \exp(x)-1 }
  \quad\hbox{and}\quad x = 1.44 \left( { \mu\hbox{m} \over \lambda } {10^4~\hbox{K} \over T(t)}\right).
   \label{eqn:levolve}
\end{equation}
The spectral term $G(x)$ has a maximum at $x_{max}=3.92$ corresponding to $T\simeq 6700$~K for the V band,
so these transients tend to be most efficient at producing visual light near peak.  

A recurring theme of our paper, however, will be that it is very hard to fight the
optical depth evolution because it is large and sitting in an exponential (Eqn.~\ref{eqn:levolve}). 
Most post-transient detections of surviving stars are interpreted as requiring an 
obscuring optical depth of $\tau_{e,V} \simeq 3$ (see \S\ref{sec:objects}).  If this is the
optical depth at time $t$, then at time $2t$ the optical depth will have dropped
to $\tau_{e,V}=3/4$ and the star should be almost ten times brighter.  To a limited
extent, such changes can be balanced by lowering the stellar luminosity or raising 
its temperature, but it requires fine tuning (there are no obvious physical correlations
between the optical depth and expansion of the shell with the stellar properties)
and can only be fine tuned at one wavelength.  To hold the apparent flux constant
in the face of a $\Delta \tau_e=2.3$ reduction in optical depth requires either reducing
the stellar luminosity by an order of magnitude or the equivalent of raising the
stellar temperature from that of the cool, eruptive state ($T_{peak}\simeq 7000$~K) 
to a typical quiescent temperature of a hot star, $T_* \simeq 27000$~K.  It is
essentially impossible for the star to balance significantly
larger changes in optical depth.

Once the dust has formed and the grains have grown in size, the Planck absorption
efficiencies are significantly different than those for the smallest grains that
control dust formation and we need a temperature averaged over a distribution of
grain sizes.  Individual grain temperatures are still controlled by 
$L_* \pi a^2 Q_P(T_*,a)/4\pi R^2 = 4 \pi a^2 \sigma T^4 Q_P(T,a)$, 
but for a particle size distribution $dn/da$ we can estimate a mean temperature
by averaging this equation over the grain size 
distribution so that $L_* Q_P(T_*) = 16 \pi \sigma T^4 Q_P(T)$.  It is convenient
to use polynomial approximations for the size averaged Planck factors.\footnote{For
$Q_P(T) \equiv \int da (dn/da) a^2 Q_P(T,a)/\int da (dn/da) a^2$
and the \cite{Mathis1977} distribution $dn/da \propto a^{-3.5}$ with $0.005 \mu\hbox{m} < a < 0.25 \mu\hbox{m}$ assumed by DUSTY
we can build an adequate approximation for $Q_P(T)$  as follows.  Let $Q_1(T)$ apply for $t=\log(T)<t_0$
and $Q_2(T)$ apply for $t>t_0$  with $\log_{10} Q_i = \sum_{j=0}^3 a_{ij} (t-t_0)^j$
where $a_{10} = a_{20}$ forces continuity at $t=t_0$.  
For \cite{Draine1984} graphitic dusts over the temperature range $1 < t < 5$, we find
$a_{1j} = -1.967$, $-1.379$, $-2.605$, $-0.641$,
and 
$a_{2j}=  -1.967$, $-1.384$, $3.054$, $-0.940$ 
with $t_0=2.867$, an rms error of $0.021$~dex
 and a maximum error of $0.073$~dex.  
For \cite{Draine1984} silicate dusts we find
$a_{1j} = -1.968$, $-1.388$, $-2.612$, $-0.6434$ 
and 
$a_{2j}= -1.968$,  $-1.375$, $3.050$,  $-0.940$ 
with $t_0=2.869$, an rms error in $0.035$~dex 
and a maximum error of $0.094$~dex.  
}
With these assumptions the mean dust temperature is 
\begin{equation}
        T = \left( { L_* \over 16 \pi \sigma R^2 } \right)^{1/4} \left( { Q_P(T_*) \over Q_P(T) } \right)^{1/4}
          = 270 \left( { L_* \over 10^6 L_\odot }\right)^{1/4} \left( { 500 \hbox{km/s} \over v_w } \right)^{1/2}  
                 \left( { Q_P(T_*) \over Q_P(T) } \right)^{1/4} \left( { \hbox{decade} \over t } \right)^{1/2} \hbox{K}.
         \label{eqn:tdust}
\end{equation} 
The corrections due to the Planck factors are significant and typically increase the expected
temperatures by almost a factor of two. 
Assuming a $\lambda^{-1}$ emissivity, $\lambda L_\lambda$
then peaks at $\lambda = 2.9(1000~\hbox{K}/T)\mu$m and we report this nominal peak wavelength
in Table~\ref{tab:objects2} assuming a source with $L_*=10^6 L_\odot$ and $T_*=10^4$~K.  The
oldest sources should have SEDs peaking at $24\mu$m while the youngest sources should peak in
the short wavelength IRAC bands.

\subsection{DUSTY Models}
\label{sec:dusty}

In addition to these simple scaling laws, we will model the SEDs using
DUSTY (\citealt{Ivezic1997}, \citealt{Ivezic1999}, \citealt{Elitzur2001}),
which properly solves for radiation transfer through a spherically symmetric
dusty medium.  We use models based on either pure silicate or pure graphitic
dust from \cite{Draine1984}.  DUSTY parametrizes its models by the total optical
depth from both absorption and scattering $\tau_V \equiv \tau_{tot} = \tau_{abs}+\tau_{scat}$, while
the optical depth in our analytic models is the effective absorption optical depth
$\tau_e \simeq \left( \tau_{abs} \tau_{tot} \right)^{1/2}$ representing
the net absorption. At V band, the DUSTY models
have total opacities of $\kappa_{tot} \simeq 170$ and $84$~cm$^2$/g assuming
a bulk density for the dust of $\rho_{bulk}=3$~g/cm$^3$ and a dust to gas ratio of
$X_d=0.005$ ($\kappa \propto (\rho_{bulk}X_d)^{-1}$),
scattering opacities of $\kappa_{scat} = \kappa_{tot}w \simeq 80$
and $72$~cm$^2$/g, absorption opacities of $\kappa_{abs}=\kappa_{tot}(1-w)\simeq 90$
and $12$~cm$^2$/g, and scattering albedos of $w=0.47$ and $0.86$ for
graphitic and silicate dusts, respectively.
Thus, the effective absorption
opacities we should use in the analytic models are
$\kappa_e = (\kappa_{abs}\kappa_{tot})^{1/2}\simeq 120$ and $31$~cm$^2$/g,
respectively.  The effective visual absorption optical depths for
a DUSTY model defined by $\tau_{e,V}$ are then $\tau_e = (1-w)^{1/2}\tau_V=0.73 \tau_V$ and $0.37\tau_V$
for graphitic and silicate dusts, respectively, so we should generically
find that silicate models require DUSTY model optical depths roughly
twice that of graphitic models in order to produce the same net V-band absorption.
These distinctions become unimportant in the mid-infrared because the
opacities are much smaller, with $\kappa_{tot}\simeq 5.9$ and $1.7$~cm$^2$/g
at $3\mu$m for the graphitic and silicate dusts, respectively, and because
the opacity is dominated by absorption as the albedos are only
$w\simeq 0.13$.

It is important to remember that the effects of circumstellar dust are quantitatively
different from the effects of a foreground, intervening dust screen even for dust
with identical physical properties because the geometries have different balances
between absorption and scattering (e.g. \citealt{Goobar2008}).  
As already noted, the relationship between an apparent
visual extinction of $A_V$ and the DUSTY optical depth $\tau_V$ is not
$A_V = 2.5 \tau_V/\hbox{\rm ln} 10$ but $A_V \simeq 2.5 \tau_V (1-w)^{1/2}/\hbox{\rm ln } 10$ to
first order. It is also important to remember that dust contributes to the
emission, and that for winds these contributions begin to matter in the near-IR.
These differences lead \cite{Walmswell2011} to significantly overestimate the
effects of dusty winds on the photometric properties of supernova progenitors
on three levels.  First, a given amount of mass loss produces less net absorption
(by a factor of $(1-w)^{1/2}\simeq 0.4$ for silicates). Second, for the same amount of
net absorption at V band, a silicate wind absorbs more strongly at bluer
wavelengths and less at redder.  Third, the shift to increased emission
compared to a dust free star already starts at H-band for a $1500$~K
dust formation temperature.  These issues will be important for
our discussion of SN~1961V in \S\ref{sec:objects}.

We generally modeled the dust geometry as a shell with a ratio between the inner and outer radii
$R_{out}/R_{in}=2$ and we are generally citing $R_{in}$ as the shell radius because
the optical depth is dominated by the inner edge. We experimented
with shells where the radius ratio is $1.2$ and $4$, although in most
cases changing the thickness had little effect.
The shell is heated by a simple
blackbody model of the central star with temperature $T_*$ and luminosity $L_*$.
We tabulated models for both dust types as a function of stellar temperature ($T_*=5000$, $7500$, $10000$, $15000$, $20000$,
$30000$ and $40000$~K), dust temperature ($T_d=50$ to $1500$~K in steps of $50$~K) and optical depth ($\tau_V=0$
to $6$ in steps of $0.1$ and $\tau_V=6$ to $30$ in steps of $0.5$ and $\tau_V=40$ to $100$ in steps of 10).  The dust temperature is
related to the shell radius and stellar luminosity, so we will usually set the dust temperature
based on an estimate of the shell radius given the elapsed time and an estimate of the expansion velocity (Eqns.~\ref{eqn:radius}
and ~\ref{eqn:tdust}).

\subsection{Asymmetry and Inhomogeneity}
\label{sec:caveats}

While DUSTY addresses any shortcomings of the analytic models for spherical 
systems, the ejecta probably are not spherically symmetric (as illustrated by
the structure of $\eta$~Carinae).  For dust emission, the primary consequence of 
asymmetry is that the spread in radius with azimuth produces a spread in the dust
temperature that is likely far greater than the spread introduced by the finite
shell thickness.  In our quantitative models we use shells with $R_{out}/R_{in}=2$
in large part to mimic this effect, although in many cases changes in
the shell thickness have few consequences for interpreting the available data.  
Geometry can produce differences between the apparent optical depths in
absorption, as measured by the optical/UV flux, and  emission, as
measured by the mid-IR flux.  The observed optical/UV flux is
dominated by scattered photons rather than direct emission along the 
line of sight, so the two measures will be relatively tightly coupled.  
In silicate DUSTY models with $R_{out}/R_{in}=2$, the half-light radii of 
the total V-band luminosity are roughly $0.3$, $1.2$ and $1.5 R_{in}$ 
for optical depths of $\tau_V=1$, $10$ and $100$.  Because scattering is not
strongly forward directed, the surviving optical/UV photons ``sample'' a large 
fraction of the ejecta rather than a narrow pencil beam along the line of 
sight to the source (see, for example, the models of $\eta$ Carinae by \citealt{Davidson1975}).  
As a result, we should not see enormous differences between these two 
means of estimating the optical depth even in complex geometries.  

Complex geometric structures also have only limited effects on the 
expected evolution of the emission.  Consider the optical flux
escaping a thin, axisymmetric, expanding shell under the
assumption that the photon directions are isotropic at the
surface.  At some initial time $t_0$ the shell has optical
depth $\tau_0(x)$ in direction $x=\cos\theta$, so the observed
flux at later times is essentially 
\begin{equation}
   \propto \int dx \exp\left(-\tau_0(x)t_0^2/t^2\right)
         \simeq (t/t_0) \left|\tau_0''(x_{min})\right|^{-1/2} 
          \exp\left(-\tau_0(x_{min}) t_0^2/t^2\right)
    \label{eqn:angstruc}
\end{equation} 
where $x_{min}$ is the direction in which the optical depth
is lowest (sum over multiple minima if needed).  The pre-factor, 
which is the effective area over which photons escape (and becomes
a constant in the limit that there are no significant optical
depth variations over the shell), only makes it harder to 
combat the geometric evolution of the optical depth.
Adding an anisotropic scattering function ($dx \rightarrow g(x) dx$)
or making the distribution non-axisymmetric ($\tau_0(x,\phi)$) does not
qualitatively change anything because neither changes
the temporal scaling of the exponential.  Thus, geometry
complicates the relationship between optical depth and 
ejected mass, but does not fundamentally alter the temporal
scaling of the optical depth.

Another potential problem is the effect
of small scale optical depth fluctuations on the observed ``effective''
optical depth -- when the optical depth is high, the observed optical
flux becomes exponentially sensitive to leaks through the optically thick
medium, similar to the arguments about super-Eddington winds in \cite{Owocki2004}.  
At least in simulations of a turbulent molecular cloud, \cite{Ostriker2001} 
found log-normal column density fluctuations with a variance on the order of a factor of 2.
While a model of dust radiation transfer through an expanding three dimensional medium
with large spatial variations in the optical depth is beyond the scope of
this paper, it is instructive to consider a simple model.  Suppose the
dust optical depth has a log normal distribution about $\tau_0$ with 
logarithmic fluctuations $\sigma$, and we estimate the effective 
optical depth as
\begin{equation}
   \tau_e = -\ln \left[ { 1 \over (2\pi)^{1/2} \sigma } \int_{-\infty}^{\infty} d\ln \tau
          \exp(-\tau) \exp(-(\ln^2(\tau/\tau_0)/2\sigma^2) \right],
    \label{eqn:inhomog}
\end{equation}
corresponding to averaging the fraction of escaping radiation over 
the probability distribution of optical depths.  If we assume
$\sigma=\ln 2$ following the simulations of turbulent molecular 
clouds, then the effective optical depths for $\tau_0=1$, $3$, $10$, $30$
and $100$ are $\tau_e=0.99$, $2.3$, $4.9$, $8.7$ and $14.9$, respectively.
Reducing the amplitude of the fluctuations $\sigma$ brings the effective 
optical depth closer to the mean optical depth, but even if we halve
the amplitude to $\sigma=(1/2)\ln 2$, we find $\tau_e=1.00$, $2.7$, $7.2$,
$15.4$ and $30.5$.  
The problem, however, with using clumping to mitigate the optical depth evolution
is that the clumping itself generally evolves towards larger density
inhomogeneities at later times (e.g. \citealt{Garcia1996}, \citealt{Toala2011}
for stellar winds/shells).  While constant inhomogeneities can slow the
evolution of the effective optical depth, their growth as the shell expands will greatly
accelerate it, to say nothing of the additional acceleration of the evolution
produced by the geometric pre-factor in Eqn.~\ref{eqn:angstruc}.

\section{Data and Analysis}
\label{sec:data}

Table~\ref{tab:log} presents our estimates of the Spitzer IRAC 
($3.6$, $4.5$, $5.8$ and $8.0\mu$m) and MIPS $24\mu$m fluxes
for our targets. 
We downloaded the Post-Basic Calibrated Data (PBCD) for these programs
from the Spitzer archive.  These IRAC images are two-times oversampled
and have a pixel scale of 0\farcs60, while the MIPS $24\mu$m 
images have a pixel scale of 2\farcs45. 
Where there were multiple epochs, we aligned and combined the data using the
ISIS (\citealt{Alard1998}, \citealt{Alard2000}) image subtraction package.  
We also used ISIS to difference image between the available epochs, to search 
for any signs of variability, and to difference image between the IRAC 
wavelengths.  When we scale and subtract the $3.6\mu$m image from 
the longer wavelength IRAC images, all normal stars vanish to leave
only the red stars dominated by dust emission and emission by the 
interstellar medium (see \citealt{Khan2010}).
This wavelength differencing procedure isolates the relatively
rare, dusty stars while eliminating the crowding from the normal stars.
Extended PAH emission at $8.0\mu$m sometimes cause the procedure to fail.
The residual fluxes in the longer wavelength bands are then a good
indicator of the dust emission.  In some cases our astrometric match
is imperfect, but dusty stars are relatively rare so we have chosen
to adopt obvious mid-IR sources as the counterparts -- in general, 
rejecting this possibility then requires the transient to have been
a true SNe with no surviving star.

We estimated the source fluxes using aperture photometry (the IRAF apphot 
package). In some cases, especially when the local environment was crowded
and the background strongly variable, we would find large variations in
the source flux for different choices of the source and sky apertures.
For this reason, we measured the fluxes of the sources $f_i$ using the nine
combinations of three source apertures ($1\farcs2$, $1\farcs8$,
and $2\farcs4$) and three sky annuli ($2\farcs4$-$7\farcs2$,
$7\farcs2$-$14\farcs4$, and $14\farcs4$-$24\farcs0$).  For the 
standard $2\farcs4$ aperture we used the standard Spitzer aperture
corrections, while the non-standard aperture corrections were derived 
from bright stars in the images.   The larger source apertures 
more securely determine the source flux while being more sensitive
to uncertainties in the sky estimate, while the smaller source
apertures require larger aperture corrections but are less sensitive
to uncertainties in the sky estimate.    We then calculated the mean flux 
\begin{equation}
\langle{f}\rangle = { \sum_i f_i\sigma_i^{-2} \over \sum_i \sigma_i^{-2} }
\end{equation}
combining all nine flux estimates $f_i$ weighted by their estimated statistical
uncertainties $\sigma_i$.  We estimated the uncertainties as a statistical
error $\sigma_0$, defined by the statistical error estimate for the $1\farcs8$ 
radius source aperture and the $2\farcs4$-$7\farcs2$ background annulus, combined
with the systematic scatter between the nine flux estimates,
\begin{equation}
   \sigma^2 = \sigma_0^2 +
   { \sum_i \left( f_i - \langle{f}\rangle \right)^2 \sigma_i^{-2} \over
          \sum_i \sigma_i^{-2} },
\end{equation}
to include an estimate of the systematic uncertainties.

In general we could use HST photometry of the sources from the literature.
There were two exceptions, the near-IR limits for SN~1999bw and the
optical limits for SN~2002bu.  Here we estimated the detection limits
presented in Table~\ref{tab:hstphot} using simple aperture photometry
and standard aperture corrections.  For the ACS/HRC data the apertures,
aperture corrections and zero points were taken from \cite{Sirianni2005},
while those for the NICMOS/NIC2 observations were taken from the
NICMOS data handbook.\footnote{http://www.stsci.edu/hst/nicmos/documents/handbooks/DataHandbookv8/nic\_ch5.9.3.html}

Finally, we have been monitoring NGC~2403 (SN~1954J and SN~2002kg/V37) and 
NGC~3627 (SN~1997bs) in the UBVR bands with the Large Binocular Telescope (LBT) 
as part of our project to set 
limits on the existence of failed supernovae (\citealt{Kochanek2008}).  
After carrying out standard data reductions, these data were also
analyzed with ISIS.  The ISIS reference images were analyzed
with DAOPHOT (\citealt{Stetson1987}).  Table~\ref{tab:lbtphot} presents 
the DAOPHOT magnitudes at the positions of SN~1954J and SN~1997bs in 
the ISIS reference images and the light curve of SN~2002kg.  There is no clearly 
detected source at the position of SN~1997bs and the source at the position
of SN~1954J is known to be composite of several unresolved stars (see
below), so these should be regarded as upper limits.
  
We provide the differential ISIS light curves of SN~1954J and SN~1997bs in
Table~\ref{tab:isisphot}.  These are the changes in flux at each epoch
relative to the reference image.  We present the differential light curves
because the absolute fluxes
of these sources in the reference image are not known due to crowding.
Fortunately, the images are not crowded with
variable stars, so any observed variability is almost certainly
dominated by a single source.
In addition to the ISIS light curves,
we can use the subtracted images to visually and quantitatively
characterize the variability.  In what follows, it is important to
realize that all the subtracted images are on the common flux 
scale of the reference image.  First, we try to convolve the
images to a standard FWHM corresponding to a Gaussian of
dispersion $\sigma_0=3$ pixels ($0\farcs672$) by convolving
each image with a Gaussian of width 
$\sigma^2=\hbox{max}(0,\sigma_0^2-FWHM^2/8\ln2)$.  We
then produce images representing the best fit linear model
of the time variability of the pixels in each image $I_i$
and the rms residual from this model.  If $\Delta t_i=t_i-\langle t_i\rangle$
is the time relative to the average epoch $\langle t_i \rangle$,
we have the mean image $\bar{\Delta I} = \langle \Delta I_i\rangle$,
the slope image
$S = \langle \Delta t_i \Delta I_i\rangle/\langle \Delta t_i^2 \rangle$     
and the rms residual image
$\langle \sum_i \left(\Delta I_i - \bar{\Delta I} - S \Delta t_i \right)^2 (N-2)^{-1}\rangle^{1/2}$.
The slope image emphasizes steady trends while the rms
image emphasizes more erratic or periodic behavior.

\section{Results}
\label{sec:objects}

We now discuss each object in turn, beginning with a short summary of the
properties of the system followed by our new results.  Table~\ref{tab:objects}
summarizes the general properties of each transient, and Table~\ref{tab:objects2}
summarizes our crude estimates of the ejected mass, the expected optical depths
and the peak wavelength of any mid-IR emission.  Some of these estimates
are very sensitive to the expansion velocity we adopted from \cite{Smith2011}
since $\tau \propto v^{-2}$ even for a velocity-independent estimate of the
ejected mass.  We start with  an examination of $\eta$ Carinae because of
the frequency with which it is used as an analogy to the SN impostors. 
We note, however, that while the physics of its expanding shell should be
similar, it had radically different energetics, time scales
and (probably) mass budgets from most of the other systems.  For example, if the impostors have radiative efficiencies
$f \simeq 1$ like $\eta$ Carinae, they are almost two orders of magnitude
less energetic ($\Delta t_{1.5} \nu L_\nu$) and have ejected far less mass
(see Table~\ref{tab:objects2}).  As a result, the optical depths expected
for the typical impostor are relatively low and it is difficult to obscure a 
surviving star for extended periods of time.  

\begin{figure}[p]
\centerline{\includegraphics[width=5.5in]{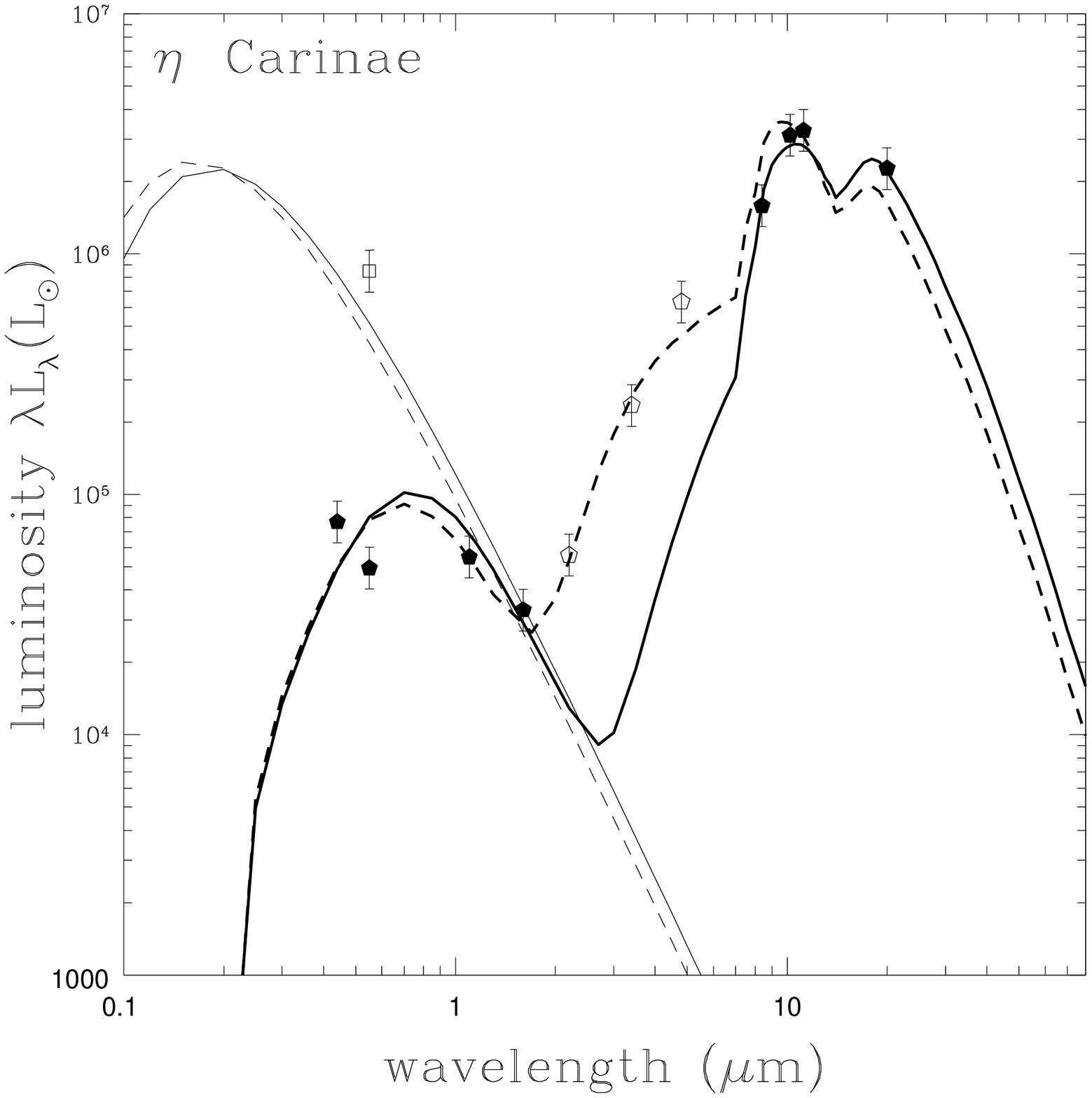}}
\caption{ The spectral energy distribution of $\eta$ Carina circa 1974.  The pentagons
  show the optical, near-IR and mid-IR luminosities from \cite{vanGenderen1984},
  \cite{Whitelock1994} and \cite{Robinson1973}.  The solid curves show the 
  standard model fits where the shell thickness $R_{out}/R_{in} \equiv 2$. 
  These models are only fit to the filled pentagons and excluded the near-IR
  data (the open pentagons).  The dashed curves show the better fits obtained 
  by allowing a thicker shell with $R_{out}/R_{in}=5.7$ in 1974.  
  These models were fit to the full SED. 
  The thick curves show the model SEDs, while the thin curves show the
  black body models of the intrinsic stellar SED fit to the
  estimated quiescent V-band magnitude (open square).
  }
\label{fig:etased}
\end{figure}

\begin{figure}[p]
\centerline{\includegraphics[width=5.5in]{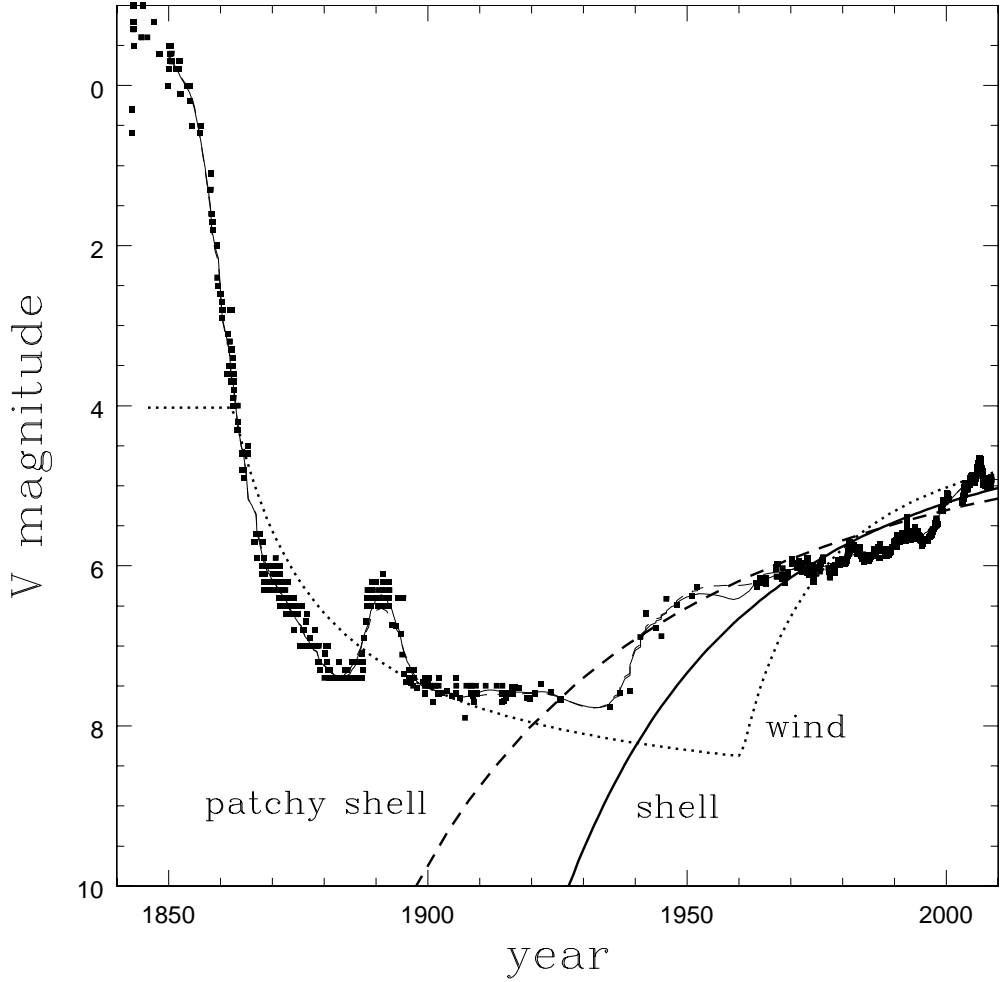}}
\caption{ V band light curve (points) of $\eta$ Carinae from the tabulation by \cite{Fernandez2009}.
  The curves show 5 models for the light curves.  The thick solid (dashed) curves
  show the evolution expected if we simply evolve the optical depth of the $T_*=20000$~K
  SED model (the $R_{out}/R_{in}=2$ model in Fig.~\ref{fig:etased}) using the geometric
  $\tau \propto t^{-2}$ scaling of the optical depth for a thin expanding shell assuming it is 
  homogeneous (inhomogeneous following Eqn.~\ref{eqn:inhomog}).  While these models assume a fixed stellar 
  luminosity and temperature, no physically plausible changes in these variables
  can overcome the optical depth evolution of these models and reproduce the observed light curve.
  The thick dotted line shows a wind model (Eqn.~\ref{eqn:twind}) with fixed luminosity 
  and temperature where dust formation (somewhat arbitrarily) begins in 1862 and 
  ends in 1943.  The differences are now small enough that they can be explained by variations
  in luminosity and temperature.  The thin solid and dashed lines, which largely
  overlap and are hidden by the data points, are the non-parametric models of the light curve 
  from Fig.~\ref{fig:etamodel} for the homogeneous and inhomogeneous absorption models. 
  }
\label{fig:etalc}
\end{figure}

\begin{figure}[p]
\centerline{\includegraphics[width=5.5in]{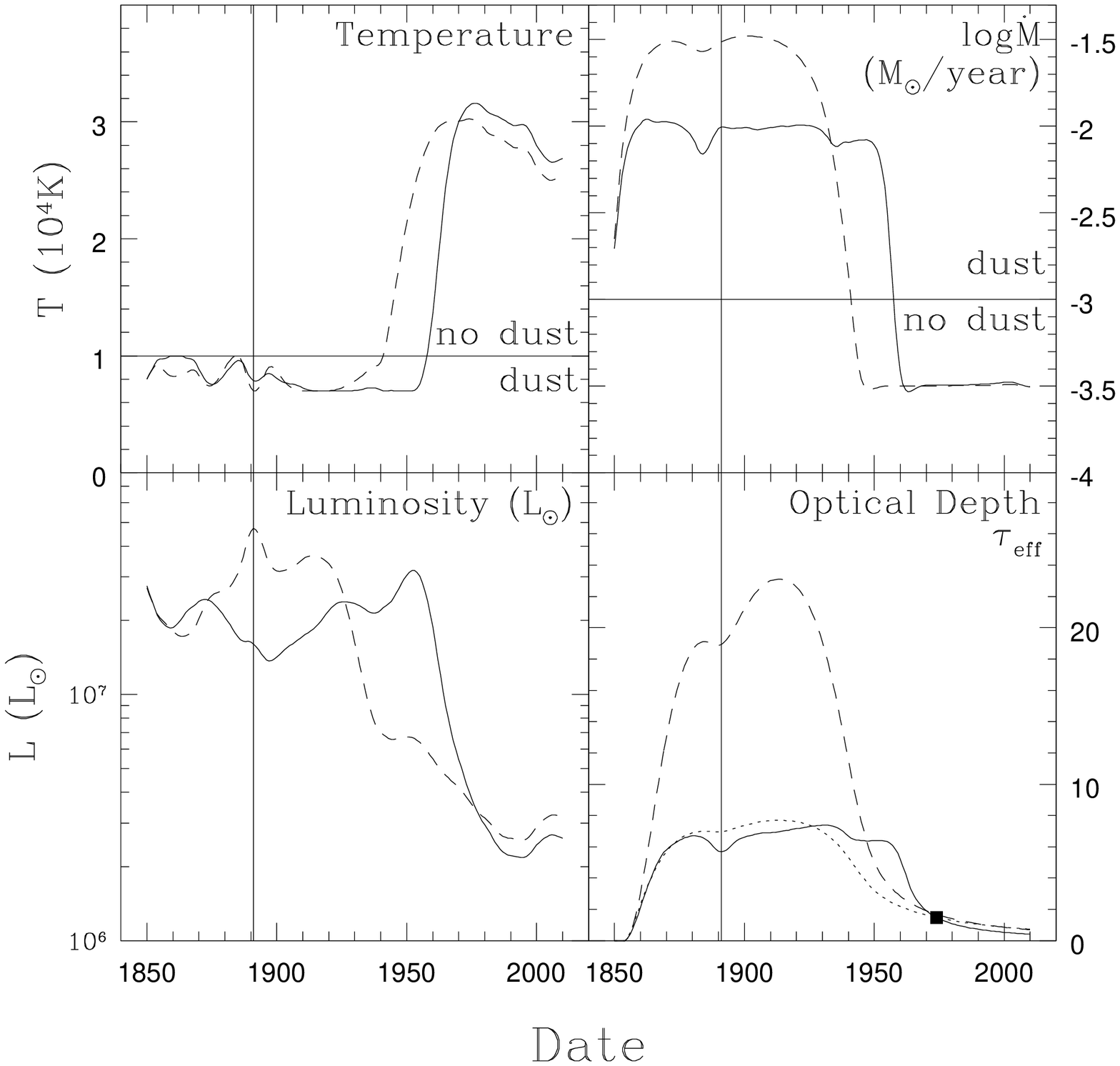}}
\caption{Physical properties of the non-parametric models for the V-band light
  curve of $\eta$ Carinae.  The panels on the left show the stellar luminosity
  (bottom) and temperature (top), while the panels on the right show the
  mass loss rate (top) and the resulting effective absorption 
  optical depth (bottom).  The solid
  (dashed) lines are for the homogeneous (inhomogeneous, Eqn.~\ref{eqn:inhomog})
  dust models.  In the optical depth panel, the dotted line shows the 
  net optical depth of the inhomogeneous model after correcting for
  the inhomogeneities.  The horizontal
  lines in the temperature and $\dot{M}$ panels show the maximum temperature
  and minimum mass loss rates for dust formation.  The vertical line at
  1890 marks the secondary peak in the light curve.  The solid point in
  the optical depth panel indicates the value inferred from the SED models
  in 1974 to which the models are normalized.  
  }
\label{fig:etamodel}
\end{figure}

\subsection{$\eta$ Carinae}
\label{sec:carina}

We had to use relatively old data to assemble a global SED for $\eta$ Carinae
similar to what can be obtained for extragalactic systems.  We combined the mid-IR
observations from $1.65$ to $20\mu$m of \cite{Robinson1973} from JD 2441517.9 (18 July 1972),
with the optical measurements by \cite{vanGenderen1984} on JD 2442125 (18 March 1974) and the
near-IR measurements by \cite{Whitelock1994} on JD 2442491 (19 March 1975).  Where they 
overlap in wavelength, the \cite{Robinson1973} and \cite{Whitelock1994} flux estimates
agree and we adopted the average value.  We define the measurement epoch by the optical
data since the near/mid-IR will evolve much more slowly than the optical, and we use 
1 January 1845 as the start of the transient.  The resulting SED is in reasonable 
agreement with that shown in \cite{Humphreys1994} for an unspecified epoch.  
We set the quiescent V band magnitude of the star to be approximately $V=3$~mag 
(see the light curve compilation of \citealt{Fernandez2009}).  This corresponds
to a luminosity of $L_*=10^{6.1}$ ($T_*=7500$~K) to $10^{6.7}L_\odot$ ($T_*=20000$~K).
We matched the SED in 1974 to the library of DUSTY models, initially ignoring
the wavelength range from $2.2$ to $5\mu$m.  These wavelengths are 
energetically less important, and the emission probably comes from 
a sub-dominant, hotter dust component  (e.g. \citealt{Robinson1987}).  
We also discuss only the silicate dust models for simplicity because
the strong silicate features in the SED cannot be fit by the graphitic models.  
Each SED is associated with a shell radius which we can turn into
an expansion velocity $v_r=r/t$ where $t$ is the elapsed time given
in Table~\ref{tab:objects}.  

The velocity for $\eta$ Carinae in Table~\ref{tab:objects} corresponds
to the long axis of the Homunculus, so we use a prior of $v_{out}=650\pm50$~km/s
over 124 years (1850 to 1974) for the outer radius.
The best fit models prefer a somewhat slower
expansion rate of $v_{out} = 460 \pm30$~km/s for $R_{out}$ 
even with the prior. The star is hot ($T_*=20000$~K)
and luminous ($L_* = 10^{6.49\pm0.02} L_\odot$).  The DUSTY
optical depth is $\tau_V = 4.5 \pm 0.3$, which corresponds
to an estimated shell mass of order 
$\log (M \kappa_{100}/M_\odot) \simeq 0.67 \pm 0.06$ that is consistent
with standard estimates (\citealt{Smith2003}).  As we see in
Fig.~\ref{fig:etased}, the fit to the SED is reasonable in both
the optical and far-IR, but fails badly in the near-IR.  Unlike
most cases we consider, the SED of $\eta$ Carinae is clearly
inconsistent with our standard assumption of a shell with 
$R_{out}/R_{in} \equiv 2$.

As an experiment, we embedded DUSTY inside a Markov Chain Monte Carlo 
(MCMC) driver and used this to fit a model and then estimate uncertainties 
for the full SED of $\eta$ Carinae including the near-IR data points.
We varied $T_*$, $\tau_V$, the dust temperature $T_{in}$ at the inner edge, and the 
shell thickness $R_{out}/R_{in}$, where the location of $R_{out}$ 
was constrained by a prior on the velocity of 
$v_{out} = R_{out}/(124~\hbox{years})=600 \pm 50$~km/s for
a time line from 1850 to 1974.  The solution has $T_*=22000\pm1000$~K,
$\tau_V=4.8 \pm 0.2$, $T_{in}=687\pm 16$~K and $R_{out}/R_{in}=5.7\pm0.5$,
where it is the greater thickness that lets the model better fit the near-IR
data compared to our standard models (see Fig.~\ref{fig:etased}).  Derived parameters are the
stellar luminosity $L_* = 10^{6.53\pm0.02} L_\odot$, $R_{in}=10^{16.61\pm0.03}$~cm
and $R_{out}=10^{17.37\pm0.03}$~cm. The inner and outer radii correspond to
expansion velocities of $100$ and $600$~km/s, respectively, for material
ejected circa 1850.  The mass estimate is now 
$\log (M \kappa_{100}/M_\odot) \simeq 0.47 \pm 0.04$.  Compared to
the more restricted models, there is little change in the stellar properties
or the optical depth, but results for velocities, radii and thicknesses
are clearly closely related and can produce factor of $\sim 2$ uncertainties
in the absence of a well-sampled SED.

Fig.~\ref{fig:etalc} shows the V band light curve compilation for $\eta$ Carinae
from  \cite{Fernandez2009}.  There is a bright peak circa 1850
followed by a steep decline to a plateau from 1850 to 1940 
punctuated by the shorter eruption from 1890-1900, followed by
a steady rise to the present day (e.g. \citealt{vanGenderen1984}).  
We now know from \cite{Rest2011} that the star was surprisingly 
cool ($T_*\simeq 5000$~K) during the Great Eruption, which is quite
different from the other sources we discuss.
This light curve represents the total optical emission of the
source rather than the direct unabsorbed emission from the star 
(see, e.g., \citealt{Martin2006}).
Superposed is the expected light curve assuming the $\tau \propto t^{-2}$
scaling of an expanding shell, and it is a reasonable model of
the trend since roughly 1960.  At earlier times it fails 
catastrophically.  The ``leaky'' dust model of Eqn.~\ref{eqn:inhomog},
where the optical flux is dominated by light leaking out through 
lower opacity channels in an inhomogeneous medium, can extend the 
agreement to roughly 1920 but then fails.  

As shown in Eqn.~\ref{eqn:levolve}, the only variables available
are the luminosity $L(t)$, temperature $T(t)$ and optical depth
$\tau_{e,V}(t)$.  The light curve $L_V(t)$ implies a strict upper 
bound on the effective absorption optical depth for a given total 
luminosity
\begin{equation}
    \tau_{e,V} < -\ln\left( { L_V \over G_{max} L } \right)
           < 6 + \ln \left( { L \over 5 \times 10^6 L_\odot } \right)
\end{equation}
because the V-band luminosity is maximized at the temperature
$T_{max}=6700$~K where $G_{max}=G(x_{max})=0.74$.  Raising (or lowering) 
the temperature from $T_{max}$ lowers the limit on the optical depth and
even an order of magnitude increase in the luminosity only
raises the limit by $\Delta \tau_{e,V} = \ln 10=2.3$.    
This limited ability to use luminosity and temperature to
balance changes in optical depth means that the V-band light 
curve very tightly constrains the history of $\eta$ Carinae.
In particular, it makes it impossible to explain the light curve
with dust formed in a thin shell created during a short-lived ($\sim$decade)
eruption in the mid-nineteenth century -- no plausible variation
in luminosity and temperature is large enough to compensate for
the changing optical depth (see Fig.~\ref{fig:etalc}).  The light curve does, however, have the
qualitative properties of a long lived wind, as illustrated in Fig.~\ref{fig:etalc}
by a model based on Eqn.~\ref{eqn:tevol} for a wind where dust formation 
commenced around 1862 and ended in 1943.  It is not a perfect fit, but
it is now close enough to the light curve that the differences can be
explained by changes in luminosity and temperature.

Fig.~\ref{fig:etalc} also shows two models that fit the light curve
almost exactly.  As detailed in the Appendix, the model uses a
non-parametric luminosity, temperature and mass loss history
to fit the light curve and the optical depth estimate from
the SED fits while trying to maximize the overall mass loss. 
We used an expansion velocity of $500$~km/s and a dust 
formation time of $t_f=5$~years corresponding to 
a dust formation radius of roughly $R_f=7 \times 10^{15}$~cm.
Fig.~\ref{fig:etamodel} shows the evolution of the physical parameters
in this model -- it is a long-lived wind with 
$\dot{M} \simeq 10^{-2}M_\odot$/year
ejecting roughly $1 M_\odot$ of material in the dust creating
phases.  If we use our ``leaky'' dust model (Eqn.~\ref{eqn:inhomog}),
we can triple the mass lost, but the general structure of the
solution is unchanged. As discussed in \S\ref{sec:caveats}, inhomogeneities
in the ejecta primarily affect this balance between optical
depth and associated mass rather than the temporal scaling of
the optical depth and the observed flux.
  
The star in outburst is luminous but relatively cool, with conditions that favor
the formation of dust in the wind (\citealt{Kochanek2011b}).  
While not imposed on the models, dust formation ceases just as the star reverts to its
lower luminosity, quiescent hot state, and this transition
in luminosity and temperature removes the mismatch between
the light curve and the simple wind optical depth model seen
in Fig.~\ref{fig:etalc} -- the initial rapid drop in optical depth is
balanced by the rapid drop in the V-band flux as the star returns
to a higher temperature.   Available evidence
is that the spectrum of the star was consistent with this
relatively cool (F star) state over the dust formation period, and while
the implied B$-$V colors of our model are somewhat redder than interpretations
of the historical data, they come far closer to these estimates
of the observed color than any model with the mass ejection
occurring over a short period circa 1850 (see the discussion in \citealt{vanGenderen1984}).
The brightening circa 1890 can reproduced using either a fluctuation in the
luminosity or in the rate of mass loss.  While geometry is clearly an important factor, we 
note that a constant velocity wind starting in 1850 and ending 
in 1950 has $R_{out}/R_{in}=5.2$ in 1974, matching the best 
fitting SED models (but only $R_{out}/R_{in}=2.7$ in 2010).

No attempt to drive the solution to a significantly shorter
duration was successful because no physically plausible luminosity and 
temperature variations can balance the optical depth evolution. 
As we discussed in \S\ref{sec:caveats}, this conclusion should
hold for radially expanding shells of arbitrary angular structure
short of a long lived hole through the dust directly to the star 
(which is known not to have been present from the 1950's to the present day,
see \citealt{Martin2006}).  The only remaining possibility is to
identify a process which causes the (effective)
opacity to increase with time in order to balance the expansion.
It cannot be continued dust formation because particle growth
shuts off very quickly (within $2R_f$) and dust in $\eta$ Carinae appears
to have grown to sizes where further growth reduces rather
than increases the opacity (\citealt{Smith2003}). Forming
very large grains ($a \simeq 10\mu$m) initially and then 
steadily breaking them up into smaller dust grains, would
provide a mechanism,  because the opacity scales as
$\kappa \propto 1/a$ provided $a \gtorder 0.3\mu$m 
(e.g. \citealt{Draine2011}).  The dynamic range in opacity 
from steadily shrinking $a$ from $10\mu$m to $0.3\mu$m is large enough to balance
the expansion, but there is no obvious mechanism to drive
this evolution.  Alternatively, if the ejecta evolved from
being inhomogeneous to homogeneous, the effective optical
depth for a given mean optical depth would rise and could
balance the expansion.  This makes little sense because ejected material
generally evolves from (locally) homogeneous to inhomogeneous 
(e.g. \citealt{Garcia1996}, \citealt{Toala2011}), and 
thus exacerbates rather than ameliorates the optical
depth evolution problem.

In summary, our models can fit the properties
of $\eta$~Carinae to the accuracies we
need for the extragalactic sources.  The SED models 
are roughly consistent with both the physical 
scale and ejected masses derived in more detailed
studies.  As was already known, the visual light
curve is consistent with the declining optical depth
of an expanding shell at late times but not at 
early times.  We are able to successfully model
the complete light curve, but are driven to models with
a relatively steady mass loss rate from 1850-1950
that are at variance with most interpretations
of $\eta$~Carinae's history.  Since the only
real assumptions of the model are steady, smooth
expansion and dust mass conservation, it is not
easy to escape this conclusion.  It is not obvious to 
us that this conflicts with any other observations of $\eta$
Carinae, but even if it did, we would really simply face a stalemate of contradictions
since the standard model provides no plausible, quantitative explanation of the light curve.  
Further investigation of these
issues is beyond the scope of our present goals,
but they inform analysis of our subsequent results. 

\begin{figure}[p]
\centerline{\includegraphics[width=5.5in]{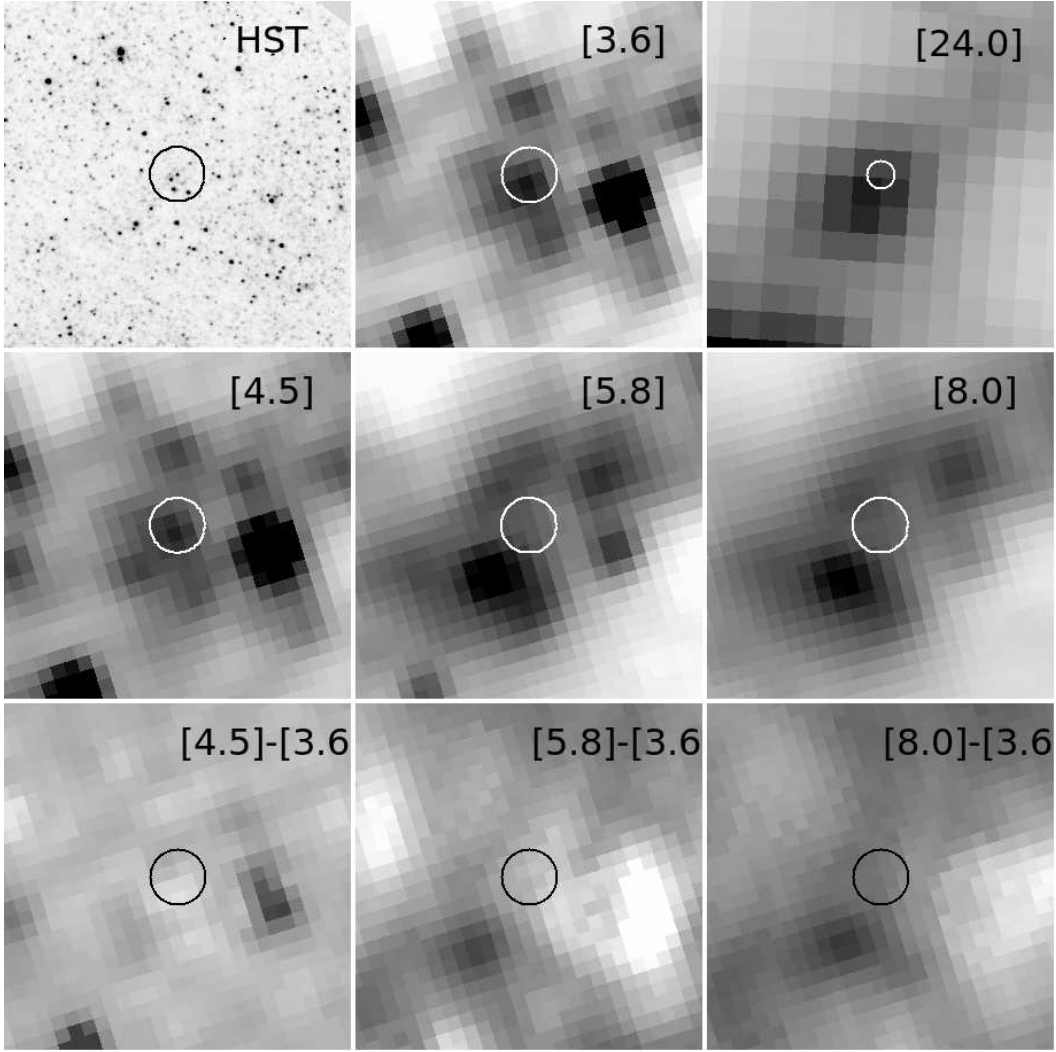}}
\caption{ The environment of SN~1954J.  The top left, middle and right panels show the
  HST, $3.6$ and $24\mu$m images of the region.  The center left, middle and right panels
  show the $4.5$, $5.8$ and $8.0\mu$m images.  The bottom left, middle and right panels
  show the $[4.5]-[3.6]$, $[5.8]-[3.6]$ and $[8.0]-[3.6]$ wavelength-differenced images.
  The circles in all panels have a radius of 1\farcs2 in this and all subsequent 
  figures.  Similarly, the $24\mu$m panel shows a region two times larger
  than the other panels.  Star \#4, which \cite{Vandyk2005}
  identify with SN~1954J, is the upper
  right star of the trapezoid of four bright stars seen inside the circle on the HST
  image.
  }
\label{fig:image54j}
\end{figure}

\begin{figure}[p]
\centerline{\includegraphics[width=5.5in]{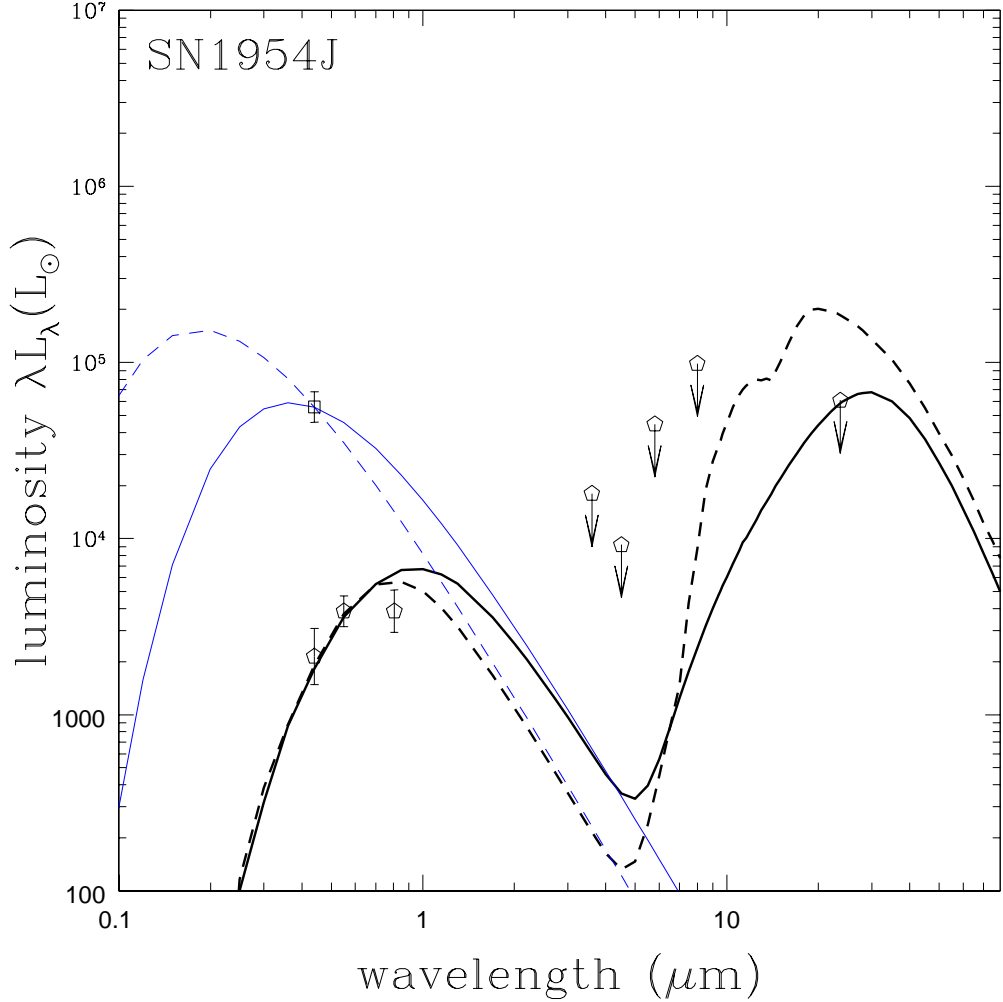}}
\caption{ The spectral energy distribution of SN~1954J.  The open pentagons are the
  optical magnitudes of star \#4 from \cite{Vandyk2005} and our mid-IR limits.
  The open square is the estimated magnitude in quiescence prior to 1949. 
  The thin solid (dashed) curve shows the 
  the SED of a $T_*=15000$~K, $L_*=10^{4.9}L_\odot$ ($T_*=20000$~K, $L_*=10^{5.3}L_\odot$) progenitor normalized 
  to match the pre-transient B band magnitude.  The thick solid (dashed) curves
  show graphitic (silicate) DUSTY models where the optical depths of $\tau_V=3.0$
  ($\tau_V=5.3$) were selected to fit the optical magnitudes of star \#4 while 
  remaining under the mid-IR limits for an expansion velocity of roughly $820$~km/s.
  }
\label{fig:sed1954J}
\end{figure}

\begin{figure}[p]
\centerline{\includegraphics[width=5.5in]{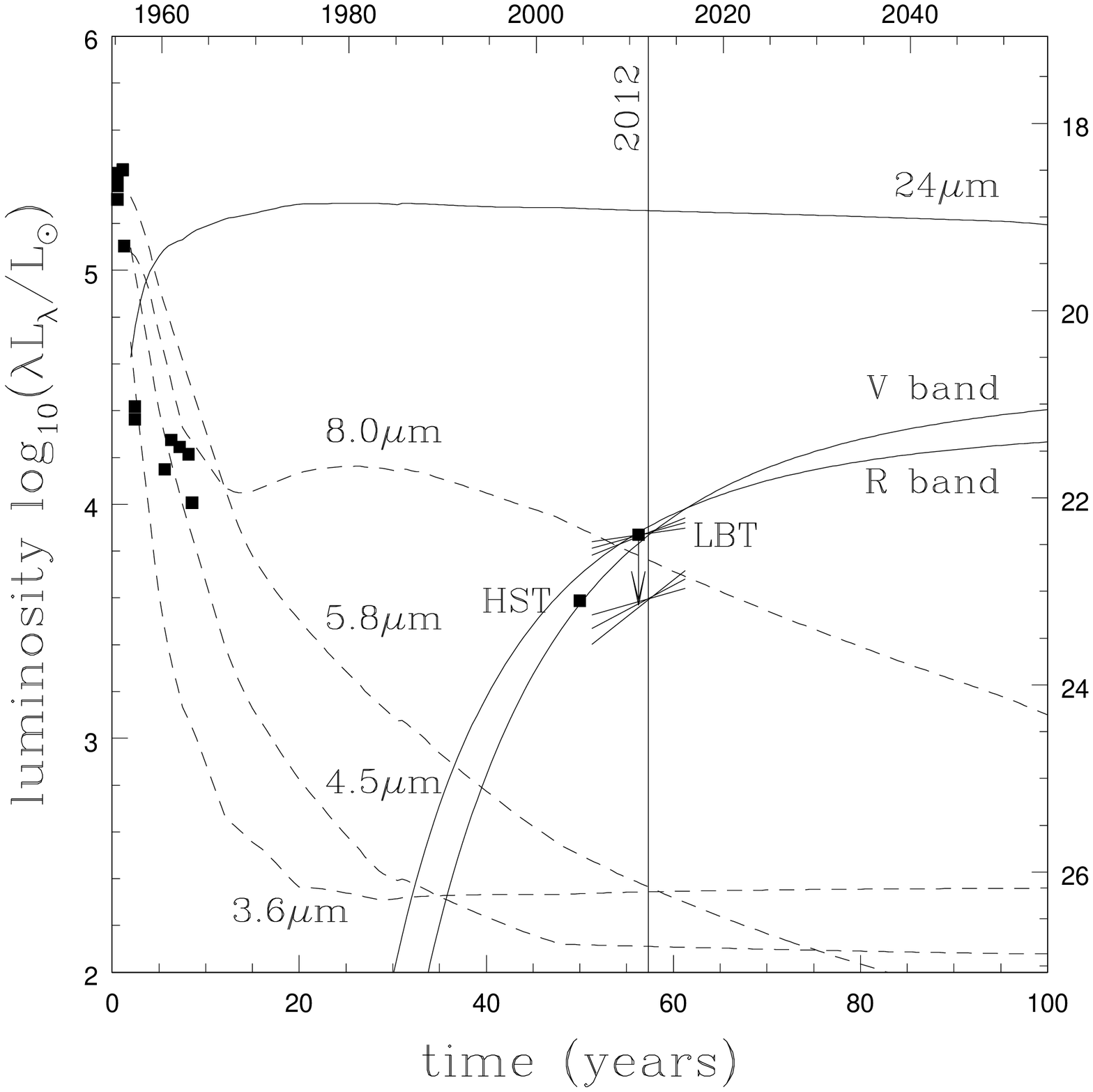}}
\caption{ Expected light curves for SN~1954J.  The curves show
  the expected V and R band (solid), IRAC bands (dashed) and $24\mu$m (solid) 
  light curves normalized by the best fit silicate SED model.  The early time
  points are the  B band data from \cite{Tammann1968}, the point labeled
  HST is the V-band flux from late 2004 from \cite{Vandyk2005}, and the
  arrow labeled LBT is the R-band upper limit from our LBT observations.
  The ``wedges'' at the LBT epoch show the formal slope of the light 
  curve and its uncertainties when normalized either at the LBT or HST
  magnitude -- the lower the reference flux, the larger the implied 
  fractional change.  
  }
\label{fig:lc1954J}
\end{figure}

\begin{figure}[p]
\centerline{\includegraphics[width=6.5in]{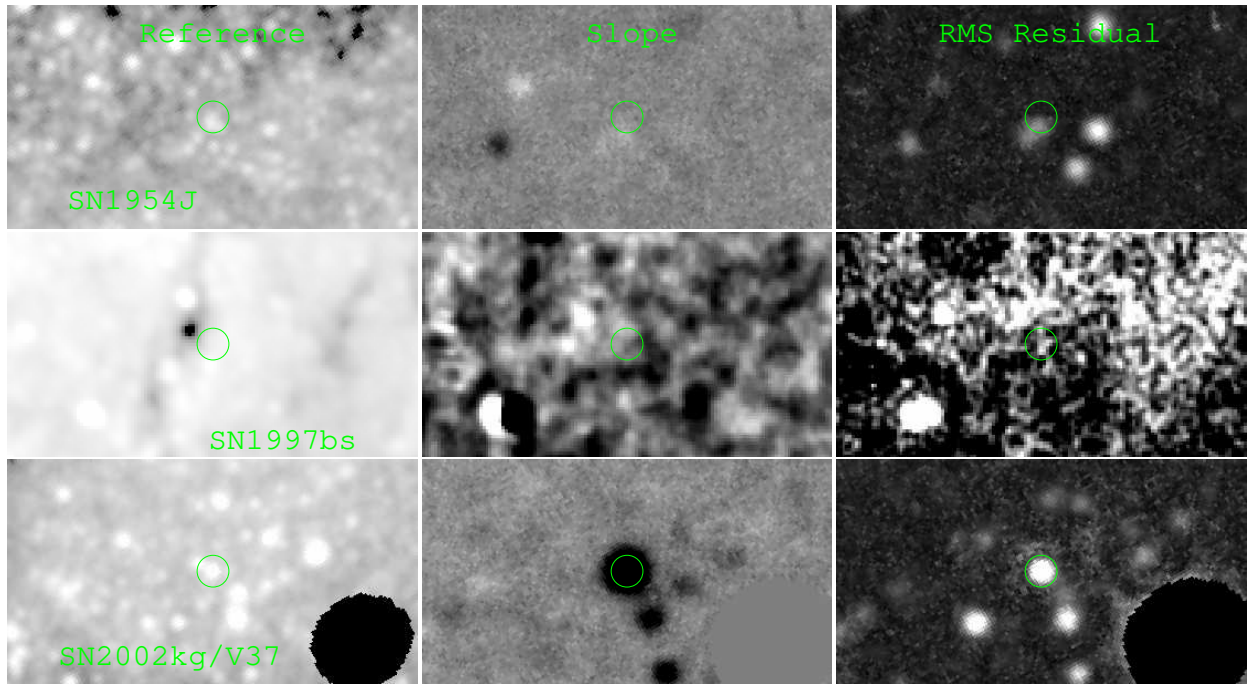}}
\caption{
  LBT R band data for SN~1954J (top), SN~1997bs (middle) and SN~2002kg/V37 (bottom).
  The left panel shows the R band reference image for each source, the middle
  panel is the Slope image, and the right panel is the RMS image as defined in
  \S\ref{sec:data}. The 2\farcs24 (10 pixel) diameter circle in each panel marks the source
  position.  Each type of image is on the same gray scale.  The reference image 
  is on a logarithmic scale. The Slope image is on linear scale covering $\pm 10^3L_\odot$/year
  (about 10\% the slope observed for  SN~2002kg/V37).  Similarly, the 
  RMS residual image is on a linear scale from $0$ to $10^3 L_\odot$ in 
  (the rms residuals for SN~2002kg/V37 are about twice this).
  }
\label{fig:lbt}
\end{figure}

\subsection{SN~1954J/V12 in NGC~2403}

SN~1954J \citep{Tammann1968, Humphreys1999} is considered a classic example of an
extragalactic $\eta$ Carinae-like eruption.  Originally identified as the luminous
blue variable star V12, it was relatively quiescent prior to 1949 with a blue
photographic magnitude of $m_{pg} \sim 21$.  
It became increasingly variable
from 1949 to 1954 with a largely unobserved eruption in late 1954, mainly noted
as a steep decline from an observed peak magnitude of $\sim 16.5$ to be about
a magnitude fainter than its pre-eruption flux by 1960.  We adopt 1 September
1954 (JD 2434987) as the nominal date of the peak. \cite{Smith2001} 
searched for a surviving star using ground-based optical and near-IR images,
and identified a candidate source roughly 10 times fainter than the pre-outburst
star with signs of a near-IR (K-band) excess (U, B, V, R, I, J, H and K 
of $22.5\pm0.3$, $22.7\pm0.2$, $21.9\pm0.3$, $21.1\pm0.2$, $20.9\pm0.2$,
$20.3\pm0.2$, $19.9\pm0.2$, and $19.0\pm0.2$~mag).  The optical observations
were taken in February 1999 and the near-IR in April 2000.
They modeled the spectral energy distribution (SED)
as a roughly $10^4$~K black body with $\sim 1.5$~mag of visual extinction. \cite{Vandyk2005}
found that this candidate source broke up into 4 stars at the resolution of HST,
and they identified star \#4 as the surviving star based on its H$\alpha$ emission  
and an SED consistent with a hot ($T_*=30,000$~K), luminous ($L_* \simeq 10^6 L_\odot$)
supergiant obscured by $A_V \simeq 4$~mag of dust.  The optical magnitudes of star \#4 are 
$B=24.3 \pm 0.4$, $V=23.1\pm 0.2$ and $I=22.2\pm 0.3$, respectively.  
The HST observations were taken 17 August 2004,
and \cite{Vandyk2005} also had optical observations from 20 October 1996 
and 13 October 1997.  Both \cite{Smith2001} and \cite{Vandyk2005} attribute the
extinction to material ejected during the eruption.
A spectrum of the H$\alpha$ emission from the region shows broad
H$\alpha$ emission, with a line width of order $700$~km/s (\citealt{Smith2001}). 
For comparison, if we correct the progenitor magnitude following \cite{Smith2001}
and then subtract the fluxes from the contaminating stars found by \cite{Vandyk2005},
the progenitor luminosity is $L_*=10^{4.9}$ ($T_*=7500$~K) to $10^{5.4} L_\odot$
($T_*=20000$~K).

The mid-IR observations of SN~1954J/V12 were made over a 5 year period from
2004 to 2009.  We found no evidence for time variability associated with the
source, so we analyzed the combined data.
As we can see in Fig.~\ref{fig:image54j}, the mid-IR emission in the vicinity of
SN~1954J comes from multiple sources and there is a ridge of PAH 
emission extending over the region at the longer wavelengths.  There appears
to be a (probably composite) source at the position of SN~1954J in the shorter
wavelength $3.6$ and $4.5\mu$m bands and no identifiable source at the longer 
bands.  The ratio of the $3.6$ and $4.5\mu$m fluxes is very close to that for a
Rayleigh-Jeans spectrum, $F_{4.5}/F_{3.6} = 0.64 \pm 0.26$ compared to
$(3.6/4.5)^2=0.64$.  If we scale and subtract the $3.6\mu$m flux from the
$4.5\mu$m flux, the estimated dust emission at $4.5\mu$m is only 
$0.00\pm0.02$~mJy, consistent with the absence of any sources in
the wavelength-differenced images in Fig.~\ref{fig:image54j}.  
Thus, the observed emission is probably dominated by the composite
emission of the normal stars seen in the HST images and we simply
treat the mid-IR fluxes as upper bounds on the emission from SN~1954J.  

Fig.~\ref{fig:sed1954J} shows the SED of this system along with two representative SEDs
for graphitic and silicate dusts.  Since there was no evidence of time variability
in the mid-IR data, this should be viewed as the SED at the time of the HST
observations in late 2004, 50 years after the peak.  We constrained the unobscured SED to match the 
pre-1949 B-band ($m_{pg}$) luminosity, and the obscured SED to fit the HST photometry
of star \#4 and the mid-IR limits.   We used a velocity prior of $v_{in}=700\pm70$~km/s
for the expansion of the inner edge of the shell, although the results change little
if we apply it to the outer edge instead.  Fig.~\ref{fig:sed1954J} shows the two
best fits including the velocity prior.  The graphitic models fit somewhat better
because the silicate peak tends to exceed the limit on the $24\mu$m luminosity
The graphitic models prefer $T_*=10000$ or $15000$~K with $L_* = 10^{5.0\pm0.1}L_\odot$,
a DUSTY optical depth of $\tau_V = 3.1\pm0.2$ and an inner dust radius of $R_{in}=10^{16.99\pm0.04}$~cm,
corresponding to a velocity of $620 \pm 60$~km/s.  The silicate models prefer a 
somewhat hotter $T_*=15000$ or $20000$~K star with $L_* = 10^{5.3\pm0.1} L_\odot$,
$\tau_V = 5.2 \pm 0.3$ and an inner dust radius of $R_{in}=10^{17.05 \pm0.06}$~cm
corresponding to a velocity of $720 \pm 100$~km/s.  The effective absorption
optical depths of the two models are the same, with $\tau_{e,V} \simeq 2$ (see the
discussion in \S\ref{sec:dusty}).  If we use the limits on the mid-IR emission
from the wavelength-subtracted images, the graphitic models are in conflict
with the $4.5\mu$m limit.  The star cannot be as hot or luminous as proposed by
\cite{Vandyk2005} without being in gross conflict with the mid-IR limits -- the
present day luminosity of the star has to be well below the minimum luminosity of 
$L_* \gtorder 10^{5.8} L_\odot$ sometimes used for LBVs (e.g. \citealt{Smith2004}).  The ejecta masses
needed to have these optical depths 50 years after the transient are high, with 
$\log ( M \kappa_{100}/M_\odot) \simeq 0.57 \pm 0.08$ and $0.92 \pm 0.12$ for the graphitic
and silicate models.  These are an order of magnitude
or more larger than the mass scales estimated in Table~\ref{tab:objects2}, and
would require a radiated to kinetic energy ratio of $f \ll 1$ that is indicative 
of an explosively driven transient.  Producing the absorption with a 
$\dot{M} \sim 10^{-3} M_\odot$/year wind and the same velocity could produce 
these optical depths using far less mass, but it would produce an SED peaking
in the short wavelength IRAC bands that is ruled out by the data.

Fig.~\ref{fig:lc1954J} shows the expected visual and mid-IR light curves of this
system normalized by the best fit silicate model to the SED.  For each epoch we
interpolated through our tabulated sequences of $R_{out}/R_{in}\equiv 2$ DUSTY models 
assuming fixed stellar properties, a $700$~km/s expansion velocity, and the 
$\tau \propto 1/r^2$ geometric scaling of the optical depth.  At the present
time, only the $8\mu$m band should show any variability in the mid-IR (2\%/year), 
while the $24\mu$m flux is nearly constant and the shorter wavelength IRAC bands 
are too faint compared to the level of contamination from other sources.  Thus,
it is not surprising that we found no mid-IR variability.   
Fig.~\ref{fig:lc1954J} also shows the B band photometry of
the early phases from \cite{Tammann1968}.  Even though these fluxes presumably
represent sums over the unresolved stars, the star seems to take 
almost ten years to fade, which seems long compared to reasonable dust formation
time scales of less than one year (Eqn.~\ref{eqn:tform}).  While the data are too sketchy for a detailed
model, the early light curve is only consistent with forming enough dust to 
produce the proposed obscuration in 2004 if the mass loss occurred over a 
very extended period of time (decades). 

If the absorption is now due to an expanding shell, the optical fluxes 
should be rising by 5\% (R-band) to 15\% (U-band) per year as shown in
Fig.~\ref{fig:lc1954J}.  We have been monitoring NGC~2403 with the LBT,
and Fig.~\ref{fig:lbt} shows the reference, slope and rms images 
(see \S\ref{sec:data})
of the
region around SN~1954J.  At the estimated position of SN~1954J, DAOPHOT 
estimates $R=22.4$~mag, consistent with the earlier ground
based observations by \cite{Smith2001}.  For an expanding shell,
this is also roughly the expected magnitude of the star given the
expected brightening from the epoch of the HST observations, so the
LBT observations are probably inconsistent with such a brightening 
because the measured flux includes the contaminating emission from the
nearby unresolved stars (see Fig.~\ref{fig:image54j}).  There is 
little evidence for variability at this location in either the
slope or the rms images.  Formally, we estimate an R-band slope
of $(180 \pm 30) L_\odot$/year with an rms residual of $740 L_\odot$,
but the detections are not overwhelmingly convincing in this or
the other bands.  As shown in Fig.~\ref{fig:lc1954J}, the rate
of change is also flatter than expected for an expanding shell of
material.  Thus, the LBT data argues against an expanding shell.
Note that this argument is independent of the actual expansion velocity as the
optical depth evolution depends only on the elapsed time
once the scale of the optical depth is set by the photometric model
(see Eqn.~\ref{eqn:tevol}).

The preponderance of the evidence for SN~1954J/V12 is inconsistent with
the standard picture.  \cite{Vandyk2005} make a clear case for identifying star \#4 as 
the counterpart to SN~1954J/V12 because it seems to be the only plausible
source of the broad H$\alpha$ line emission.  The mid-IR flux limits allow
star \#4 to be a surviving star obscured by an expanding, dusty shell, but the 
intrinsic stellar luminosity and temperature have to be quite low, 
$T_*\simeq 15000$~K and $L_* \simeq 10^{5.0}$--$10^{5.3}L_\odot$.  The SED does not
allow the present day obscuration to be due to a wind, but the lack of optical
variability and the early-time light curve appear to be inconsistent with 
the presence of an expanding shell formed in a short transient.  
Moreover, the $\gtorder M_\odot$ of ejecta needed to produce the present
day optical depth is difficult to reconcile with a radiatively driven
ejection mechanism given the low luminosity and short duration of the
transient.  Fully characterizing the spectral energy distribution of the
candidate star and continued monitoring should clarify the nature of this
source.

\begin{figure}[p]
\centerline{\includegraphics[width=5.5in]{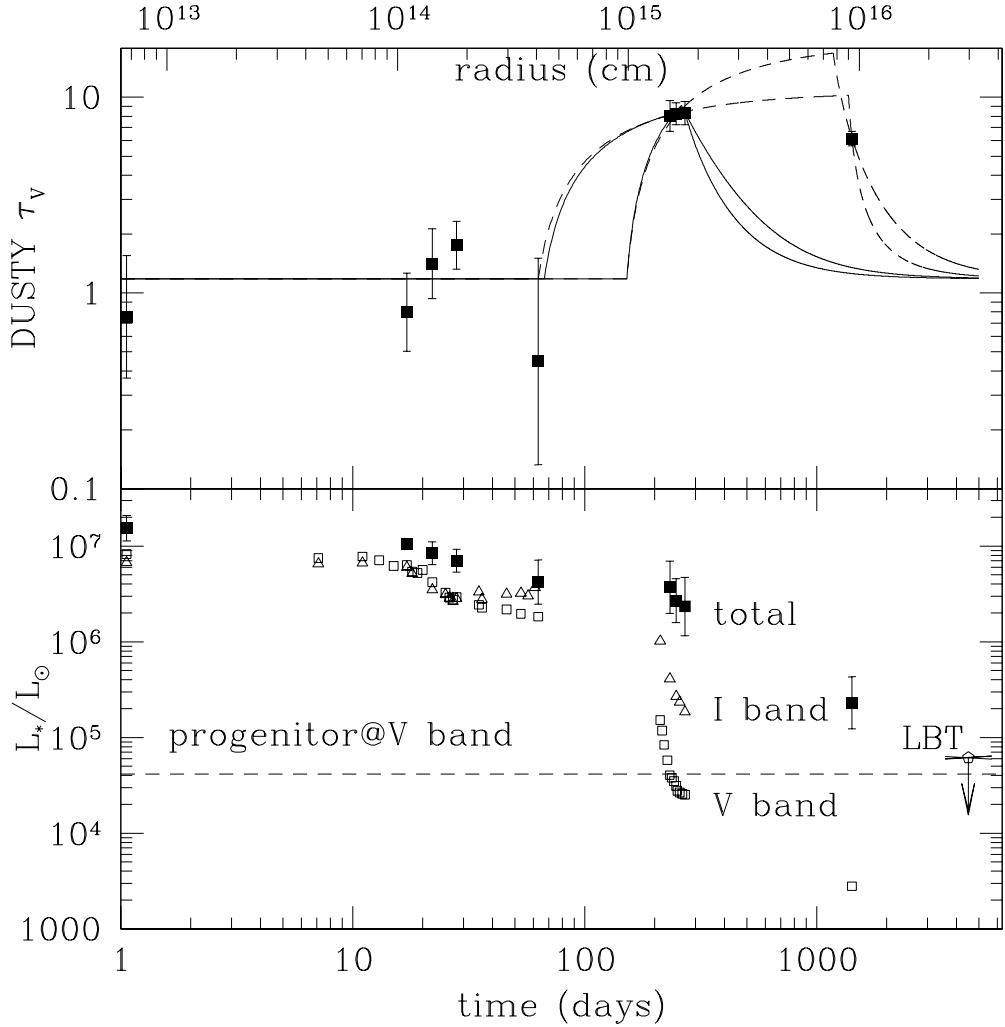}}
\caption{ 
  Evolution of SN~1997bs.  The lower panel shows the 
  evolution of the estimated total (filled squares), I band
  (open triangles) and V band (open squares) luminosities
  as compared to the V band luminosity of the progenitor (dashed line).
  The LBT point should be viewed as an upper limit.  The lines
  through the point indicate the limits on the rate of change
  of the V-band luminosity from the LBT data.
  The upper panel also shows the evolution of the silicate optical
  depth $\tau_V$.  The points are the results of DUSTY fits to the
  SED which are then modeled as a constant plus a finite duration
  wind (Eqn.~\ref{eqn:twind}).  The solid curve ignores the final HST epoch and terminates
  the mass ejection as soon as possible, while the dashed 
  curve includes the final HST epoch.  Two time scales for the
  onset of dust formation are shown, where the earlier time
  scale corresponds to a very high dust formation temperature
  and the later time scale corresponds to normal assumptions.
   Note the rapidity of the
  decline in the optical depth when the wind ends.  The upper
  radius axis assumes an ejecta velocity of $765$~km/s following
  Table~\ref{tab:objects2}. 
  }
\label{fig:od1997bs}
\end{figure}

\begin{figure}[p]
\centerline{\includegraphics[width=5.5in]{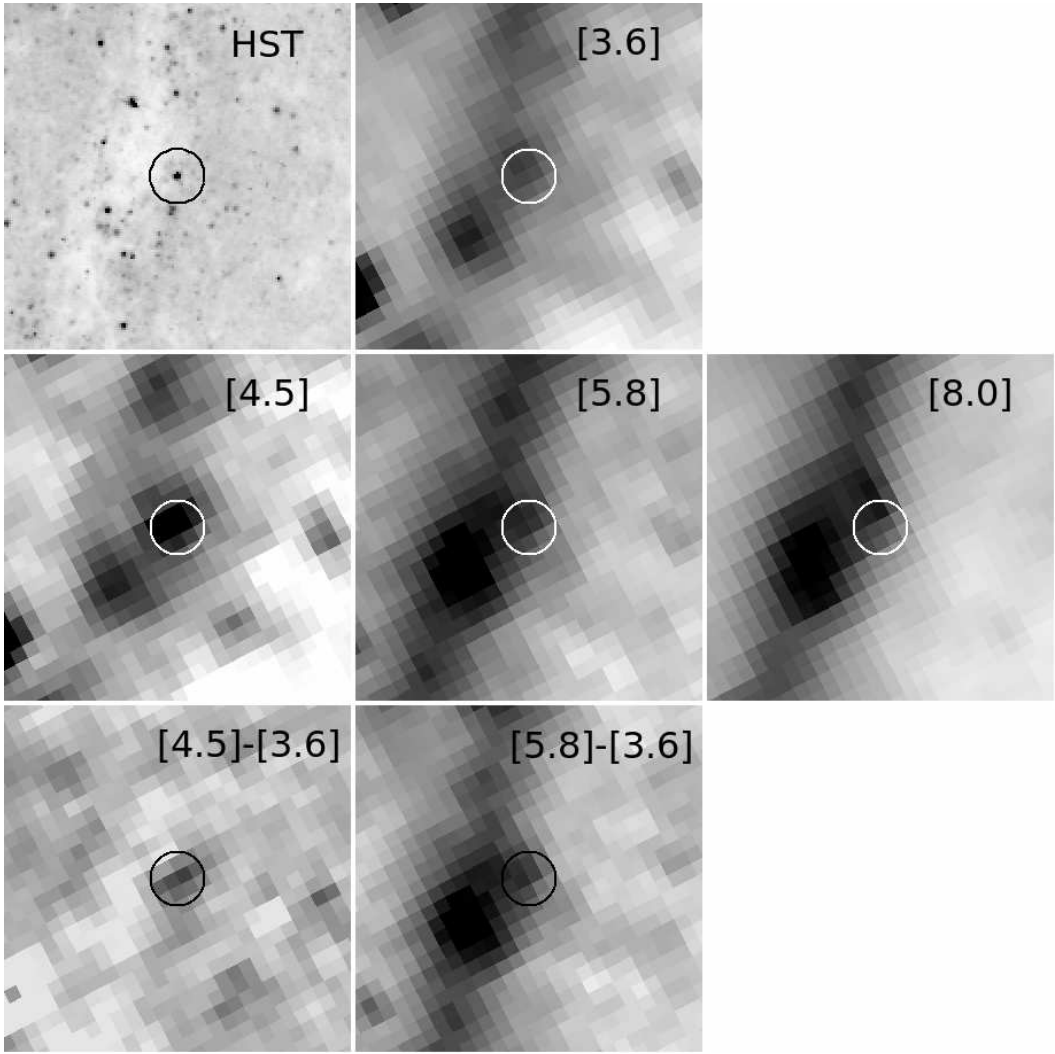}}
\caption{ The environment of SN~1997bs.  The top left and middle panels show the
  HST and $3.6\mu$m images of the region.  The HST image was taken during the
  transient. The center left, middle and right panels
  show the $4.5$, $5.8$ and $8.0\mu$m images.  The bottom left, middle and right panels
  show the $[4.5]-[3.6]$ and $[5.8]-[3.6]$ wavelength-differenced images.  There
  is no $24\mu$m data and the wavelength-differencing procedure failed for 
  the $[8.0]$ image.  A 1\farcs2 radius circle marks the position of the transient.
  }
\label{fig:image97bs}
\end{figure}

\begin{figure}[p]
\centerline{\includegraphics[width=5.5in]{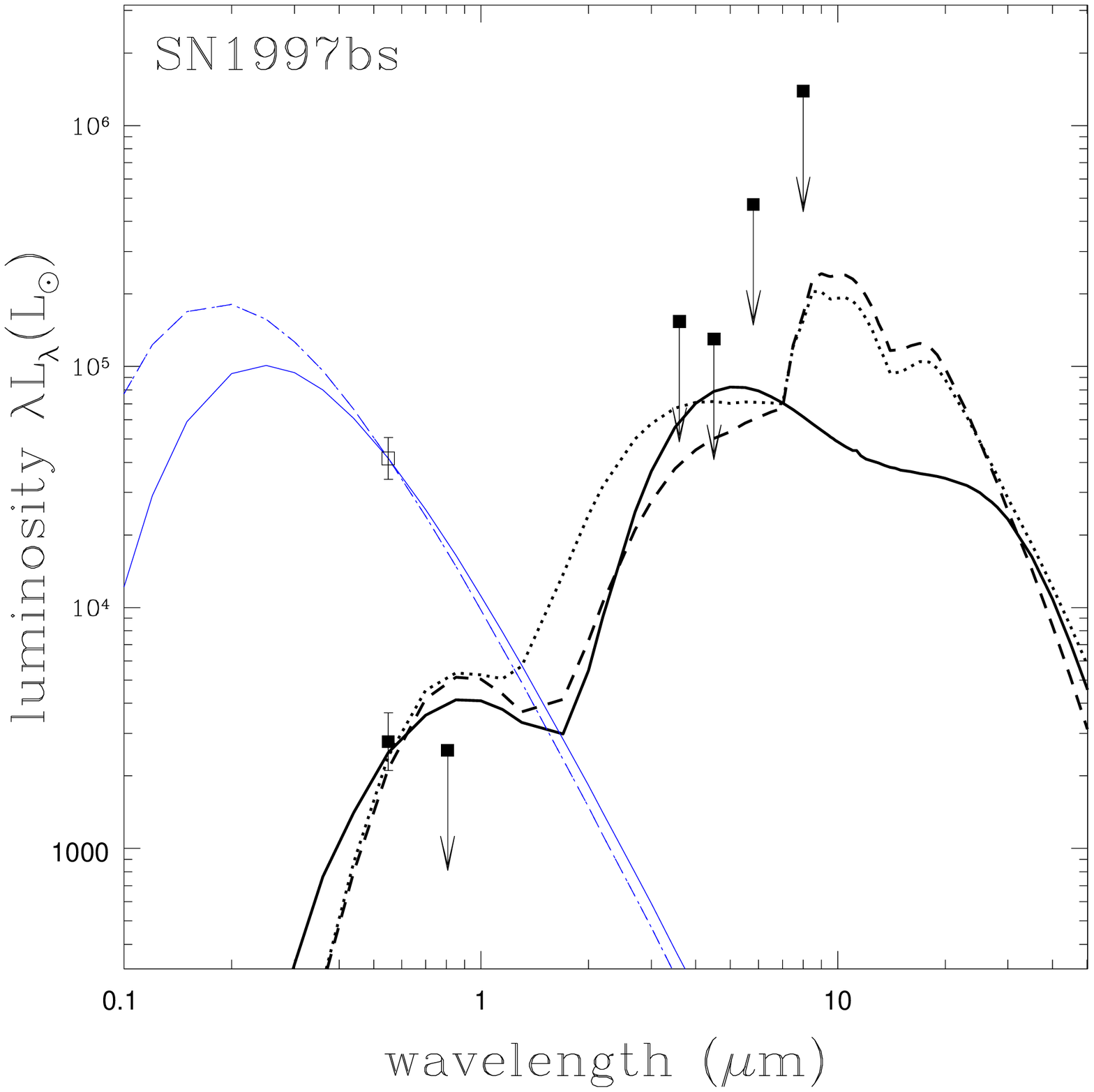}}
\caption{ The spectral energy distribution of SN~1997bs.  The open square shows
  the progenitor flux estimate from \cite{Vandyk1999}. The filled squares show
  the optical fluxes in 2001 from \cite{Li2002} and our limits on the mid-IR
  fluxes.  The graphitic shell model (solid lines) has a $T_*=15000$~K, 
  $L_*= 10^{5.1}L_\odot$ progenitor (thin line) obscured by a $\tau_V=3.5$ 
  shell with the inner edge expanding at $740$~km/s.  The silicate shell
  model (dashed lines) has a $T_*=20000$~K, $L_*= 10^{5.4}L_\odot$ progenitor (thin line)
  obscured by a $\tau_V=6.5$ shell expanding at $680$~km/s.  The dotted curve
  shows a silicate wind model with an inner edge temperature of $1000$~K.
  All the models have some conflict with the \cite{Li2002} I-band upper limit,
  and these are severe for wind models with hotter inner-edge dust temperatures.
  }
\label{fig:sed1997bs}
\end{figure}

\subsection{SN~1961V in NGC~1058: Dead and Loving It!}

In \cite{Kochanek2010} we carried out a similar study of SN~1961V in NGC~1058, arguing that the
lack of any mid-IR emission indicated that SN~1961V was a real, if peculiar, Type~IIn supernova,
a conclusion supported by \cite{Smith2011} on energetic grounds.  \cite{Vandyk2011} recently
argued against this conclusion, so we include a brief discussion of it here.  
Up to minor differences, \cite{Vandyk2011} obtain the same mid-IR limits as the analysis in
\cite{Kochanek2010}.  They significantly revised the earlier HST data analyses, settling
on a V band magnitude of $V\simeq 24.7 \pm 0.2$~mag for the surviving star.  They then
model the available data assuming low metallicity, hot (30000~K) stellar models with 
foreground Galactic extinction to derive $A_V \simeq 2$~mag (beyond the Galactic contribution)
and a luminosity $L\simeq 10^{6.0} L_\odot$.  Thus, one part of their argument is to make the surviving
star roughly three times fainter than the typical estimates for the progenitor star used by 
\cite{Kochanek2010}, arguing that the star was already in outburst in the decades before
the transient.  Their estimate of the absorption is completely consistent with our
results since producing an absorption corresponding to $A_V=2$ with the standard
silicate DUSTY models requires exactly the value used by \cite{Kochanek2010}, $\tau_V=4.5$
(see the discussion in \S\ref{sec:dusty}).  Thus, the luminosity and optical depth estimates are
basically mutually consistent, and in reality the only way \cite{Vandyk2011} found to 
rescue the impostor hypothesis was to have essentially no dust in the ejecta.  

They still, however, require dust, so they put it all in the foreground as an additional
intervening $A_V \simeq 2$ dust screen,\footnote{They put it so close to the star (1~pc) that 
it contributes the long wavelength emission in their Figure~4, but this is unimportant to
our discussion.} but they do not then consider the consequences of this model for
the inferred properties of the progenitor and the transient.  First, the
progenitor moves from having $m_B \simeq 18$ and $B-V \simeq 0.6$ ($T \simeq 6000$~K)
to $m_B^{corr} \simeq 15.4$ and $(B-V)_{corr} \simeq 0$ ($T > 10000$~K).
This corresponds to a shift from $M_B \simeq -12$ and $M_{bol} \simeq -12$
to $M_B^{corr} \simeq -14.6$ and $M_{bol}^{corr} \simeq -15$ at a temperature that is no longer
characteristic of a (great) LBV eruption ($T \sim 7000$~K, \citealt{Humphreys1994}).  Moreover, if  
the progenitor properties were already extreme, they are now bizarre even if interpreted 
as a star in a multi-decade long outburst ($L_* \simeq 10^8 L_\odot$ emitting $10^{50}$~ergs/decade).
Similarly, SN~1961V at peak was already as bright as a typical Type~II SN, $M_B \simeq -17$, 
but with the additional foreground extinction it actually peaked at $M_B^{corr} \simeq -20$, making
it unusually bright even for a Type~II supernova (see \citealt{Li2011}).  There is also a similar bluewards shift in
the estimated temperature/color.  The total radiated energy rises from $\simeq 10^{49.6}$~ergs
to $10^{50.8}$~ergs, well over an order of magnitude larger than $\eta$ Carinae 
(e.g. \citealt{Humphreys1994}).  In short, forming no dust in the ejecta and instead
placing it in the foreground certainly allows an impostor to evade the arguments of
\cite{Kochanek2010}, but at the price of enormously strengthening the energetic arguments
of \cite{Smith2011} that it must have been a supernova.  We discussed having no dust
as an alternative (if unlikely) solution in \cite{Kochanek2010}, but argued this then 
required a very hot ($T_* \sim 10^5$~K) surviving star so that it would be faint in the optical due to 
bolometric corrections rather than dust in order to avoid these problems.  As a result, 
we see no reason to revisit our conclusions in \cite{Kochanek2010} in light of 
\cite{Vandyk2011} -- this is an ex-star.

\subsection{SN~1997bs in NGC~3627}

SN~1997bs was the first SN transient (1997 April 15, JD 2450693)
discovered by the LOSS survey (\citealt{Treffers1997}) and 
was classified as a Type~IIn (\citealt{Vandyk2000}). % cannot find reference in the vandyk paper to this claim
It peaked at $V \simeq 17$ and then faded to $V\simeq 23.5$ over 9 months.
\cite{Vandyk1999} identified a candidate progenitor with $V \simeq 22.86 \pm 0.16$~mag, 
making the late time emission significantly fainter than that of the progenitor.  
The progenitor magnitude corresponds to a luminosity of $L_*=10^{4.8}$ to $10^{5.4}L_\odot$
for the temperature range $T_*=7500$~K to $20000$~K.
As discussed in \cite{Vandyk2000},
the color evolution of the transient is peculiar, with a rapid evolution to the red
that is indicative of dust formation.
In Fig.~\ref{fig:od1997bs} we show DUSTY silicate dust estimates of the optical depth and
luminosity evolution based on the optical fluxes and limits, where we
have used a $\log T_*=3.875\pm0.2$ ($\pm0.4$) prior on the temperature during (after)
the transient.  At the peak, the models require a modest optical depth, $\tau_V\simeq 1$,
but after a few months the optical depth starts to rise rapidly, almost reaching $\tau_V\simeq 10$
after 9 months before starting to fall.  Also note that the duration and energetics
of the transient are very different after correcting for the dust extinction -- far
from fading rapidly to below the flux of the progenitor in the first year, as suggested
by the optical light curves, the total luminosity is roughly constant at $3 \times 10^6 L_\odot$.
After almost 4 years, \cite{Li2002} found that the source had continued to fade while becoming 
bluer rather than redder, with $V \simeq 25.8 \pm0.3$ and $I > 25$ in early 2001.   

Fig.~\ref{fig:image97bs} shows the mid-IR images of the region from 2004. 
There are no clear mid-IR detections, although
the proximity of the source to a dust lane and the extended PAH emission makes 
it difficult to search for faint dusty point sources even in the wavelength-differenced 
images.  There may be a point source present in the $4.5\mu$m images, but most of its flux vanishes
in the wavelength differenced images.  We do not feel there are any significant 
detections and we view our aperture photometry results as upper limits.  
This leads to the late time SED of SN~1997bs shown in Fig.~\ref{fig:sed1997bs}.  Here we 
combine the optical fluxes in 2001 from \cite{Li2002} with the limits from the
2004 Spitzer observations.  This should be viewed as characteristic
of the optical epoch. When the optical depth is large, the visual fluxes
are exponentially sensitive to the optical depth while the mid-IR fluxes are
relatively insensitive to the optical depth and primarily affected by the 
weaker and slower dependence of the dust temperature on radius.  The optical
fluxes in 2001 are both fainter than the progenitor at V band, and bluer than
the transient in their V$-$I color.  

We modeled this SED by normalizing the DUSTY models to the $V=22.86\pm0.16$ magnitude of the progenitor,
and then fitting the fainter \cite{Li2002} optical magnitudes.  There is no
difficulty finding models with expansion velocities of $v_w=765$~km/s as
illustrated in Fig.~\ref{fig:sed1997bs}.  The
star is not very luminous, with $L_*=10^{5.1\pm0.2}$ and $10^{5.3\pm0.2}L_\odot$
for the graphitic and silicate models, and preferred temperatures of 
$T_*=15000$ and $20000$~K, respectively.  If we drop the flux of the
progenitor from the fits, still lower luminosities are allowed.  The
DUSTY optical depths are $\tau_V=3.5\pm0.3$ and $6.0\pm0.5$ where, as usual,
the silicate models have higher total optical depths but the two models
have comparable effective optical depths $\tau_{e,V}$.   The results 
including the mid-IR limits are consistent with the earlier estimates
using only the optical data.  For our standard
opacities, these correspond to ejected masses of $\log( M \kappa_{100}/M_\odot)=-1.4\pm0.1$
and $-1.2\pm0.1$, respectively, that are somewhat larger than the estimates
in Table~\ref{tab:objects2}.  Unlike SN~1954J, the mid-IR limits are weak enough 
to allow reasonable fits with a steady wind rather than an ejected shell 
provided the dust temperature at the inner edge is rather cool
($T \ltorder 1000$~K), and we show an example in Fig.~\ref{fig:sed1997bs}.
 
This latter point may be crucial because the optical depth history in
Fig.~\ref{fig:od1997bs} is not consistent with a short duration transient
in 1997.  In our discussion of $\eta$ Carinae, the problem was that a shell of
material with the present day optical depth would be too optically thick at earlier
times.  Here the problem is that a shell with the optical depth observed in the
first year would be too optically thin to obscure the star in 2001.  For a steady
mass loss rate, the optical depth is dominated by the inner edge, $\tau \propto 1/R_{in}$
with $R_{in} \simeq R_f$ set by the dust formation radius (Eqn.~\ref{eqn:opdepth2}).  When dust formation
stops, the time scale for the optical depth to drop is only $t_f \simeq R_f/v_w$
which is far too rapid.  We can illustrate this by fitting the optical depth
model from Eqn.~\ref{eqn:tevol} to the optical depth history, including an   
additional constant optical depth $\tau_c$ to model any additional foreground
or distant circumstellar dust.  Fig.~\ref{fig:od1997bs} shows two examples
producing reasonable fits.  The first model has
$\Delta t = 202$~days, $t_f = 67$~days, $\tau_0=9.8$ and $\tau_c=1.2$,
while the second has
$\Delta t = 109$~days, $t_f =152$~days, $\tau_0=18.9$ and $\tau_c=1.2$.
We do not include error estimates because we only seek to make a qualitative
point.  The key point point is the rapid collapse of the optical depth at time
$\Delta t+ t_f$ when the last of the ejecta reaches $R_f$.  By the time of the later HST,
LBT and/or Spitzer observations, the optical depth should be negligible. 
Sustaining the optical depth requires sustaining the mass loss.  For example, the evolution
including the optical depth estimate for 2001 can be fit using
$\Delta t = 1309$~days, $t_f=63$~days, $\tau_0=9.5$ and $\tau_c=1.2$, 
or
$\Delta t = 1026$~days, $t_f=152$~days, $\tau_0=17.8$ and $\tau_c=1.2$, 
as also shown in  Fig.~\ref{fig:od1997bs}.   Once again, the optical 
depth goes into a steep decline on time scale $t_f$ after time $\Delta t + t_f$.

We illustrated this for two different values of $t_f$ in order to discuss whether
the possible rise in optical depth after one month can also be due to dust formation
in the ejecta.  The models with $t_f =152$~days were chosen to have $R_f = t_f v_w \simeq 10^{15}$~cm
for an expansion velocity of $v_w=765$~km/s (Table~\ref{tab:objects}) and a dust 
formation temperature near $T_d \simeq 1500 (L/10^6 L_\odot)^{1/4}$~K. The
problem for the models with early dust formation at $t_f=67$~days is that
they must form dust at unphysically high temperatures 
$T_d \simeq 2200(L/10^6 L_\odot)^{1/4}$~K.  Dust formation in the ejecta after only a few
weeks implies still higher temperatures and must be impossible.  Since only
progenitors heavily obscured by dust have high enough wind densities to reform
dust in a pre-existing wind (see \citealt{Kochanek2011}), this strongly suggests that
the very early-time color changes are due to changes in temperature or the presence of
emission lines.  The larger color changes of the later phases are too large to
be explained by anything other than dust.

We have also been monitoring NGC~3627 with the LBT.  As with SN~1954J, we lack
a clear detection of the individual star and report the DAOPHOT photometry for
the estimated position of SN~1997bs in Table~\ref{tab:lbtphot} and the 
difference imaging light curves in Table~\ref{tab:isisphot}.  With 
$V=22.44$~mag, the DAOPHOT magnitude is consistent with the progenitor
flux but really represents an upper bound since the measured flux cannot
represent that of a single isolated star (compare Fig.~\ref{fig:image97bs}
and Fig.~\ref{fig:lbt}).  If we treat the estimates as measurements, the
best fits are for relatively low luminosities $L_* \simeq 10^{4.8}L_\odot$ 
and temperatures $T_* \simeq 8000$~K with no dust (except Galactic), but 
none of the models
are statistically consistent with the data.  All we can be certain of 
is that if the star is not obscured, it is not very luminous.
With five R-band epochs spread over 3.7 years, we see no evidence for
variability, as shown in Fig.~\ref{fig:lbt}.  The estimated slope
of the R-band light curve is $(0\pm 260) L_\odot$/year with an rms residual
of $1100 L_\odot$ -- if the star is there, its variability is
remarkably low ($ \ltorder 2\%$).  The two models shown in Fig.~\ref{fig:od1997bs} 
predict changes at V-band of 1-4\% over the period of the LBT observations,
so the lack of variability is marginally consistent with the model.  

In short, the standard picture of SN~1997bs as a brief eruption leading
to a surviving star obscured by the ejecta from the transient cannot
be correct in its details.  First, while the optical peak of the transient
was short, 45 days (Table~\ref{tab:objects}), the luminosity remained
high for at least 9 months, as did the period of high mass loss.  Second,
keeping the progenitor obscured either at the time of the HST observations
in 2001 or during our later LBT observations requires the high mass
loss rate ($\dot{M} \sim 10^{-3} M_\odot$/year) to have continued for
a still longer period of time, possibly even to the present day.
In this scenario,   the star identified
with SN~1997bs by \cite{Li2002} probably has to be an unassociated star
because matching it to the progenitor from \cite{Vandyk1999} requires
the SED of a hot star ($T_*> 10^4$~K) and dust cannot form in the 
presence of the soft UV radiation produced by such hot stars 
(see \citealt{Kochanek2011}). A cooler $T_* \simeq 7500$~K star that would 
facilitate dust production is strongly inconsistent with the limit 
on the V$-$I color from \cite{Li2002}.  The true optical depth would
be higher, and the true surviving star is faint and hidden by the
\cite{Li2002} source.   

If the star found by \cite{Li2002} genuinely is the surviving star, then
the most likely scenario is that the eruption did end sometime between
1998 and 2001, the optical depth did collapse as seen in Fig.~\ref{fig:od1997bs},
and the star is actually unobscured in 2001.  It is fainter and bluer than 
the progenitor because of a change in photospheric temperature
rather like the scenario proposed by \cite{Goodrich1989} for SN~1961V
where an LBV switches from a cooler S Doradus phase to its normal hot
state.  We should note, however, that the progenitor of SN~1997bs  
appears to be significantly fainter than the typical LBV.
One can generate additional scenarios, but resolving the nature of 
SN~1997bs depends on obtaining better data to tightly measure or
constrain the present day SED in the optical, near and mid-IR.

\begin{figure}[p]
\centerline{\includegraphics[width=5.5in]{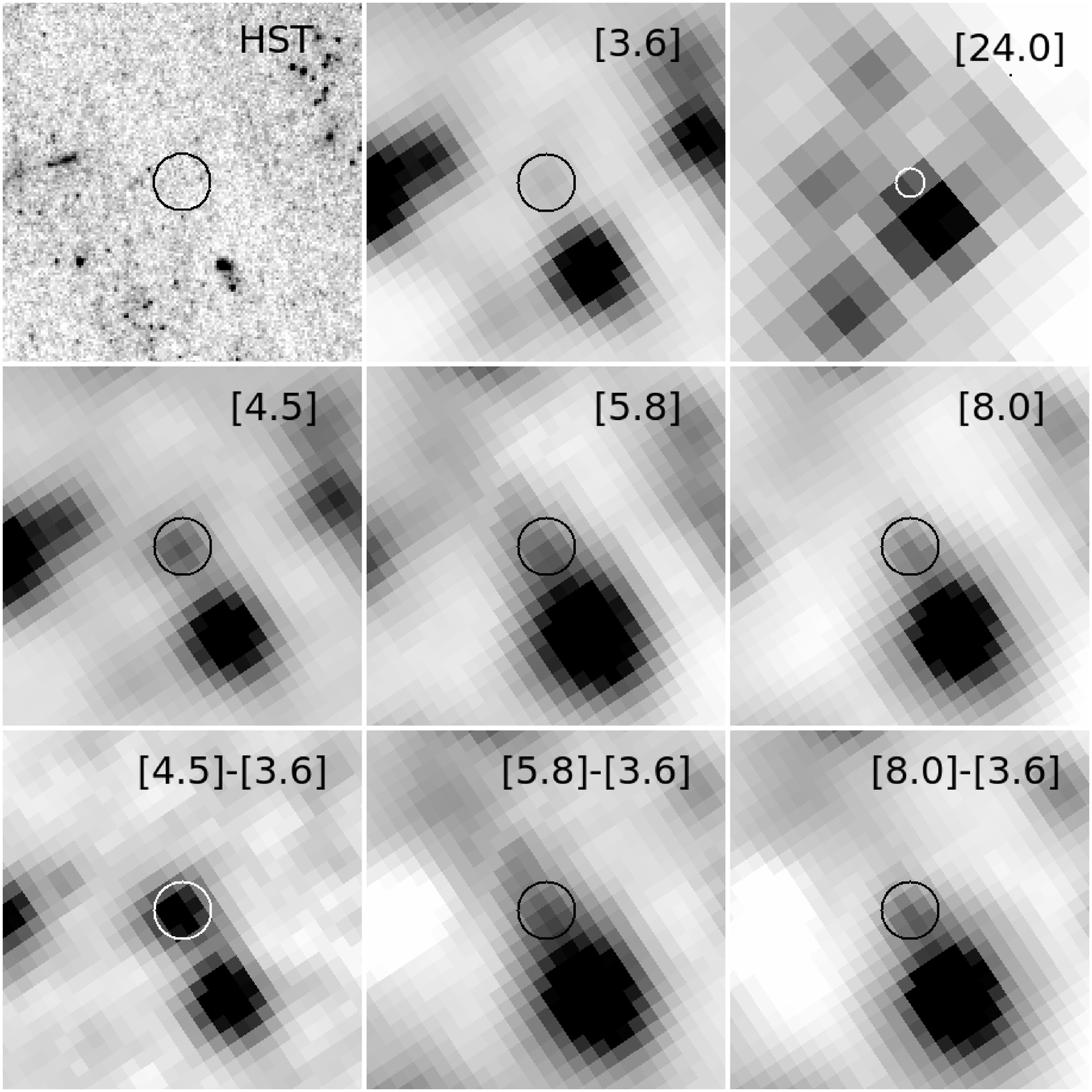}}
\caption{ The environment of SN~1999bw.  The top left, middle and right panels show the
  HST, $3.6$ and $24\mu$m images of the region, where the angular scale of the $24\mu$m
  image is twice that of the other panels.  The center left, middle and right panels
  show the $4.5$, $5.8$ and $8.0\mu$m images.  The bottom left, middle and right panels
  show the $[4.5]-[3.6]$, $[5.8]-[3.6]$ and $[8.0]-[3.6]$ wavelength-differenced images.
  A 1\farcs2 radius circle marks the position of the transient.  The HST image is from
  January 2001 while the mid-IR images are the averages of the available epochs.
  }
\label{fig:image99bw}
\end{figure}

\begin{figure}[p]
\centerline{\includegraphics[width=5.5in]{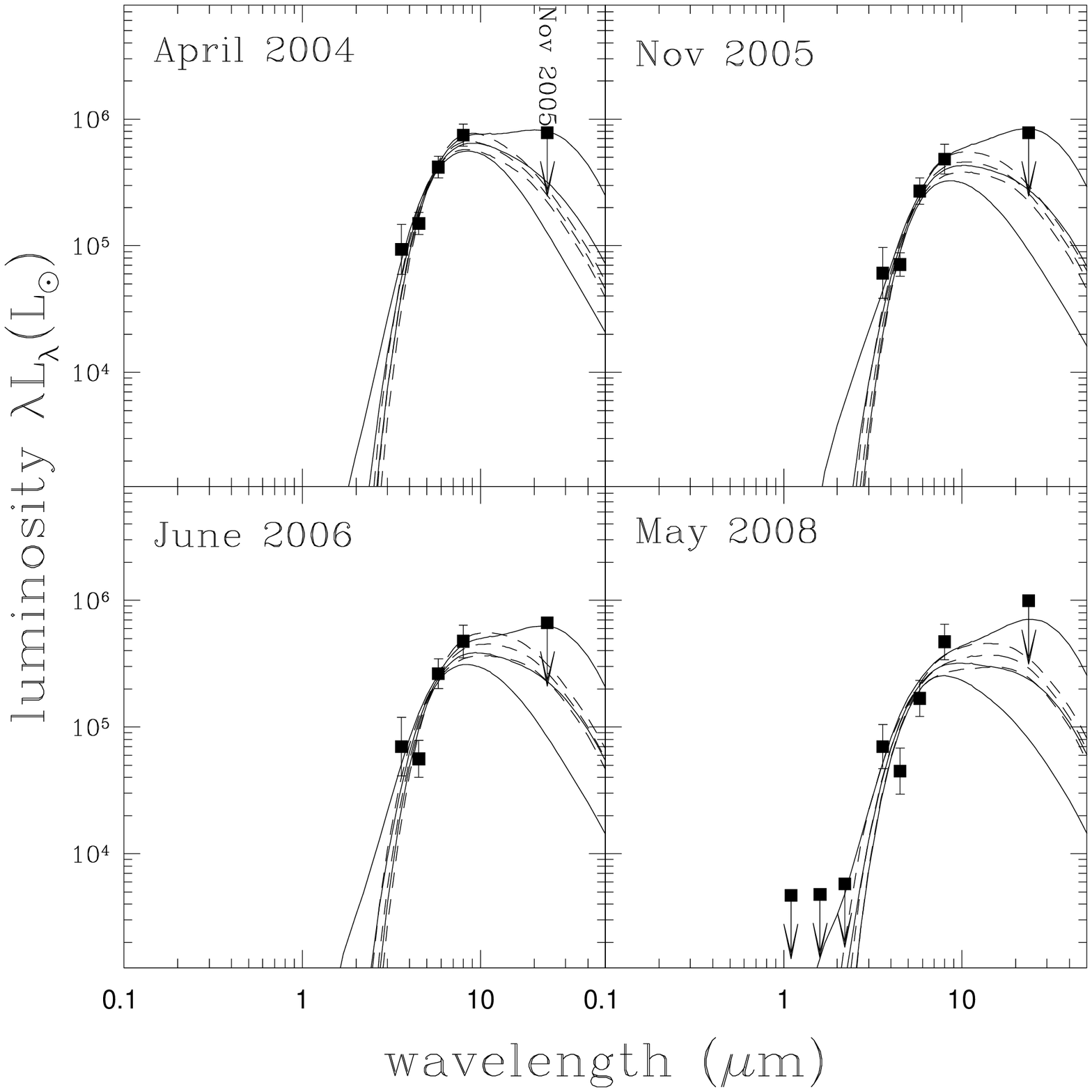}}
\caption{ The spectral energy distributions of SN~199bw in 2004, 2005, 2006 and
  2008.  The 2001 optical detection from \cite{Li2002} is used as an upper limit
  in 2004 and 2005, and the 2005 $24\mu$m upper limit is used as an upper
  limit in 2004.  
   The solid curves show the probability-weighted 
  average graphitic SED models and their dispersions with no constraint on the shell
  radius, while the dashed curves constrain the inner radius by a $630\pm30$~km/s
  expansion velocity prior.   
  The near-IR limits are shown at $1\sigma$, as appropriate for constraining
  models. The optical flux limits in 2008 are fainter than $10^3 L_\odot$ and 
  are not shown.
  }
\label{fig:sed1999bw}
\end{figure}

\subsection{SN~1999bw in NGC~3198}

There are few details on SN~1999bw.  It was discovered by LOSS (\citealt{Li1999}, 15 April 1999,
JD 2451284) and 
classified as a Type~IIn SN similar to SN~1997bs based on the narrow Balmer line
widths (\citealt{Garnavich1999}, \citealt{Filippenko1999}).  The transient peaked
at $V \simeq 18.4$~mag and then faded to $V \simeq 24$ after roughly 600~days
(\citealt{Li2002}).  \cite{Smith2011} appear to reject the \cite{Li2002} detection due to an 
un-discussed change in the estimated position, and instead assign a limit of $F555W >26.7$~mag ($3\sigma$)
with further non-detections
in October 2006 ($F606W>27.7$~mag) and April 2008 ($F606W>27.8$ and $F814W>26.8$~mag).
Very few observations were obtained near the peak, although \cite{Smith2011} note that 
the B$-$V$\simeq 0.8$ color was suggestive of foreground or circumstellar extinction.
\cite{Sugerman2004} reported detecting it
in all 4 IRAC bands in May 2004, at flux densities of $0.02\pm0.01$, $0.04\pm0.01$,
$0.11\pm0.02$ and $0.19\pm0.04$~mJy in the $[3.6]$, $[4.5]$, $[5.8]$ and $[8.0]$
bands, corresponding to a 450~K black body with luminosity of $10^{6.2}L_\odot$
(their reported flux at our adopted
distance from Table~\ref{tab:objects}) and a black body radius of $1.6 \times 10^{16}$~cm that corresponds to 
an expansion velocity $v_{ej}\simeq 1000$~km/s.  If we fit their estimates with
a DUSTY model of a hot star completely obscured by silicate dust, so there is no 
escaping flux from the star,  the best fit is for a dust radius of $2 \times 10^{16}$~cm
with a luminosity of $10^{6.0}L_\odot$ and inner/outer edge dust temperatures of
800/200~K.   Our fits to the Spitzer data are broadly consistent with \cite{Sugerman2004}.
There are no constraints on the progenitor of SN~1999bw.  As shown in Fig.~\ref{fig:image99bw},
the source is easily seen in the mid-IR (2004-2008) but with no signs of an optical 
counterpart (2001-2008).

In agreement with \cite{Smith2011}, if we fit the optical SED obtained at
the transient peak, the models require moderate extinction independent of
the assumed temperature.  For silicate DUSTY models we find $\tau_V \simeq 2.5$
for $T_* = 10^4$~K and $L_* \simeq 10^{7.3}L_\odot$.  The temperature 
is not well-constrained by the data, so cooler temperatures allow lower
optical depths and luminosities with the reverse for higher temperatures.
Without near/mid-IR data to constrain the dust temperature, this can
be either foreground or circumstellar extinction.  If this dust was
circumstellar, then SN~1999bw peaked in the mid-IR rather than the 
optical, similar to the 2008 NGC~300 transient (\citealt{Kochanek2011}).

Fig.~\ref{fig:sed1999bw} shows the evolution of the SED from 2004 to
2008.  Here we combine the optical limits from \cite{Smith2011}
with our infrared estimates from Tables~\ref{tab:log} and \ref{tab:hstphot}.
We use the 2001 optical limit as a bound
in 2004 and 2005 and the 2005 $24\mu$m limit as a bound in 2001,
as the most likely evolution given the available data is to fade
in the optical and brighten in the far-IR.  Clearly at some point between the 
transient peak and 2001 the system formed large quantities of 
dust and became a mid-IR dominated transient even if it was not one 
at peak.  The graphitic models provide modestly better fits, but
the parameters for the two dust types are generally similar. 
Fig.~\ref{fig:sed1999bw} also shows the probability-weighted 
mean graphitic SED models and the dispersion around them.  

Without any optical detections or a progenitor flux, the models
simply demand that the optical depth is high at all epochs with
the limit determined by the stellar temperature -- for $T_* = 10000$~K
we find $\tau_V \gtorder 50$ in 2008, while a hotter  
$T_*= 30000$~K star only requires $\tau_V \gtorder 20$. The 
limits largely hold for all epochs, so we have no strong 
constraints on the evolution of the optical depth.  Either dust
model fits the data well. The
luminosity was roughly constant from 2004 to 2008 at
$L_* \simeq 10^{6.0\pm 0.2} L_\odot$  depending on the
specific model, possibly with a modest fading of 0.2~dex. There
is no strong evolution in the dust temperature. 
This means
that the energy radiated in the later phases, $E \simeq 10^{48}$~ergs,
significantly exceeds that from the peak, 
$E \simeq 2 \times 10^{45} (\Delta t_{1.5}/10~\hbox{days})$~ergs
(Table~\ref{tab:objects}), unless the un-observed duration of the peak
was quite long, $\Delta t_{1.5} \gtorder 1$~year.  The best
models prefer dust radii somewhat larger than expected for 
an inner radius expanding at $630$~km/s and are certainly
inconsistent with the outer radius expanding at this velocity
for our standard $R_{out}/R_{in}=2$ models.
The ejected mass of
$ M_E \simeq 0.4 (\tau_V/10)(100~\hbox{cm}^2/\hbox{g}/\kappa_V)(v_w/630~\hbox{km/s})^2 M_\odot$
required to have a high optical depth in 2008, 9 years after the peak,
is far more than expected from the observed short duration transient
(Table~\ref{tab:objects2}).  Either SN~1999bw was an explosive transient
with $f\ll 1$ or the true duration was far longer, with $\Delta t_{1.5} \simeq 0.3$~years 
for $f\simeq 1$.  The SEDs are also well-fit by a dusty graphitic wind.

Since the SEDs are marginally consistent with an expanding
shell, it is possible that SN~1999bw was a simple eruption
ejecting $M \sim 1M_\odot$ of material at 630~km/s from a
luminous $L_* \simeq 10^6 L_\odot$ star.  The amount of mass
loss needed to explain the SED and its evolution seems to 
imply transient energies that are incompatible
with a radiatively driven ejection, although this is not certain
given the incomplete early-time light curve.  It seems far 
more likely that SN~1999bw is similar to SN~2008S and the 2008
NGC~300 transient as originally suggested by \cite{Thompson2009}.
The dust probably present at peak, the complete, long-lived 
shrouding with a dust radius that is somewhat large for the
velocity estimate, and the mid-IR luminosity are all very
similar to these two transients (see \citealt{Kochanek2011}).

\begin{figure}[p]
\centerline{\includegraphics[width=5.5in]{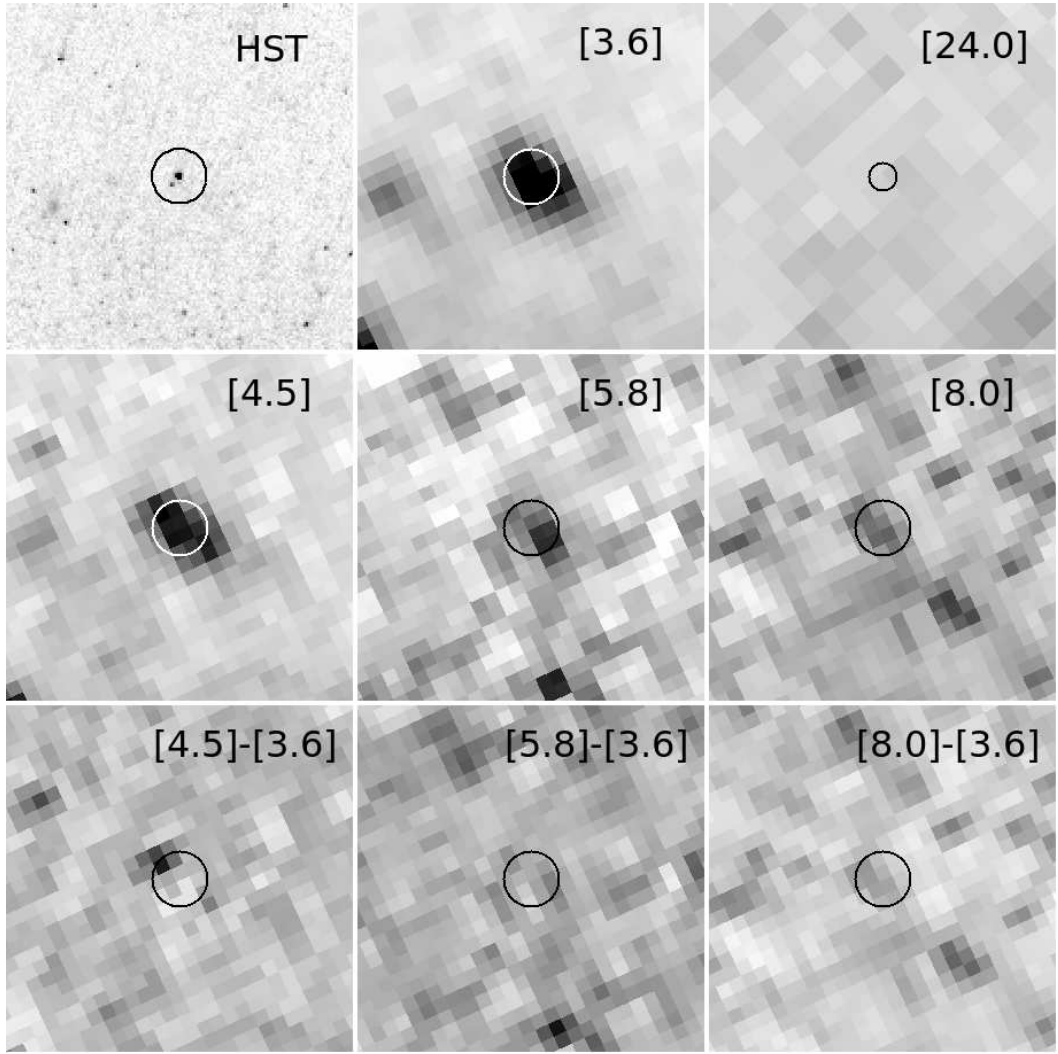}}
\caption{ The environment of SN~2000ch.  The top left, middle and right panels show the
  HST, $3.6$ and $24\mu$m images of the region, where the angular scale of the $24\mu$m
  image is twice that of the other panels.  The center left, middle and right panels
  show the $4.5$, $5.8$ and $8.0\mu$m images.  The bottom left, middle and right panels
  show the $[4.5]-[3.6]$, $[5.8]-[3.6]$ and $[8.0]-[3.6]$ wavelength-differenced images.
  A 1\farcs2 radius circle marks the position of the transient.
  }
\label{fig:image00ch}
\end{figure}

\begin{figure}[p]
\centerline{\includegraphics[width=5.5in]{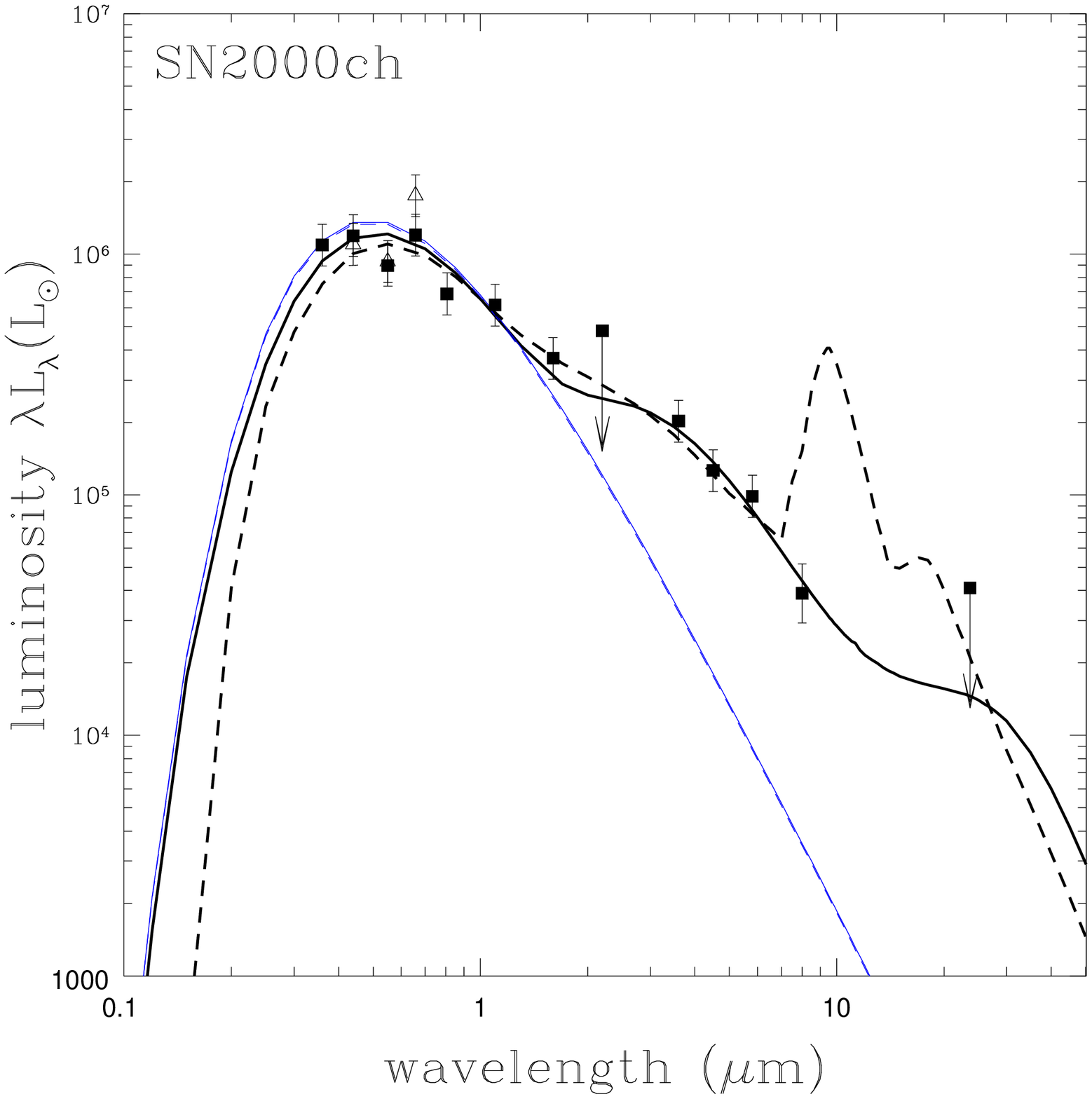}}
\caption{ The spectral energy distribution of SN~2000ch.  The mid-IR SED
  combines the IRAC data from December 2007 with the MIPS data from May 2008,
  although the star seemed to be in quiescence during this period based on
  \cite{Pastorello2010}.  The open triangles show the B, V and R data from
   near the midpoint of this period (March 2008) from \cite{Pastorello2010},
   and they are largely hidden by the filled squares showing a very similar SED measured 
   in May 2000 by \cite{Wagner2004} that that also included near-IR measurements.
   The solid (dashed) curves are DUSTY wind fits for $\tau_V=0.2$ of 
   graphitic ($\tau_V=1$ of silicate) dust around a $T_*=7500$~K, 
   $L_* \simeq 10^{6.3}L_\odot$ star.  The thick curves show the 
   model of the observed SED and the thin curves show the model SED of
   the unobscured star.
  }
\label{fig:sed2000ch}
\end{figure}

\subsection{SN~2000ch in NGC~3432}

\cite{Papenkova2000} identified a new variable star in NGC~3432 in May 2000 and \cite{Filippenko2000}
argued that it was a sub-luminous Type~IIn analogous to SN~1997bs or SN~1999bw and likely to be
an LBV outburst rather than a SN.  The initial outburst was studied in detail by \cite{Wagner2004},
and \cite{Pastorello2010} report on three subsequent transients in 2008 and 2009. The first
peak, on 3 May 2000 (JD 2451668) is the only one prior to the Spitzer observations.  
This is clearly a variable star.
In quiescence, the star is very luminous,
$R \simeq 19.5$ ($M_R \simeq -10.7$) with peaks in the transients at $R \simeq 18$ ($M_R \simeq -12$)
and post-transient minima near $R \simeq 21$ ($M_R \simeq -9$) that are interpreted as obscuration
by newly formed dust.  In outburst, the SEDs are relatively well fit by $T_* \sim 8000$ to $10000$~K
with excesses in the U-band and due to H$\alpha$ emission in the R-band.  In quiescence it seems to have a 
cooler temperature, $T_* \simeq 5000$~K.  SN~2000ch is easily detected both in the optical
and in the mid-IR, as shown in Fig.~\ref{fig:image00ch}.  

SN~2000ch was observed with Spitzer only twice, with IRAC in December 2007 and MIPS in May 2008,
where \cite{Pastorello2010} measured B, V and R magnitudes of $20.10\pm0.21$, $19.75\pm0.12$ and 
$18.67\pm0.20$~mag on 2008 Mar 12, roughly between the observations.  We picked an epoch from 
\cite{Wagner2004} (2000 May 21) with similar B, V, R magnitudes that also had U, I, J, H and
a K limit as an extended model of the SED.  As we see in Fig.~\ref{fig:sed2000ch}, the star
probably has a modest mid-IR excess.  Shell models expanding at $1400$~km/s fail to fit the SED
well.  More slowly expanding ($460$~km/s) graphitic shells can fit the SED well, but 
silicate shells always have problems because of the strong $8\mu$m feature. The models
generically have a $T_*=7500$~K, $L_*\simeq 10^{6.3}L_\odot$ star, small optical 
depths and relatively hot dust.  All the shell models have the further problem that an optical depth of 
$\tau_V \simeq 0.1$ in 2008 would be $\tau_V \simeq 1$ in 2003.  For comparison,
one of the faintest epochs in \cite{Pastorello2010} (24 November 2008) is well fit
with no circumstellar dust, $T_*=7500$~K and $L_* \simeq 10^{5.3} L_\odot$.  Formally,
the ejecta masses in the best fit shell models are $\log( M_E\kappa_{100}/M_\odot)=-1.8\pm0.2$
and $-3.2\pm0.1$ for the graphitic and silicate models. While \cite{Wagner2004}
propose that the  post-eruption flux minima are due to dust formation, they are
too close to the transient and too short lived to be consistent with the optical
depth evolution of a geometrically expanding shell.  

The simplest way to have a slowly varying optical depth  and 
an effective dust radius different from that predicted by 
the apparent velocities is to produce the dust in 
a relatively steady wind rather than an impulsive ejection.  Fig.~\ref{fig:sed2000ch}
shows the silicate and graphitic DUSTY wind models fit to the SED rather than 
the usual shell models.  The graphitic model fits well, with $\tau_V=0.2$. The
silicate model, with $\tau_V=1$, fits less well because of the $8\mu$m peak.
Correlations between the variability of the star and dust formation in the wind
can then help to explain some of the correlations.  For example, the dust
formation radius $R_f \propto L_*^{1/2}$ (Eqn.~\ref{eqn:rform}) while the
optical depth of a wind is $\tau \propto R_f^{-1}$ (Eqns.~\ref{eqn:opdepth2} and 
\ref{eqn:twind}), so increasing the luminosity drives the dust formation radius
outwards and the optical depth down to make the star bluer independent 
of any change in the stellar temperature.  There may also be correlated changes in
in the mass loss rate, and the dust opacity will change due to both changes in the
gas density at the dust formation radius and changes in the stellar spectrum
(see, \citealt{Kochanek2011b}).  In short, we propose that the dust-related
behaviors of SN~2000ch are due to modulated dust formation in a quasi-steady
wind rather than strongly transient mass ejections associated with the
luminosity peaks.  Near-IR (particularly K-band) and mid-IR light curves  
would test this scenario in detail.

\begin{figure}[p]
\centerline{\includegraphics[width=5.5in]{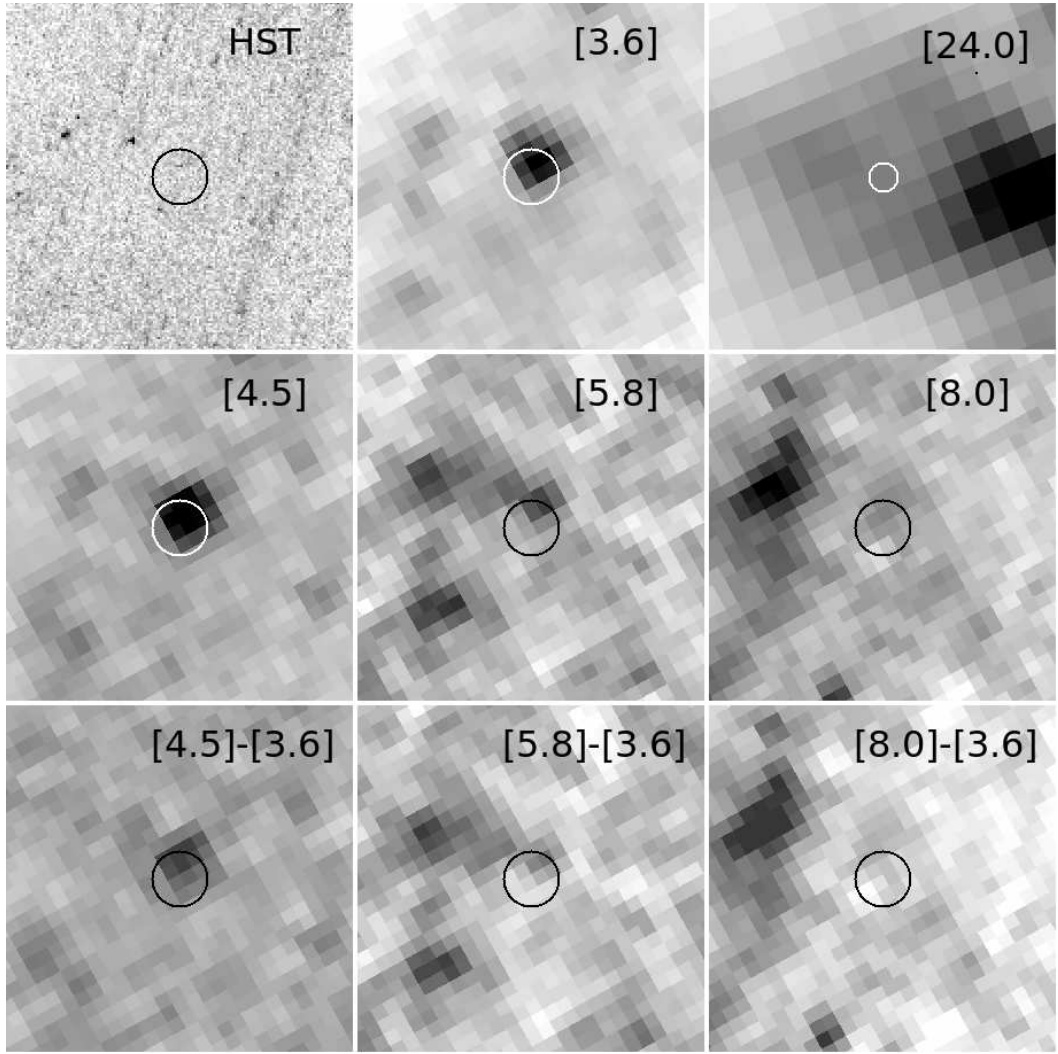}}
\caption{ The environment of SN~2001ac.  The top left, middle and right panels show the
  HST, $3.6$ and $24\mu$m images of the region, where the angular scale of the $24\mu$m
  image is twice that of the other panels.  The center left, middle and right panels
  show the $4.5$, $5.8$ and $8.0\mu$m images.  The bottom left, middle and right panels
  show the $[4.5]-[3.6]$, $[5.8]-[3.6]$ and $[8.0]-[3.6]$ wavelength-differenced images.
  A 1\farcs2 radius circle marks the position of the transient.
  }
\label{fig:image01ac}
\end{figure}

\begin{figure}[p]
\centerline{\includegraphics[width=5.5in]{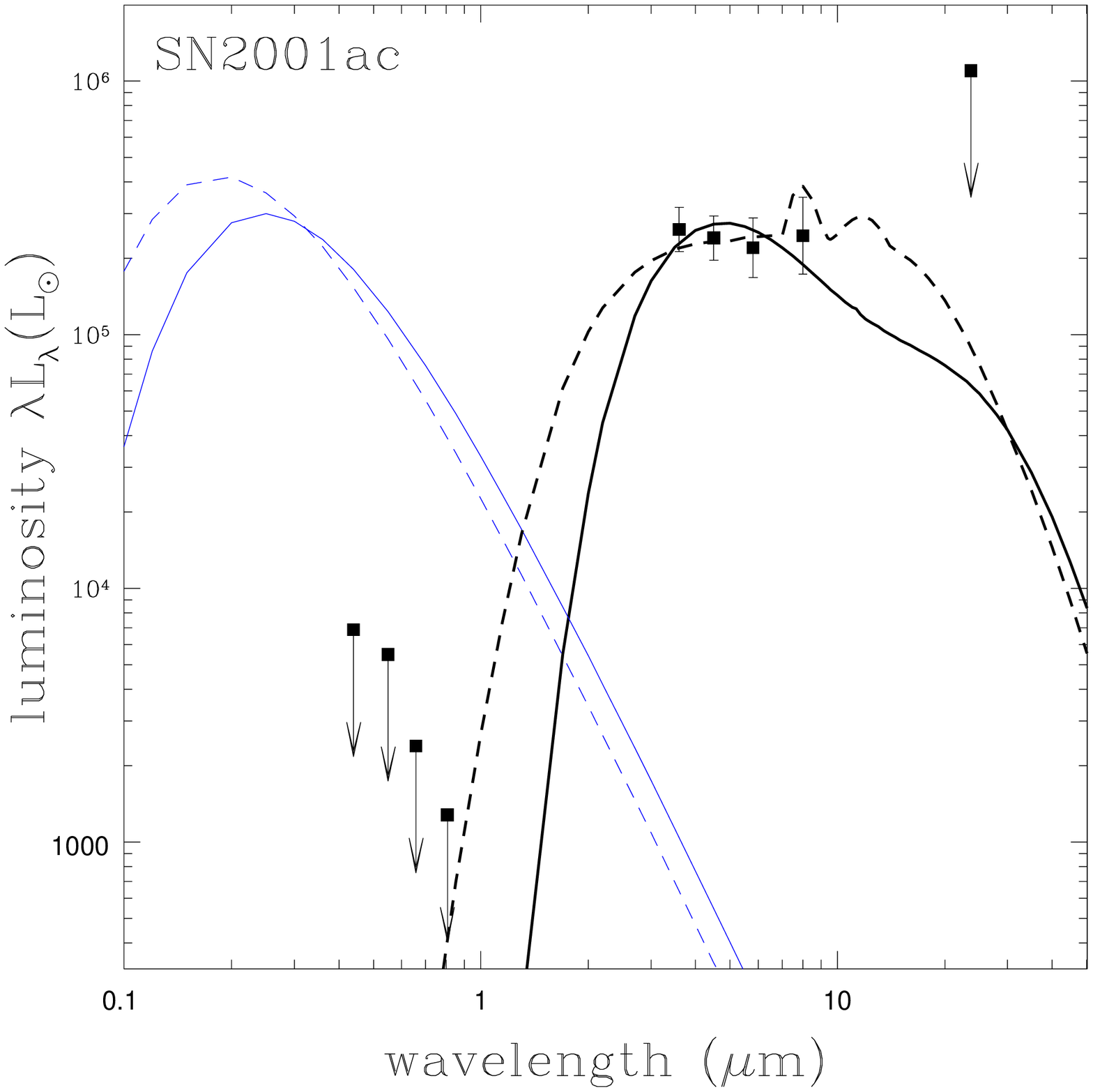}}
\caption{ 
  The spectral energy distribution of SN~2001ac. The November 2008
  optical limits are from \cite{Smith2011}. The Spitzer observations 
  are from June 2008 (IRAC) and January 2008 (MIPS). The optical limits 
  have been converted to $1\sigma$ so that model offsets can be 
  properly interpreted.  These models show relatively high
  optical depth ($\tau_V=25$ for graphitic and $18.5$ for silicate dust), slowly
  expanding shells around a $T_*=20000$~K star.  The stellar luminosities are 
  $L_*=10^{5.6} L_\odot$ (solid, graphitic dust) and $L_*=10^{5.8}L_\odot$ (dashed, 
  silicate dust).  There are acceptable fits for moderately lower optical depths, particularly
  if we increase the (unconstrained) stellar temperature.  The thin curves
  show the model for the unobscured source. 
  }
\label{fig:sed2001ac}
\end{figure}

\subsection{SN~2001ac in NGC~3504}

SN~2001ac was discovered by \cite{Beckmann2001} on 12 March 2001 (JD 2451981)
at about 18.2~mag and \cite{Matheson2001} reported that it had a Type~IIn spectrum 
similar to SN~1997bs and SN~1999bw.  Observations prior to the
transient in 1995 show no source, but the limits are of limited use because the 
position is near the chip edge in V band (F606W) and the other filter is
in the UV (F218W).  There is no source apparent in the post-transient
HST images from 2008 November 18/22, with optical ($3\sigma$) limits of  
$26.5$, $26.3$, $25.8$ and $26.1$~mag ($3\sigma$, \citealt{Smith2011}) 
The epoch of the HST band is close to those of the Spitzer data 
(2008 June 20 for IRAC and 2008 January 5 for MIPS).

Fig.~\ref{fig:image01ac} shows the HST I band and Spitzer images of the region.
There appears to be a relatively bright, red mid-IR source at the transient
position in the IRAC images, and no obvious source at $24\mu$m.  We have no
trouble modeling the SED with a $290$~km/s expansion velocity, and in Fig.~\ref{fig:sed2001ac}
we show models with $T_*=20000$~K and $L_*=10^{5.6}L_\odot$ for graphitic
dust and $L_*=10^{5.8}$ for silicate dusts.  High stellar temperatures are
not required, but the stellar luminosity does not depend strongly on the
temperature.  The nominal optical depths are high, $\tau_V=25$ and $18.5$ for the
graphitic and silicate models shown in Fig.~\ref{fig:sed2001ac}, with hotter
models allowing modestly lower optical depths.  Averaging over the
models we find $\log \tau_V \simeq 1.3 \pm 0.1$.  The ejected masses
for these optical depths are of order $\log( M_E\kappa_{100}/M_\odot) =-1.0\pm0.2$
for either model, consistent with $f \gg 1$
and a radiatively driven mechanism.  This is the only example where the
silicate optical depths are similar to the graphitic, but this is
driven by the lack of optical measurements for either the progenitor or
the post-transient source.  Both types of dust fit the SED reasonably
well, but the dust temperature at the inner edge has to be high,
with $T_d=1000$~K and $1300$~K for the models in the figure.  If the
Spitzer source is not related to SN~2001ac, a surviving, un-obscured
star can be luminous only if very hot, with $L_* \simeq 10^{3.9}L_\odot$ for
$T_*=10000$ and $L_* =10^{5.3} L_\odot$ for $T_*=40000$~K.

Complete understanding of this system requires either detecting an optical 
counterpart or
observing the time evolution of the mid-IR properties.  If there is an ejected
shell of dusty material, the warm Spitzer mid-IR fluxes
should be falling relatively rapidly because of the expansion of the shell,
and the optical depth will be constrained by the reappearance of the star
in the optical.  For example, 
the V band luminosity will reach the present limit circa 2015 for the
silicate model and 2037 for the graphitic model
for the models in Fig.~\ref{fig:sed2001ac}.

\begin{figure}[p]
\centerline{\includegraphics[width=5.5in]{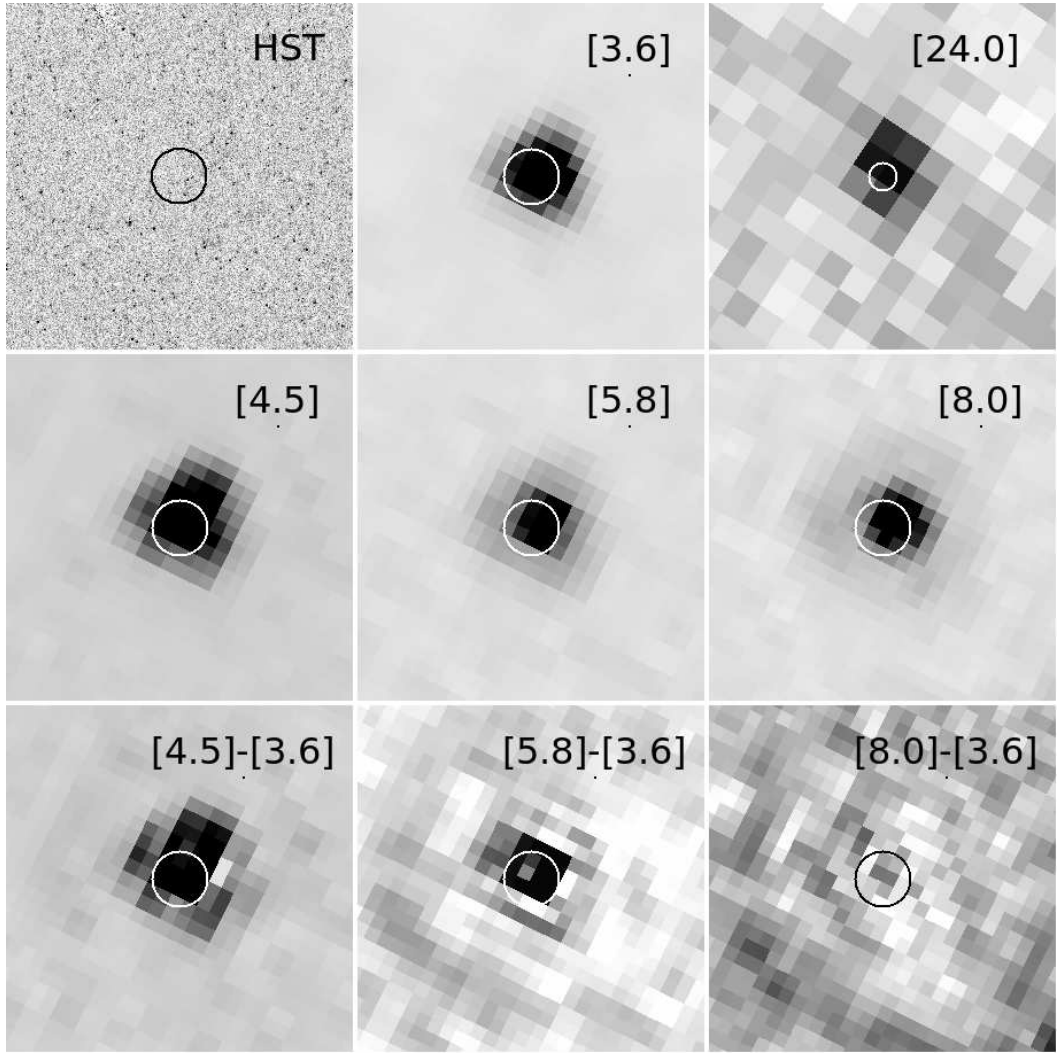}}
\caption{ The environment of SN~2002bu.  The top left, middle and right panels show the
  HST, $3.6$ and $24\mu$m images of the region, where the angular scale of the $24\mu$m
  image is twice that of the other panels.  The center left, middle and right panels
  show the $4.5$, $5.8$ and $8.0\mu$m images.  The bottom left, middle and right panels
  show the $[4.5]-[3.6]$, $[5.8]-[3.6]$ and $[8.0]-[3.6]$ wavelength-differenced images.
  A 1\farcs2 radius circle marks the position of the transient.
  }
\label{fig:image02bu}
\end{figure}

\begin{figure}[p]
\centerline{\includegraphics[width=5.5in]{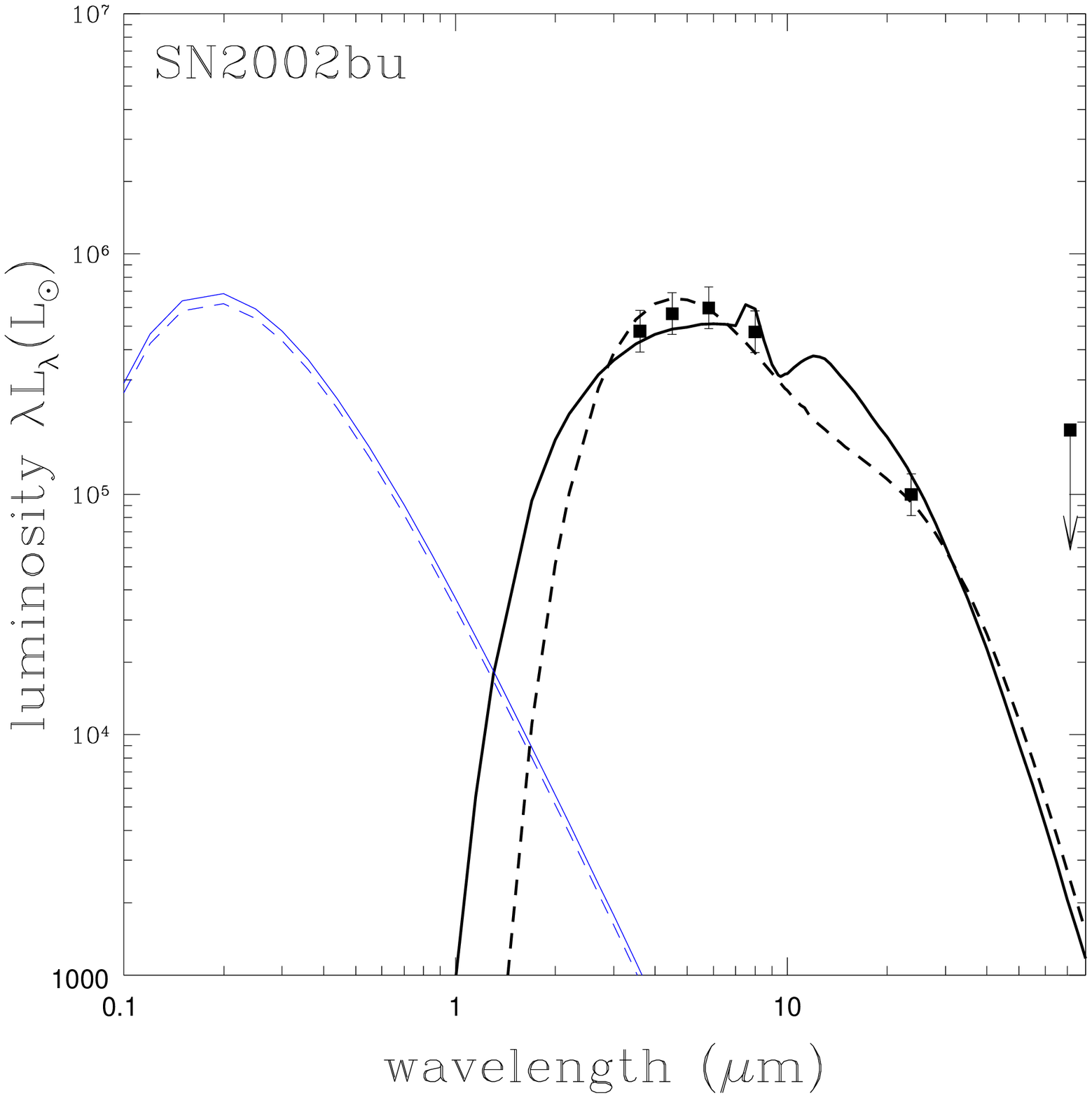}}
\caption{ The spectral energy distribution of SN~2002bu.  The mid-IR SED
  combines the Spitzer data from April/May 2004 with the HST upper
  bounds from March/April 2005.  These models have optically thick
  $\tau_V=30$ shells expanding at roughly $800$~km/s around a 
  $T_*=20000$~K, $L_* \simeq 10^{5.9}L_\odot$ star with 
  graphitic (solid) or silicate (dashed) dust.  The thick black
  curves show the observed SED and the thin curves show the
  SED of the unobscured star.  Note that the wavelength
  range has been expanded to include the limit at $70\mu$m.
  The $1\sigma$ optical limits are fainter than $10^3 L_\odot$ and
  are not shown.
  }
\label{fig:sed2002bu}
\end{figure}

\subsection{SN~2002bu in NGC~4242}

SN~2002bu was discovered by \cite{Puckett2002} on 2002 March 28 (JD 2452362)
and was classified as a Type~IIn (FWHM $\sim 1100$~km/s) by \cite{Ayani2002}.  
\cite{Foley2007} show a light curve covering the first 70 days, noting that 
it declines much more slowly than a typical Type~II and is relatively red 
($B-R\simeq 1.5$) at the end of this period.
The early light curve from \cite{Smith2011} is consistent with no extra
extinction if the transient is initially cool ($7500$~K at peak) and then 
cools further ($5000$~K after 3 months).  If we fix the temperature at
$10000$~K, then the best fit models have modest optical depths at peak
($\tau_V \simeq 1$), which then increases to $\tau_V \simeq 3$ after
three months.  There is not, however, enough spectral coverage to
choose between changes in temperature and extinction.   At the last
optical epoch, the ejecta should still be close enough to the star
($R \sim 7 \times 10^{14}$~cm) to avoid dust formation (Eqn.~\ref{eqn:rform}), so any new
dust would have to be reformed in a preexisting dense wind.

By 2004/2005 this has clearly changed, as illustrated by the 
HST I band image and the Spitzer images
of the region shown in Fig.~\ref{fig:image02bu}.
No optical counterpart is apparent, but, as noted by \cite{Thompson2009}, there is a bright,
red mid-IR source.  The lack of any bright sources in the HST images means
we could make no independent check of the header astrometry, but no source
is apparent at the nominal position with limits of roughly $\gtorder 25$~mag 
in all four bands
(Table~\ref{tab:hstphot}).  Fig.~\ref{fig:sed2002bu} shows the SED, combining the
HST upper bounds from March/April 2005 with the Spitzer observations from
April/May 2004.  We have little difficulty matching the SED with expansion
values similar to the estimate by \cite{Smith2011} in Table~\ref{tab:objects2}.
Fig.~\ref{fig:sed2002bu} uses expansion speeds of $825$ and $790$~km/s for
the graphitic and silicate models, respectively, and a $T_*=20000$~K,
$L_*=10^{5.9}$ and $10^{6.0}L_\odot$ star.  The particular models
have optical depths of $\tau_V=30$, but the optical depth is not
well constrained without an optical detection.  Producing this 
optical depth requires $\log (M_E\kappa_{100}/M_\odot) \simeq -0.9 \pm 0.2$ of ejected material
depending on the parameters and the dust type, which is consistent
with our estimates for a radiatively driven process (Table~\ref{tab:objects2}).

The available data are consistent with a transient ejecting a shell,
although they are also consistent with the different phenomenology
of SN~2008S, the 2008 NGC~300 transient and SN~1999bw.
If the mid-IR emission is due to an illuminated, expanding shell, the 
$3.6$ and $4.5\mu$m fluxes should be very different today because the
shell radius will quintuple and the dust temperature will be halved 
between 2004 and 2012. At $3.6\mu$m the source should be roughly ten
times fainter than in 2004.  Similarly, if the optical depth in
2004 was $\tau_V \sim 30$, then the source should be reappearing
in the optical as well because the optical depth will have dropped to
$\tau_V \sim 1$.  These issues can be easily addressed by new 
observations. 

\begin{figure}[p]
\centerline{\includegraphics[width=5.5in]{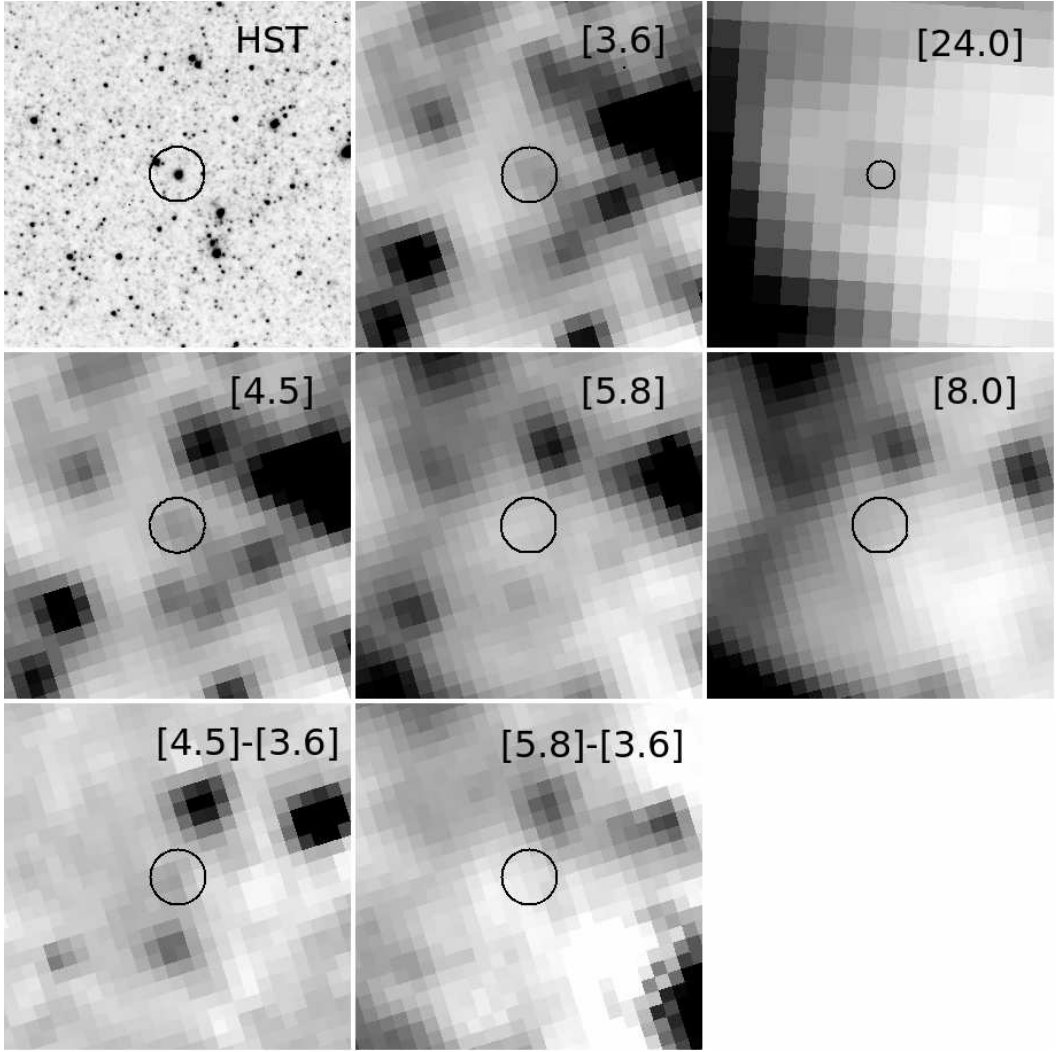}}
\caption{ The environment of SN~2002kg/V37 in NGC~2403.  The top left, middle and right panels show the
  HST, $3.6$ and $24\mu$m images of the region, where the angular scale of the $24\mu$m
  image is twice that of the other panels.  The center left, middle and right panels
  show the $4.5$, $5.8$ and $8.0\mu$m images.  The bottom left, middle and right panels
  show the $[4.5]-[3.6]$ and $[5.8]-[3.6]$ wavelength-differenced images. The subtraction
  procedure failed for the [8.0] micron image.
  A 1\farcs2 radius circle marks the position of the transient.
  }
\label{fig:image02kg}
\end{figure}

\begin{figure}[p]
\centerline{\includegraphics[width=5.5in]{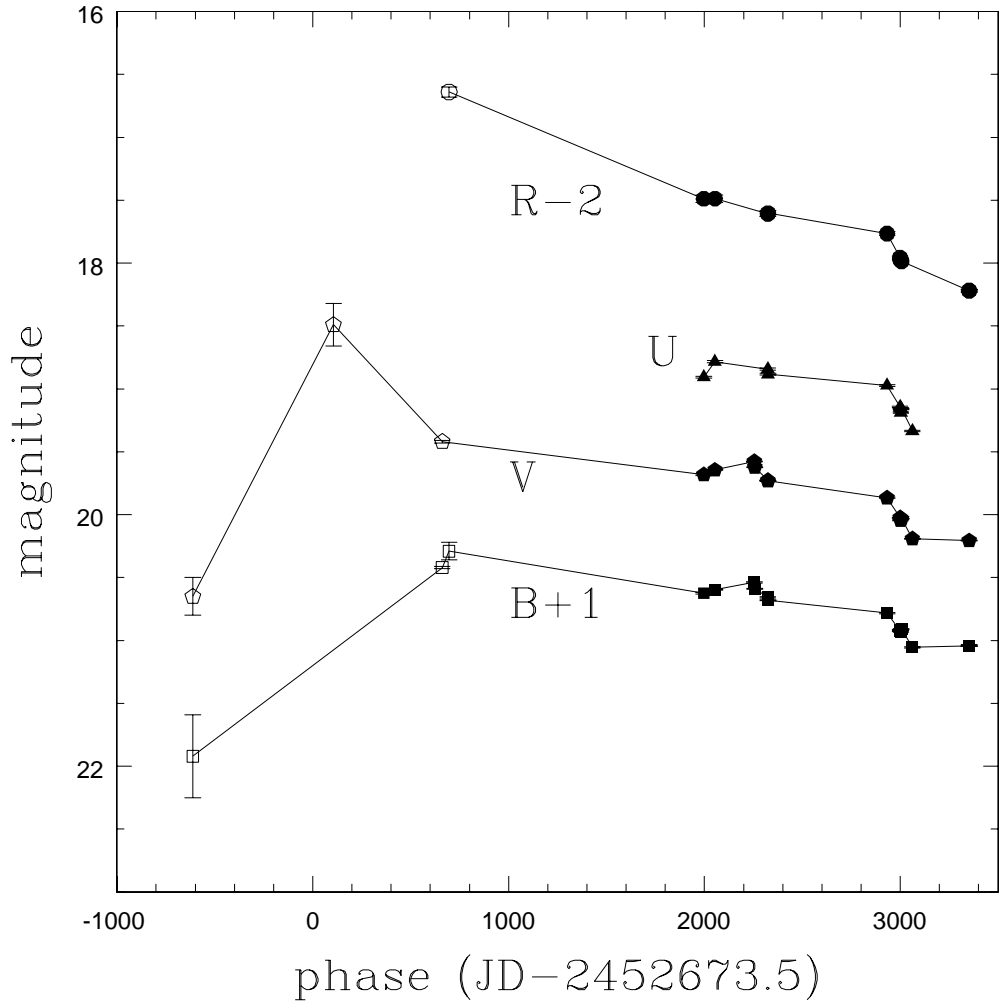}}
\caption{ Optical light curves of SN~2002kg.  The data prior to
  1000 days after the transient is from \cite{Maund2006} (open points)
  and the data afterward is from the LBT (filled points).  The source is 
  slowly fading with roughly constant optical colors.
  }
\label{fig:opt2002kg}
\end{figure}

\begin{figure}[p]
\centerline{\includegraphics[width=5.5in]{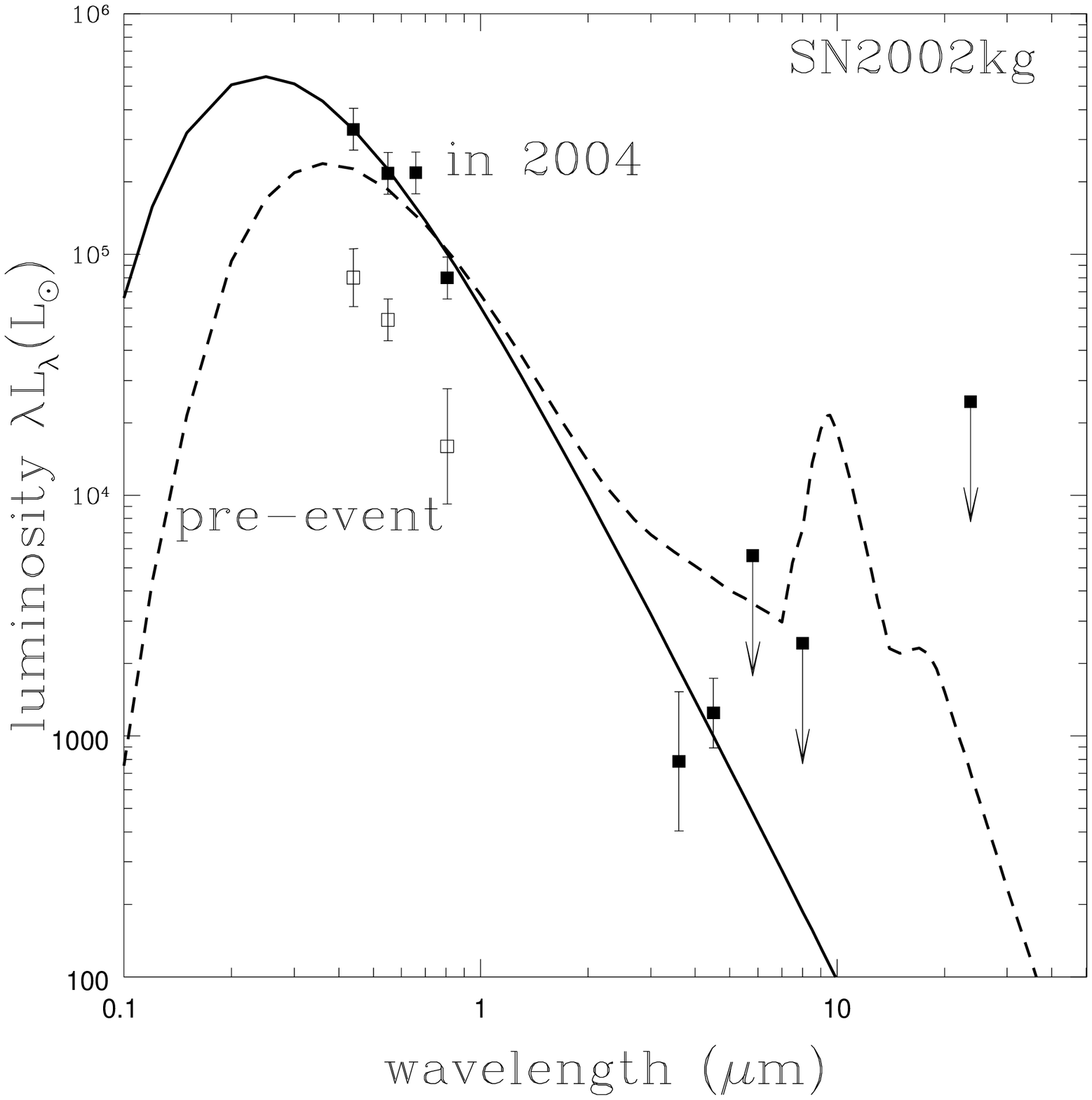}}
\caption{ The spectral energy distribution of SN~2002kg in 2004,
  two years post-peak.  The solid line is a model 
  with $T_*=15000$~K, $L_*=10^{5.9} L_\odot$ and no dust, while
  the dashed line shows a silicate model with $T_*=10000$~K,
  $L_*=10^{5.6}L_\odot$ with $\tau_V=0.1$ of dust at 
  $R_{in}=2 \times 10^{15}$~cm ($350$~km/s for 2~years).
  The filled squares show the
  SED in late 2004 where the optical points are from \cite{Maund2008}.
  The open squares show the pre-transient SED.  
  }
\label{fig:sed2002kg}
\end{figure}

\subsection{SN~2002kg/V37 in NGC~2403}

SN~2002kg was discovered on 26 October 2003 (JD 2452827) and initially classified as a Type~IIn SN based
on its narrow emission lines (\citealt{Schwartz2003a}).  \cite{Weis2005} identified it with
the LBV V37 in \cite{Tammann1968}, reconstructed its historical light curve, showed it was
a strong H$\alpha$ source and also found it in post-transient HST observations.  \cite{Maund2006}
found that V37 had B, V and I magnitudes of $20.9\pm0.3$, $20.65\pm0.15$ and $20.9\pm0.6$
roughly two years before the transient (the correct SN~2002kg magnitudes for \cite{Maund2006}
are found in \cite{Maund2008}), similar to its typical historical fluxes.  This corresponds
to a luminosity of $10^{5.0}L_\odot$ for $T_*=10^4$~K (cooler black bodies fit poorly)
or $10^{5.5}L_\odot$ for $20000$~K.  The transient 
only brightened by $\Delta V \simeq -2.2 \pm 0.2$ magnitudes, from $M_V\simeq -8.2$ to
$M_V \simeq -10.4$, a level which it seems to have achieved several times since its discovery (\citealt{Weis2005}).      
\cite{Maund2006} estimate that the local extinction is of order $E(B-V) \simeq 0.17\pm0.02$.

Fig.~\ref{fig:image02kg} shows the region around SN~2002kg.  The HST V band (F606W) image
is from August 2004.  The mid-IR images combine the available epochs and there is no
evidence for mid-IR variability.  There appears to be
a counterpart to SN~2002kg in the IRAC bands, which subtracts well in the wavelength
differenced images, indicating that the source is not dominated by dust.  There is a
$24\mu$m peak slightly East of our estimate of the position which does not seem to be
associated with the source.  In fact, it is likely an artifact of some kind since it only
appears in one of the SST epochs.  

While there is no good evidence for mid-IR variability, the source is clearly varying
in the optical, as illustrated in Fig~\ref{fig:lbt}.  The source is fading, since there
is a strong Slope detection, but relatively steadily, since there is little signal in the RMS image.
Such tricks are hardly needed for SN~2002kg/V37, and it is included in Fig.~\ref{fig:lbt}
as a contrast to the behaviors of SN~1954J and SN~1997bs.  Fig.~\ref{fig:opt2002kg} 
shows the combined light curves from \cite{Maund2006} and our LBT observations (Table~\ref{tab:lbtphot}). The
source is clearly fading, and at relatively constant colors except for the R-band 
flux, which seems to drop faster than the other bands.  The SED, shown in Fig.~\ref{fig:sed2002kg},
suggests that the R band flux in \cite{Maund2006} is significantly enhanced by
H$\alpha$ emission.  At the present mean rate of decline, $\simeq 0.14$~mag/year at V band,
it will return to its quiescent magnitude in another $\sim 3$~years.  In units of
luminosity, the R-band decline seen in Fig.~\ref{fig:lbt} is about $13000 L_\odot$/year
with rms residuals of $2300 L_\odot$.

Fig.~\ref{fig:sed2002kg} shows our models of the SED, where we match the last \cite{Maund2006}
epoch, two years post peak, to the mid-IR fluxes measured in the same year.  The best fit
models have no dust, and $\tau_V=0.1$ models grossly over-predict the mid-IR fluxes
These limits imply a negligible amount of ejected mass ($M \ltorder 10^{-4}M_\odot$).
The SED is reasonably well fit with $T_*=15000$~K, no dust, and $L* \simeq 10^{5.9}L_\odot$,
as shown in Fig.~\ref{fig:sed2002kg}.  In our limited sampling of temperatures, it
can be as cool as $T_*=10000$~K with $L_*\simeq 10^{5.8} L_\odot$, and hotter models
are somewhat preferred, although the luminosities become unreasonable, with 
$L_* \simeq 10^{6.9}L_\odot$ by the time $T_*=40000$~K.  The LBT data is
also consistent with a hot ($T_*>10000$~K) star and no dust.

\begin{figure}[p]
\centerline{\includegraphics[width=5.5in]{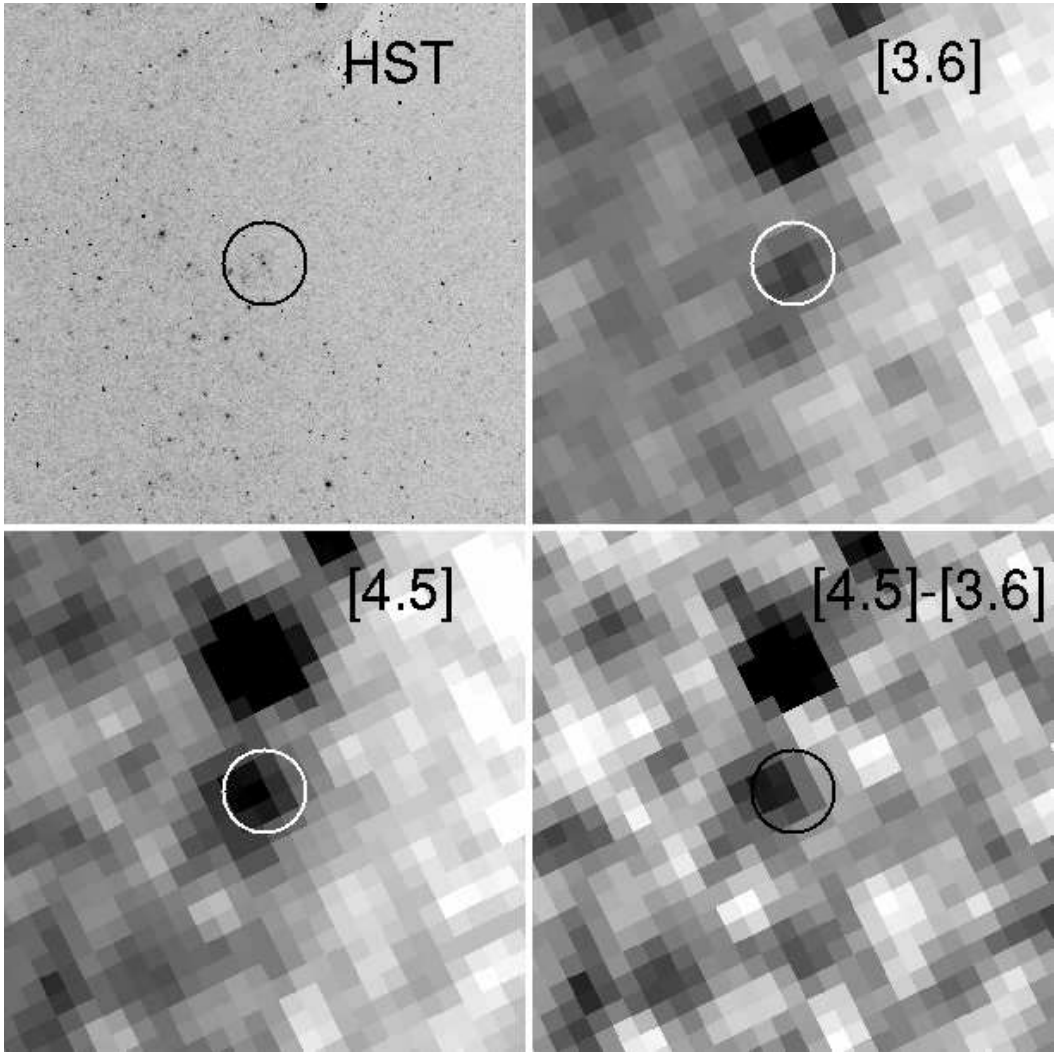}}
\caption{ The environment of SN~2003gm.  The top left and right panels show the
  HST and  $3.6\mu$m images of the region, while the lower left and right panels
  show the $[4.5]$ and $[4.5]-[3.6]$ wavelength-differenced images.
  A 1\farcs2 radius circle marks the position of the transient.
  }
\label{fig:image03gm}
\end{figure}

\begin{figure}[p]
\centerline{\includegraphics[width=5.5in]{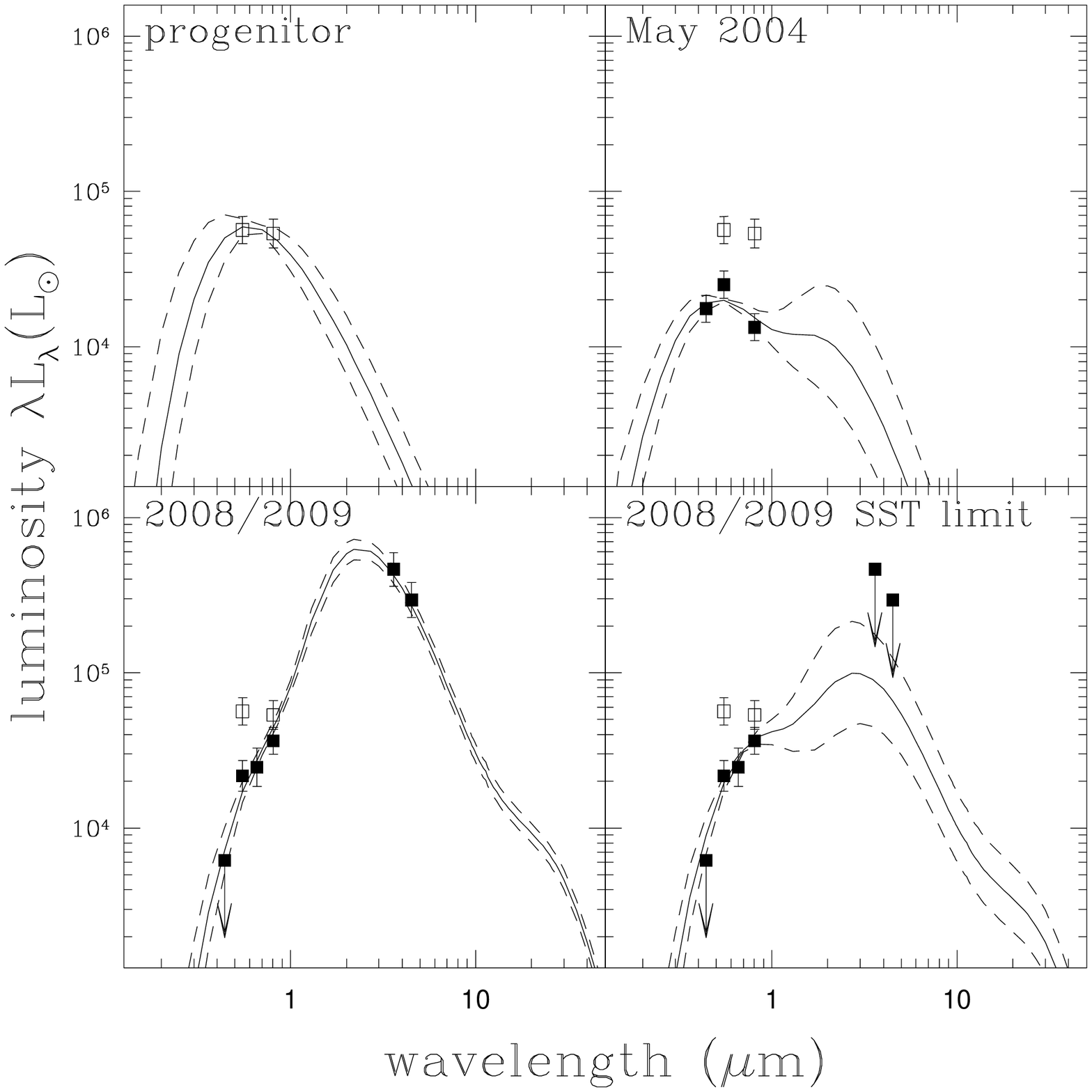}}
\caption{ Spectral energy distributions for SN~2003gm.  The progenitor
  (open squares) properties from \cite{Maund2006} are shown in
  each panel.  The solid lines are the probability weighted mean 
  SED fits and the dashed lines are the rms dispersion about the
  mean SED for the graphitic dust models.  The upper left panel
  shows fits to the progenitor fluxes from \cite{Maund2006} with
  no additional circumstellar dust.  The upper right panel shows
  the SED in May 2004 (solid points), also from \cite{Maund2006}.
  The lower panels show the SED combining the optical data
  from December 2008 (\citealt{Smith2011}) with the mid-IR
  data from August 2009.  The models in the lower panels
  treat the mid-IR fluxes as either a measurement (left) or an
  upper bound (right).  In May 2004 the source seems to be
  fainter than the progenitor but it is difficult to be 
  certain without any near/mid-IR data.  
  }
\label{fig:sed2003gm}
\end{figure}

\subsection{SN~2003gm in NGC~5334}

SN~2003gm was discovered by \cite{Schwartz2003b} on 6 July 2002 (JD 2452884) at 17~mag,
and \cite{Patat2003} reported that
it had a Type~IIn spectrum similar to SN~1997bs.  \cite{Maund2006} identified the 
progenitor as a yellow supergiant with $M_V \simeq -7.5 \pm0.2$ and $V-I\simeq 0.8\pm0.3$
($V=24.2\pm 0.1$ and $I=23.4\pm0.2$) assuming a (kinematic) distance of $20.05$~Mpc.  
The measured, but poorly sampled, transient peak had $B \simeq 18.5$, $R\simeq 17.8$ and
$I\simeq 18.0$, so the peak brightness of $M_I \simeq -13.6$ was faint compared to a normal
SN.  HST observations roughly a year later identified the source with $B \simeq 26.07 \pm0.09$,
$V\simeq 25.12\pm 0.07$ and $I\simeq 24.90 \pm 0.08$ (\citealt{Maund2006}), significantly fainter 
than before the transient.  \cite{Smith2011} find fluxes of 
with $F450W >26.0$ ($3\sigma$), $F555W=25.28\pm 0.25$, $F675W = 24.73 \pm 0.31$ and 
$F814W=23.81 \pm 0.13$~mag in December 2008, which seems to imply a significant fading
at B, little change at V and a significant brightening at I over the intervening four
years.  If we assume no additional circumstellar dust, the progenitor is consistent
with a yellow ($T_* \simeq 7500$~K) star with a luminosity of $L_* \simeq 10^5 L_\odot$,
as found by \cite{Maund2006}.

Fig.~\ref{fig:image03gm} shows the F814W HST image from a year after the 
outburst (2004 May 24, JD 2453150), the IRAC $3.6$ and $4.5\mu$m images and the
$[3.6]-[4.5]$ wavelength differenced image.  There is a source in the IRAC images
whose position is consistent with the location of SN~2003gm, but there are multiple
HST sources, so the match is uncertain.  Since the optical source is far fainter
than the progenitor and the mid-IR source is consistent with dust emission, we
proceed on the assumption that they are the same source. 
Fig.~\ref{fig:sed2003gm} shows the SED in 2008/2009,
combining the \cite{Smith2011} HST fluxes from December 2008 with the 
Spitzer fluxes from August 2009.  Fits simultaneously constrained by
the progenitor fluxes from \cite{Maund2006} are poor, so we fit only
these transient fluxes.  There are solutions that match the slow
expansion rate in Table~\ref{tab:objects} with a $T_*=10000$ to $20000$~K
temperature, luminosity $L_* \simeq 10^{5.9 \pm0.1} L_\odot$, and very
hot $T_d \simeq 1500$~K dust at the inner edge.  Because the dust is
so close to the formation radius at $R_{in} \simeq 10^{15.4}$~cm, little
mass is needed, with $M_E \simeq 0.01 M_\odot$.  Models that simultaneously
try to match the progenitor work poorly because for reasonable temperatures
and little extinction of the progenitor, it is significantly less luminous
than the transient, with $L_* \simeq 10^{5.1}$ for $T_*=10000$~K and
$L_* \simeq 10^{5.7}$ for $T_*=20000$~K.  The luminosity seen in 2008
can only be reached for a very hot star, but the SED in 2008 is
inconsistent with such high temperatures.  In 2004, the transient
seems to have been considerably bluer and lower luminosity.  The
preferred models are probably cooler, less luminous and little obscured 
compared to 2008, with $T_*$ in the range from $7500$ to $20000$~K,
luminosity $L_* \simeq 10^{4.8 \pm 0.3}L_\odot$.  The optical depths are 
$\tau_V \simeq 2.0 \pm 0.1$ and $4.6\pm 0.3$ for the graphitic and
silicate models, and the two models fit equally well.  The ejected mass
need only be $\log (M_E\kappa_{100}/M_\odot) \simeq -2.5 \pm 0.1$ for the graphitic
models and $-2.3\pm 0.1$ for the silicate, consistent with a low efficiency radiatively
driven transient.  The inner radius of the dust, $R_{in} \simeq 10^{15.3}$~cm,
roughly corresponds to the dust formation radius (Eqn.~\ref{eqn:rform}) for
its luminosity, consistent with the high inner edge dust temperatures,
(1500 and 2000~K) of the best models.  Neither fit is perfect, but with
the observations separated by a year this may not be surprising.

There are two problems with this model arising from the fact that the optical 
fluxes changed little between May 2004, one year post peak, and December 2008, 
over 5 years post-peak. The first is simply that at the earlier epoch the inner edge
of the ejecta would lie over 5 times closer to the star and would have over twice the temperature,
but dust simply cannot form/survive at temperatures of 3000-4000~K.  This
suggests that the velocities must be higher than $130$~km/s. Making it
circumstellar dust that was not associated with the transient is ruled
out by the luminosity and color of the progenitor.  
The SED does allow higher dust radii and ejecta velocities, although the fits are
poorer.   The second problem is the familiar one that the optical depth
should have been over 25 times higher in May 2004, which is inconsistent
with the modest changes in the optical fluxes.   If the dust is associated
with the transient, then the dust must be forming in a relatively steady,
low optical depth wind, which is consistent with $R_{in} \simeq R_f$
and dust temperatures close to the destruction temperature.  The wind 
velocity may have to be somewhat higher than $130$~km/s in order for 
dust to form and grow since $R_f/130~\hbox{km/s}\simeq 5$~years is 
significantly larger than the elapsed time from the peak to
May 2004.

\begin{figure}[p]
\centerline{\includegraphics[width=5.5in]{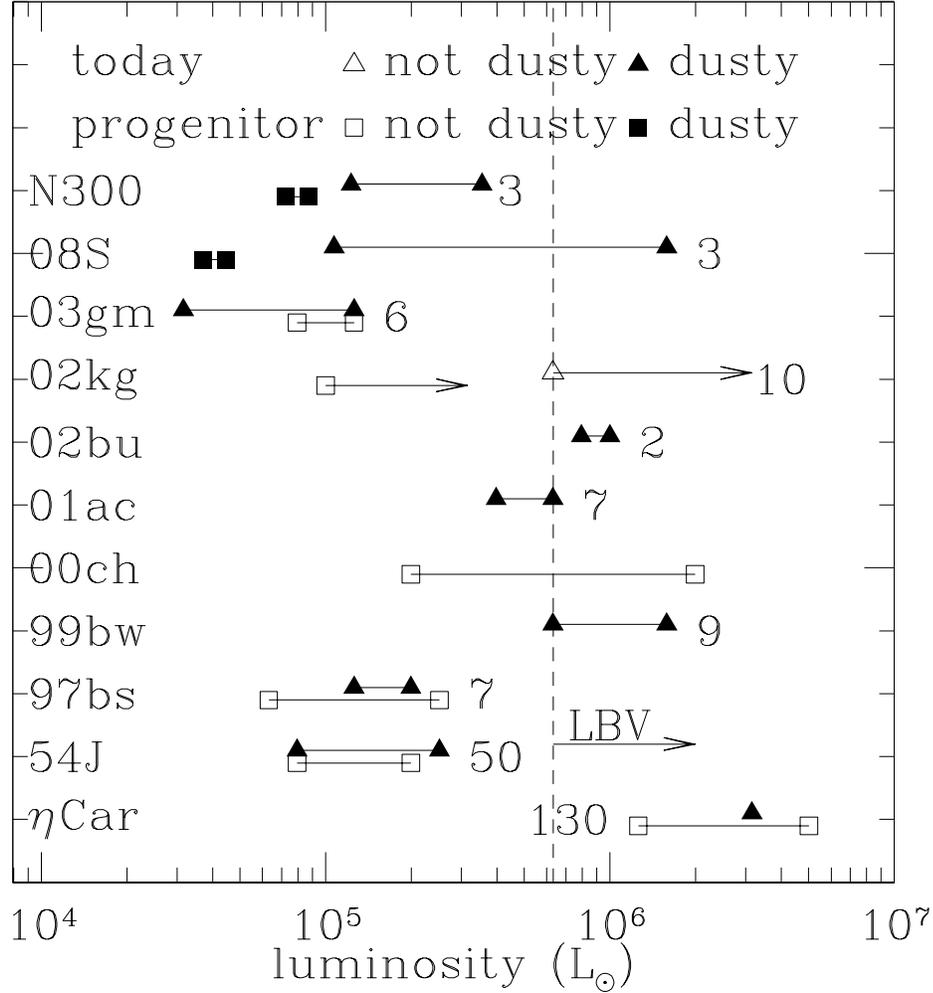}}
\caption{ Progenitor and present day luminosities of the transients.
  For each object, the lower bar with squares shows the allowed range 
  of luminosities for the progenitor stars while the upper bar with
  triangles shows the allowed range of luminosities at the time of
  the most recent mid-IR observations.  The number gives the number
  of years between the transient peak and the mid-IR observations.
  The symbols are filled if most of the energy is radiated in the
  mid-IR by dust, and open if most is direct emission from the 
  photosphere.  Except for the two variable stars, SN~2000ch and
  SN~2001kg/V37, the present day emissions of all the other systems
  are dominated by circumstellar dust.  For SN~1954J and SN~1997bs,
  the dust emission is not directly observed.  The vertical dotted
  line shows the minimum luminosity for a LBV (e.g. \citealt{Smith2006}).
  See the discussions of the individual objects for any caveats.
  }
\label{fig:lsum}
\end{figure}

\begin{figure}[p]
\centerline{\includegraphics[width=5.5in]{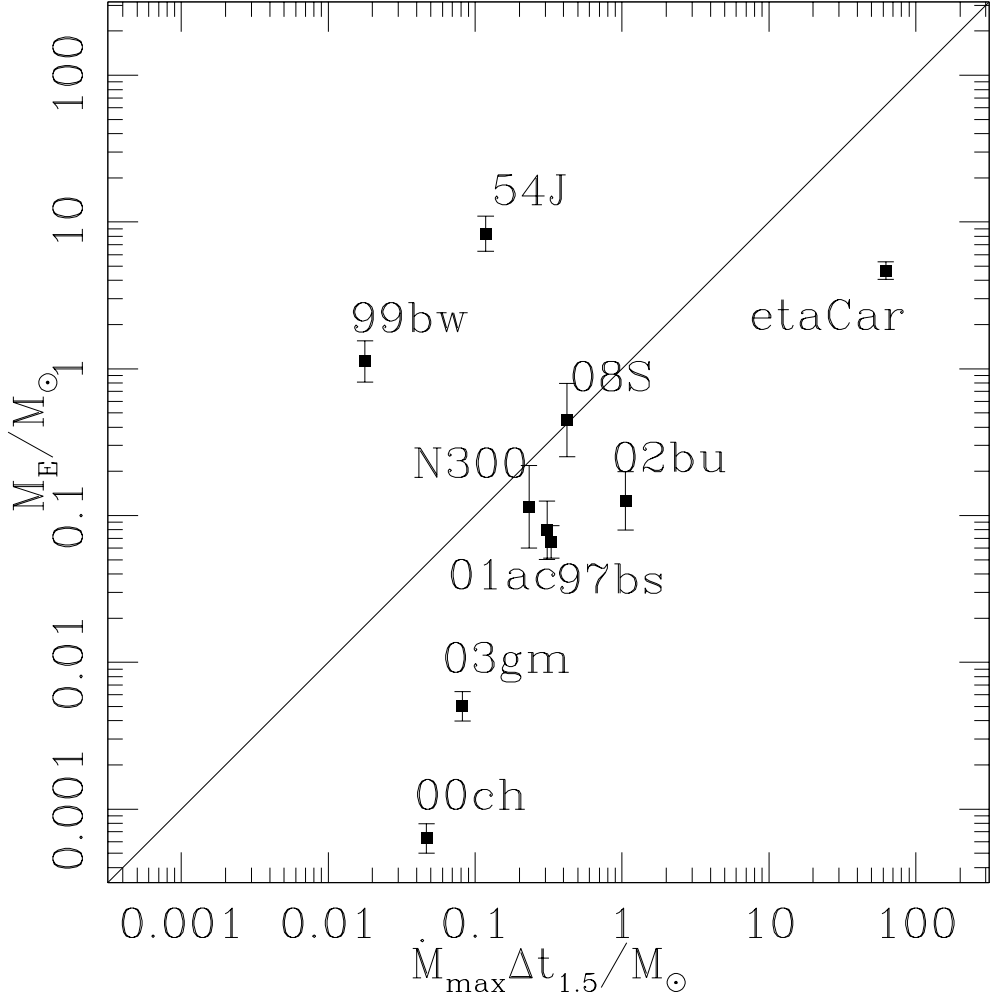}}
\caption{ The maximum ejected mass for radiative processes, $\dot{M}_{max}\Delta t_{1.5}$ (Eqn.~\ref{eqn:mdotmax},
  Table~\ref{tab:objects2}),
  as compared to the mass estimated from the DUSTY model optical depths, $M_E$, for silicate dusts.
  The estimates of $M_E$ are proportional to $M_E \propto \kappa_{100}^{-1} R_{in}^2$ where $R_{in}$
  is constrained by a combination of the SED and the velocities from Table~\ref{tab:objects}.  Radiatively driven transients
  obscured by the ejected material should lie below the diagonal line. 
   Because there is only 
  a lower bound on $\Delta t_{1.5}$ for SN~1999bw, it can be shifted to the right 
   as $\log( \Delta t_{1.5}/10\hbox{days}$) by  extending the transient to be longer
  then the lower limit of $\Delta t_{1.5} > 10$~days (see Table~\ref{tab:objects}). 
  }
\label{fig:masssum1}
\end{figure}

\begin{figure}[p]
\centerline{\includegraphics[width=5.5in]{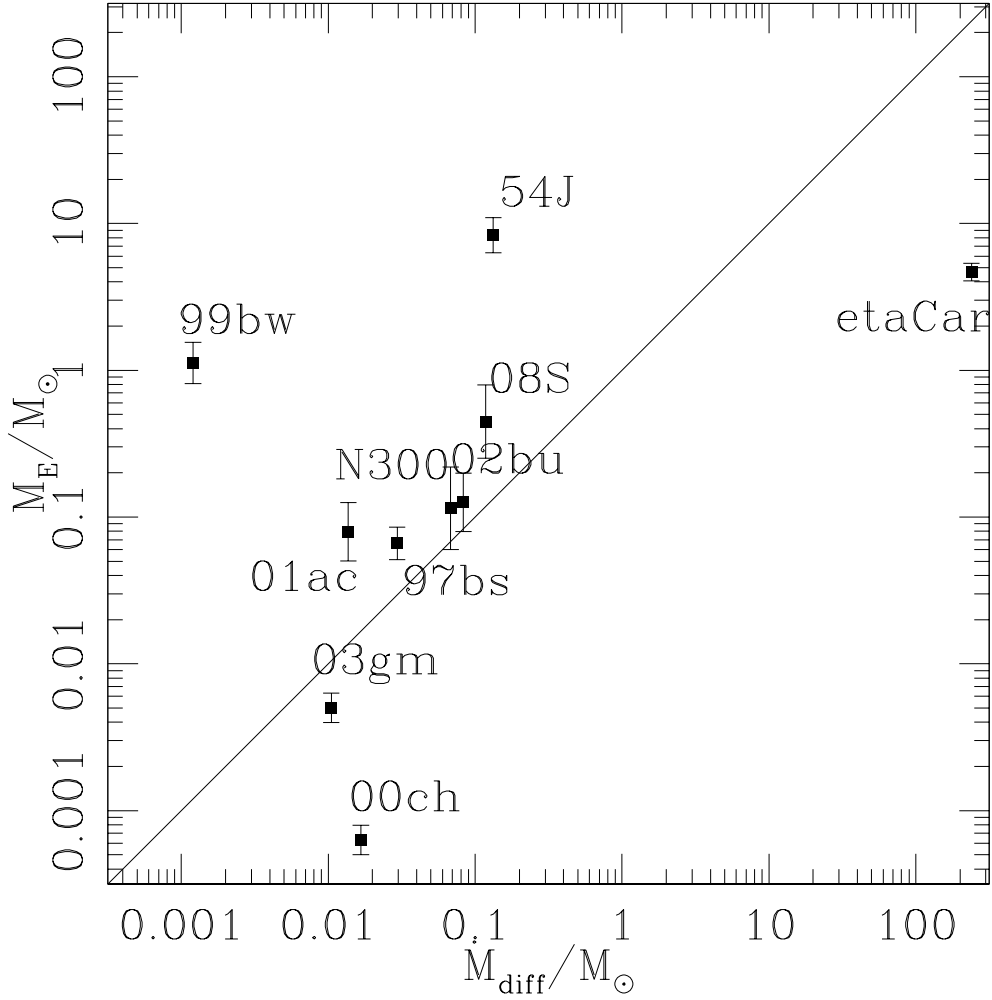}}
\caption{ Estimates of the ejected masses for explosive processes, $M_{diff}$ (Eqn.~\ref{eqn:mdiff},
  Table~\ref{tab:objects2}),
  as compared to the mass estimated from the DUSTY model optical depths, $M_E$, for silicate dusts.
  The estimates of $M_E$ are again proportional to $M_E \propto \kappa_{100}^{-1} R_{in}^2$.
    Explosive transients obscured
   by the ejected material should lie close to the diagonal line.
   Because there is only a lower bound on $\Delta t_{1.5}$ for SN~1999bw, it can be shifted to the right 
  as $\log( \Delta t_{1.5}/10\hbox{days})$  by extending the transient duration. 
  }
\label{fig:masssum2}
\end{figure}

\begin{figure}[p]
\centerline{\includegraphics[width=5.5in]{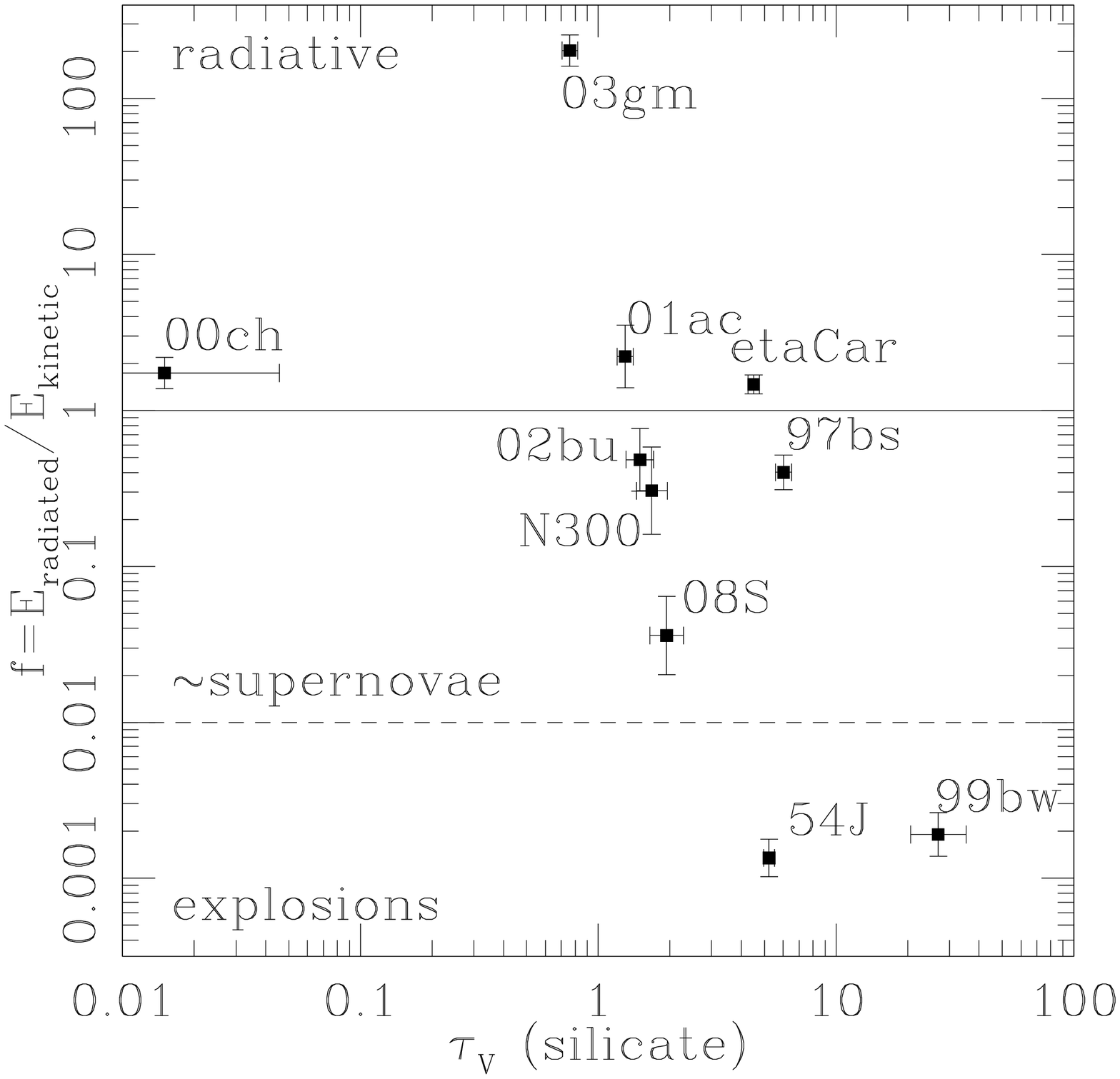}}
\caption{ The mechanism diagnostic $f=E_{radiated}/E_{kinetic}$ as a function of
  the DUSTY silicate optical depth $\tau_V$ for the most recent mid-IR epoch.
    Radiatively driven models should generally have $f \gg 1$
  and explosions should have $f\ll 1$, where $f=1$ is indicated by the solid line.
  Typical supernovae have $f\simeq 0.01$, which is shown by the dashed line.  
  SN~2002kg is not shown since it shows no evidence of dust.  SN~1999bw can be 
  shifted upwards by $\log(\Delta t_{1.5}/10~\hbox{days})$ because the true event 
  duration is unknown. 
  }
\label{fig:mechanism}
\end{figure}

\section{Discussion}
\label{sec:discussion}

The standard picture of these transients is that the stars undergo
a relatively short duration eruption with high mass loss rates, 
forming a shell of material that forms dust and then obscures
the star at later times.  The most striking result of our survey 
is that we find no source (with adequate data to draw a conclusion)
that is consistent with this standard picture.  The problem for
the standard picture is that expanding shells of material impose
a tyranny of geometry on the evolution of the system because
the optical depth drops as $\tau \propto 1/r^2 \propto 1/t^2$.
Optical fluxes depend exponentially on the optical depth and
it generally requires physically unreasonable changes in luminosity 
and temperature to balance the effects of the evolution in the optical 
depth on the optical fluxes. The 
problem is generic to transient radial outflows, not just to spherical
shells.  It may be a failure of imagination, but we have
found no plausible, generic means of avoiding this basic
scaling and some, such as the growth of inhomogeneities in the 
ejecta, that will make the problem worse.
  
To make these events stellar transients, in many cases the simplest solution
is to make them long lived.   The optical transient is simply
a sign post that the star is transitioning from a low mass 
loss rate to a high mass loss rate.  For these typically hot
stars, a mass loss rate of order $\dot{M} \gtorder 10^{-2.5} M_\odot$/year
leads to dust formation because the wind shields the dust
formation region from the UV emissions of the generally hot
stars and the high densities allow particle growth 
(see \citealt{Kochanek2011}).  Due to a combination of 
the fading of the initial transient and the formation of
dust, the star fades in the optical to be fainter than
the progenitor {\it but the star remains in the high mass 
loss state!}  The advantage of a steady wind is that it
has a roughly constant optical depth once the dust starts
to form, which is far more consistent with the available data.
When the star leaves the high mass loss state and drops
below the threshold for dust formation, the optical depth
initially drops on the time scale to cross the dust formation
radius and then settles onto the $\tau \propto 1/t^2$ 
scaling of a shell. 
 
In particular, this is the only solution we found to explain
the historical visual light curve of $\eta$ Carinae -- 
$\eta$ Carinae went into a high mass loss state circa 1850
and came out of it circa 1950.  As it exited the high 
mass loss phase, it dropped in luminosity and reverted to
a hotter temperature.    This is by far the simplest
explanation for the long period of roughly constant visual
luminosity and color.   If all the dust was produced
in a short lived period circa 1850, there is no 
physically reasonable evolution in the stellar luminosity
and temperature that can reproduce the observed light curve --
the luminosity must change by an order of magnitude just to
compensate for a change in optical depth of $\Delta \tau_{e,V} = \ln 10=2.3$,
while the optical depth of the ejecta should have changed by almost two orders
of magnitude over that period!  The only effective counter
to the effects of expansion on the optical depth is to 
steadily raise the opacity as the shell expands, but we have been unable to
identify any realistic mechanism for doing so.  It is
more likely that the effective opacity is diminishing
as the shell becomes less homogeneous with paths through
the ejecta of lower optical depth than the mean.

For the extragalactic sources, the existing data suggest this
solution for SN~1997bs, SN~1954J and SN~2003gm if 
they are to be explained as stellar transients.  Of these 
three sources, only SN~2003gm is clearly a bright mid-IR
source, while we only obtained upper limits for SN~1997bs
and SN~1954J.  The optical light curves of SN~1997bs
require dust formation with an optical depth of order
$\tau_V \simeq 10$ after 9 months strongly indicating
that the mass loss rate was high for an extended 
period after the transient faded in the optical.  
SN~1999bw, SN~2001ac, and SN~2002bu (and possibly SN~2003gm) seem
more likely to have the physics of SN~2008S (see \citealt{Kochanek2011})
as previously noted for most of these cases in \cite{Thompson2009}.  
They generally show signs of extinction at peak, become completely obscured
post-transient, have long lived luminosities of $10^{5.5}$
to $10^{6.0}L_\odot$, and show some signs that the dust
initially appears at radii larger than allowed by the expansion
speed.  Like SN~2008S, these systems were probably obscured by
dust forming in a dense pre-existing wind.  The dust is then
largely destroyed by the transient but then starts to reform
as the transient fades because of the very high wind densities.
The reformed dust again cloaks the system and the late time 
luminosity is powered by the continued expansion of the shock
wave through the wind.  The case is strongest for SN~1999bw, whose mid-IR
SED changed little between April 2004 and May 2008.

Two of the sources we consider, SN~2000ch and SN~2002kg/V37, are
optically variable sources without strong mass loss or significant
dust creation.  Both sources have also had repeated episodes
of optical variability (\citealt{Weis2005}, \citealt{Pastorello2010}).
SN~2002kg shows no signs of dust, while SN~2000ch appears to have
a low optical depth dusty wind.  While they have been referred to as 
LBV's, the very luminous star associated with SN~2000ch has 
luminosity/temperature variations opposite (\citealt{Pastorello2010}) 
from those normally associated with LBVs (\citealt{Humphreys1994}), 
while SN~2002kg/V37 is of fairly low luminosity for an LBV.  The
inverted color evolution of SN~2000ch could be created by the
stellar variability modulating the dust properties of a wind
though variations in the formation radius or grain size with 
stellar luminosity and mass loss rate.

Fig.~\ref{fig:lsum} provides a summary of the luminosities of 
the progenitors and for the most recent mid-IR epoch for each
object. We also include the estimates for SN~2008S and the 2008
NGC~300 transient from \cite{Kochanek2011}.  There are many
caveats to these estimates (as discussed in the previous 
sections), but there are some general patterns.  First, of the
systems where we can estimate the progenitor luminosity, 
only the two sources known to simply be variable stars, 
SN~2002kg/V37 and SN~2000ch, can be above the frequently used 
luminosity limit of $10^{5.8}L_\odot$ for LBVs (e.g. \citealt{Smith2006}).   
In general, these transients do not seem to be associated
with very massive, very luminous stars, so the entire analogy
with the (great) eruptions of LBVs seems weak since they are
generally lower luminosity stars far from their Eddington limits.  
These are largely transients from stars with $L_* < 10^{5.3}L_\odot$,
which corresponds to $M_* \ltorder 25 M_\odot$ in the \cite{Marigo2008}
models.   As noted in
\cite{Kochanek2011b}, they also have far higher expansion velocities
than is typically observed for the shells around Galactic LBVs.
This could be related to the observation by \cite{Dessart2010}
that the envelopes of lower mass ($M_* \ltorder 15 M_\odot$) stars   
are far less tightly bound, which makes it far easier for them
to produce lower luminosity and energy (explosive) transients.
Broadly speaking, there is no convincing evidence that there
is any relationship between the progenitors of the supernova impostors and LBVs. 

Second, in all cases where we have information on both the progenitor
luminosity and the late time luminosity of a (potentially) surviving  star, there is
no evidence that the stars ever become significantly  
sub-luminous compared to their progenitors.  Physically, this is expected because 
stars are self-gravitating systems and essentially cannot be
significantly sub-luminous at the surface without also being
out of hydrostatic equilibrium.   If significantly sub-luminous, 
a star must start to dynamically contract, and the potential energy release
returns the star to hydrostatic equilibrium.  The recovery
time discussed by \cite{Smith2011}, $t_R = \Delta E/L_*$,
is not really relevant for the stellar luminosity
because the dominant means of reheating an envelope with
too little thermal energy is Kelvin-Helmholtz contraction rather
than the energy radiated from the interior.  Moreover, the sources
observed at late times are frequently brighter rather than fainter
than their progenitors.

Figures~\ref{fig:masssum1} and \ref{fig:masssum2} compare the mass loss
estimates from the SED models to the predictions given in Table~\ref{tab:objects2}
for radiatively ($\dot{M}_{max}\Delta t_{1.5}$ from Eqn.~\ref{eqn:mdotmax})
and explosively ($M_{diff}$ from Eqn.~\ref{eqn:mdiff}) driven transients.
In most cases, either scenario is arguably consistent with the mass 
required to support the observed optical depths given all the other
uncertainties.  If the transients are driven radiatively and all
the mass has to be ejected in $\Delta t_{1.5}$, then a large fraction
of the available energy/luminosity has to be used to accelerate the ejected
material.  If, however, much of the mass loss occurs over a period
significantly longer than the optical transient, as seems to be
the case for many of the systems (e.g $\eta$~Carinae and SN~1997bs
in particular), there should be little difficulty radiatively driving
the necessary mass loss.  The two problematic systems in either scenario
are SN~1954J and SN~1999bw, where far more mass is needed to produce
the late-time optical depths than can be accounted for in either model.
Extending the transient in SN~1999bw can help, but seems unlikely to
solve the problem.  SN~1954J is still more problematic because the 
$50$~year time scale makes the dust radius very large and hence 
requires a great deal of ejected mass to retain a finite optical depth.
These are the clearest cases where the mass estimates appear to
require different physics, where the simplest solution is to use
a longer lived wind for SN~1954J and to make SN~1999bw an object
like SN~2008S and the 2008 NGC~300 transient.  It is not a coincidence
that the oldest systems other than $\eta$ Carinae show the greatest 
problems, because systems with different physics that are then 
modeled as an ejected shell will have estimates of $M_E$ that
will tend to increase with the elapsed time (as $t^2$).  Observing
such a trends for the younger systems (e.g. SN~2002bu, SN~2008S, and the NGC~300 transient)
would provide additional evidence that they are a separate class
of transients.
 
Finally, Figure~\ref{fig:mechanism} shows estimates of the ratio of
radiated to kinetic energy $f=E_{rad}/E_{kinetic}$ (Eqn.~\ref{eqn:me})
as another diagnostic of the transient mechanism.  We show it as
a function of the silicate DUSTY model optical depths $\tau_V$,
which can be used as a proxy for changes in the ejecta mass since
$\tau_V \propto M_E$ and $f \propto M_E^{-1}$.  The picture is
broadly similar for the graphitic models.  
Most of these transients are more consistent
with an explosive origin for the transient, with $f<1$.  While a
radiatively driven transient can have $f<1$ when the observed
luminosity is significantly less than the true luminosity because most
of it is absorbed into accelerating the ejecta (Eqn.~\ref{eqn:mdotmax},
\citealt{Owocki2004}), it is difficult to then understand why
these systems show no significant differences in their external
appearance.  A number of systems, particularly SN~1954J,
SN~1999bw, SN~2008S have $f \ll 1$ and probably cannot be
reconciled with a short duration radiatively driven process.  We argued in
\cite{Kochanek2011} that SN~2008S and the NGC~300 transient
had to be explosive in nature in order to explain the destruction
of the dust that obscured the progenitors, so perhaps all these
transients are driven by an initial (weak?) explosion, as discussed
in \cite{Dessart2010}.  After ejecting some mass in the initial
optical transient, many of the systems then move into a high mass 
loss state as the envelope re-equilibrates following the explosion 
in order to explain the wind-like features needed to explain many
of the systems.      

For the present study we have simply assumed that there is a 
surviving star powering the present day emissions.  Aside from the variable
stars SN~2000ch and SN~2002kg, there is no solid evidence this
is correct. Most of the observed luminosities are also   
consistent with being powered by a shock expanding through a 
dense circumstellar medium as \cite{Kochanek2011} posited for 
SN~2008S and the NGC~300 transient.   Optical spectra or
X-ray observations can test this hypothesis, but for heavily
obscured sources behind a veil of dust and gas it is difficult
to make these tests.  The most powerful argument against these
sources being SNe remains the overall energetics, but it is
awkward that no compelling case can be made for any other
mechanism 
(see, for example, the discussions in \cite{Thompson2009} or \cite{Smith2011}). 

The biggest problem for characterizing these sources is the fragmentary
nature of the data.  In particular, it is almost certain that some of 
our particular conclusions are incorrect because we are interpreting such 
fragmentary data.  For example, many of the sources show signs of absorption at the
transient peak, but there are no near/mid-IR observations to detect the
emissions that result if the dust is circumstellar --  SN~1999bw and
SN~2002bu may well have had their peak emissions in the mid-IR rather
than the optical.  Since we now know that dust and dust formation in the ejecta are 
a key part of the physics, it is essential to characterize these transients beyond the time
scales on which dust can form, typically 6 months to a year, both in
the optical to observe the increasing optical depth, and in the near/mid-IR
to determine the dust temperature/formation radius.  The evolution in
the following year rapidly determines whether the mass loss was associated
with the short optical transient or if the star has entered a high mass
loss state and is generating a relatively steady wind.  At later times,
we should both see the dust temperature dropping and the reappearance
of the stars as the optical depth drops.  
For the objects we discuss, most of the questions can be answered by
coarsely monitoring their optical through mid-IR SEDs over a period 
of years.  In the optical and near-IR this is relatively straight 
forward given the capabilities of HST but will require greater
wavelength coverage and moderately deeper exposures than have generally
been obtained.   Warm Spitzer can effectively monitor the mid-IR
emission in the short term, but JWST may be required to address some
final puzzles because it will largely eliminate the problem of confusion
and will again allow measurements at the longer wavelengths (principally 
$8$-$24\mu$m) needed to fully characterize the dust emission. Whatever the failings
of our analysis, these transients are creatures of the near/mid-IR and simply cannot be 
understood without data at these wavelengths.

\acknowledgements 

We thank R.M.~Wagner for supplying images of SN~2000ch, and J.F.~Beacom,
J.-L. Prieto, and T.A. Thompson for discussions and comments.
CSK, DMS and KZS are supported by NSF grant AST-0908816.  KZS
is also supported by NSF grant AST-1108687.
Based in part on observations made with the Large Binocular Telescope.
The LBT is an international collaboration among institutions in the
United States, Italy and Germany. The LBT Corporation partners are:
the University of Arizona on behalf of the Arizona university system;
the Istituto Nazionale di Astrofisica, Italy; the LBT
Beteiligungsgesellschaft, Germany, representing the Max Planck
Society, the Astrophysical Institute Potsdam, and Heidelberg
University; the Ohio State University; and the Research Corporation,
on behalf of the University of Notre Dame, University of Minnesota and
University of Virginia. 
This work is based in part on observations made with the Spitzer Space Telescope, which is operated by the
Jet Propulsion Laboratory, California Institute of Technology under a contract with NASA and
on observations made with the NASA/ESA Hubble Space Telescope,  
and obtained from the Hubble Legacy Archive, which is a collaboration 
between the Space Telescope Science Institute (STScI/NASA), the Space 
Telescope European Coordinating Facility (ST-ECF/ESA) and the         
Canadian Astronomy Data Centre (CADC/NRC/CSA). 
%Based in part on observations made with the NASA/ESA Hubble Space Telescope, obtained from the data
%archive at the Space Telescope Institute. STScI is operated by the association of Universities for Research in Astronomy, Inc.
%under the NASA contract  NAS 5-26555.
This research has made use of the NASA/IPAC Extragalactic Database (NED) which is operated by the Jet Propulsion Laboratory,
California Institute of Technology, under contract with the National Aeronautics and Space Administration.

\facility{LBT, Spitzer, HST}

\appendix
\section{A Non-Parametric Model For The Light Curve of $\eta$ Carinae}
\label{sec:appendix}

Consider a model with mass loss rate $\dot{M}(t)$, constant
velocity $v_w$, 
luminosity $L(t)$, and temperature $T(t)$ defined 
annually from 1850 to 2010.  By an appropriate choice
of numerical variables we can restrict the luminosity
to the range  $6 < \log L(t)/L_\odot < 8$, and the 
temperature to the range $7000~\hbox{K} < T_* < 40000~\hbox{K}$.  
The primary assumption of the model is that the material
ejected at time $t_e$ can be treated as an expanding shell
with optical depth
\begin{equation}
   \tau(t|t_e) \simeq 2000 \left( { \dot{M} \over M_\odot/\hbox{year}}\right)
              \left( { 500~\hbox{km/s} \over v_w } \right)^2
              \left( { \kappa_V \over 31~\hbox{cm}^2/\hbox{g}} \right)
              \left( { \hbox{year} \over t_- } - { \hbox{year} \over t_+ } \right)
\end{equation}
at some later time $t$ where 
\begin{equation}
         t_\pm = \hbox{max}\left( t_f,t_o-t \pm \Delta t/2\right)
\end{equation}
determine the inner and outer radii of the material and the dust
formation radius is $R_f = v_w t_f$.  Essentially, each
year of the transient is treated as a shell of material evolving as
in Eqn.~\ref{eqn:tevol}.  We set $t_f=5$~years
and $v_w=500$~km/s, which is roughly correct.  The optical depth is scaled to the
effective absorption opacity for silicates ($\kappa_{e,V}=32$~cm$^2$/g). 
In these models the only effect of changing the velocity or the opacity
is to rescale the $\dot{M}$ so as to keep the optical depth fixed,
$\tau \propto \dot{M} \kappa/v_w$. 
The model also only allowed dust
formation when $T<10000$~K and $\dot{M} > 10^{-3}M_\odot$/year,
loosely following the criteria from \cite{Kochanek2011}, although
models without these criteria have the same qualitative properties.  The
V-band light curve was then fit using Eqn.~\ref{eqn:levolve}.
Priors were used to keep the luminosity, temperature and mass loss
rates smooth and to favor declining luminosities and mass loss rates
and increasing temperatures.  The optical depth in 1974 was constrained
to roughly match the SED models.  Finally, a prior was used to try to
maximize the mass loss.   By experimenting with the priors we attempted
to drive the solutions towards various scenarios, but only one class
of solutions worked.

\vfill\eject

\def\hp{\hphantom{0}}

\begin{deluxetable}{lccrrrrcr}
%\tabletypesize{\scriptsize}                        
\tablecaption{General Properties\label{tab:expect}}
\tablewidth{0pt}
\tablehead{
\multicolumn{1}{c}{Object} &
\multicolumn{1}{c}{dist} &
\multicolumn{1}{c}{$E(B-V)$} &
\multicolumn{2}{c}{$M$} &
\multicolumn{1}{c}{$\Delta t_{1.5}$} &
\multicolumn{1}{c}{$v_w$} &
\multicolumn{1}{c}{$E=\nu L_\nu(1+f^{-1})$} & 
\multicolumn{1}{c}{$\Delta t$} 
\\ % break
 &
\multicolumn{1}{c}{(Mpc)} &
 &
\multicolumn{2}{c}{(mag)} &
\multicolumn{1}{c}{(days)} &
\multicolumn{1}{c}{(km/s)} &
\multicolumn{1}{c}{($10^{48}$~ergs/s)} & 
\multicolumn{1}{c}{(years)} 
}
\startdata
  $\eta$~Carina &\multicolumn{1}{c}{$0.0023$} &$0.51$ &$-13.9$ &(B) &      $4400$  &\hp$ 650$  &$58$\hp\hp\hp &$129.3$  \\ 
       SN~1954J &\hp$ 3.1$ &$0.04$ &$-11.2$ &(B) &   \hp$ 100$  &\hp$ 700$  &\hp\hp$0.11$\hp &\hp$50.0$  \\ 
       SN~1961V &\hp$ 9.4$ &$0.07$ &$-17.8$ &(B) &   \hp$ 200$  &   $3700$  &$86$\hp\hp\hp &\hp$42.7$  \\ 
      SN~1997bs &\hp$ 9.4$ &$0.04$ &$-13.8$ &(V) &\hp\hp$  45$  &\hp$ 765$  &\hp\hp$0.31$\hp &\hp\hp$3.9$  \\ 
      SN~1999bw &   $13.5$ &$0.02$ &$-12.6$ &(R) &\hp  $>  10$  &\hp$ 630$  &\hp    $>0.017$ &\hp\hp$9.1$  \\ 
      SN~2000ch &   $10.9$ &$0.02$ &$-12.6$ &(R) &\hp\hp$  25$  &   $1400$  &\hp\hp$0.043$ &\hp\hp$7.7$  \\ 
      SN~2001ac &   $22.9$ &$0.03$ &$-13.9$ &(R) &\hp\hp$  50$  &\hp$ 287$  &\hp\hp$0.29$\hp &\hp\hp$6.8$  \\ 
      SN~2002bu &\hp$ 5.8$ &$0.01$ &$-14.9$ &(R) &\hp\hp$  70$  &\hp$ 893$  &\hp\hp$0.97$\hp &\hp\hp$2.1$  \\ 
      SN~2002kg &\hp$ 3.1$ &$0.18$ &$-10.3$ &(R) &   \hp$ 365$  &\hp$ 350$  &\hp\hp$0.075$ &\hp\hp$5.6$  \\ 
      SN~2003gm &   $20.2$ &$0.05$ &$-14.3$ &(I) &\hp\hp$  65$  &\hp$ 131$  &\hp\hp$0.35$\hp &\hp\hp$6.0$  \\ 
       SN~2008S &\hp$ 5.6$ &$0.35$ &$-13.8$ &(R) &\hp\hp$  75$  &   $1100$  &\hp\hp$0.39$\hp &\hp\hp$2.6$  \\ 
     NGC~300-OT &\hp$ 1.9$ &$0.02$ &$-13.1$ &(R) &\hp\hp$  80$  &\hp$ 560$  &\hp\hp$0.22$\hp &\hp\hp$1.6$  \\ 
\enddata   
\label{tab:objects}
\tablecomments{
  %The distance to $\eta$ Carinae is in kpc rather than Mpc.
  We use the Cepheid distances from \cite{Freedman2001} for SN~1954J/V12 and SN~2002kg/V37 in NGC~2403,
  SN~1961V in NGC~1058 (assumed to be in a group with NGC~925), SN~1997bs in NGC~3627 and SN~1999bw in NGC~3198,
  the Cepheid distance of \cite{Gieren2005} for NGC~300,  
  the distance adopted by \cite{Pastorello2010} for SN~2000ch,
   the distance adopted by \cite{Leaman2010} for SN~2001ac in NGC~3504, 
  the distance adopted by \cite{Maund2006} for SN~2003gm in NGC~5334,
  and the distance used by \cite{Sahu2006} for SN~2004et in NGC~6946 for SN~2008S.  
  \cite{Tully2008} report a Tully-Fisher distance estimate for SN~2002bu in NGC~4242.
  Galactic extinctions $E(B-V)$ are
  from \cite{Schlegel1998} except for $\eta$ Carinae where we use the foreground extinction estimate of \cite{vanGenderen1984}
  and SN~2002kg where we used the estimate based on nearby stars from \cite{Maund2006}.
  The transient magnitude $M$, duration $\Delta t_{1.5}$ and velocity $v_w$ are taken from \cite{Smith2011}.
  The elapsed time from the transient to the epoch of the mid-IR data used in the SED fits is $\Delta t$.
  }
\end{deluxetable}

\begin{deluxetable}{lccccc}
%\tabletypesize{\scriptsize}                        
\tablecaption{Mass and Optical Depth Estimates \label{tab:expect2}}
\tablewidth{0pt}
\tablehead{
\multicolumn{1}{c}{Object} &
\multicolumn{1}{c}{$\dot{M}_{max}\Delta t_{1.5}$} &
\multicolumn{1}{c}{$M_{diff}$} &
\multicolumn{1}{c}{$M_E$} &
\multicolumn{1}{c}{$\tau_V$} &
\multicolumn{1}{c}{$\lambda_{peak}$} 
 \\  %break
 &
 \multicolumn{1}{c}{($M_\odot$)} &
 \multicolumn{1}{c}{($M_\odot$)} &
 \multicolumn{1}{c}{($f^{-1} v^{-2} M_\odot$)} &
 \multicolumn{1}{c}{($f^{-1} v^{-4}$)} &
 \multicolumn{1}{c}{($L_6^{-1/4} v^{1/2} \mu$m) }
}
\startdata
  $\eta$~Carina &$ 63$\hp\hp\hp &$240$\hp\hp &$ 14$\hp\hp &\hp\hp$ 3.1$\hp &$22.5$ \\ 
       SN~1954J &\hp$0.12$      &\hp\hp\hp$0.14$ &\hp\hp\hp$0.03$ &\hp\hp\hp$0.03$ &$16.2$ \\ 
       SN~1961V &$ 95$\hp\hp\hp &\hp\hp$2.9$\hp &\hp\hp\hp$0.64$ &\hp\hp\hp$0.05$ &$28.2$ \\ 
      SN~1997bs &\hp$0.34$      &\hp\hp\hp$0.03$ &\hp\hp\hp$0.06$ &\hp\hp$ 9.4$\hp &\hp$ 5.0$ \\ 
      SN~1999bw &$>0.02$        &\hp\hp$>0.01$   &\hp\hp$>0.01$   &\hp\hp$>0.21$ &\hp$ 7.3$ \\ 
      SN~2000ch &\hp$0.05$      &\hp\hp\hp$0.02$ &\hp\hp\hp$0.01$ &\hp\hp\hp$0.04$ &\hp$ 9.8$ \\ 
      SN~2001ac &\hp$0.31$      &\hp\hp\hp$0.02$ &\hp\hp\hp$0.35$ &\hp$ 140$\hp\hp &\hp$ 3.8$ \\ 
      SN~2002bu &\hp$1.06$      &\hp\hp\hp$0.09$ &\hp\hp\hp$0.13$ &\hp\hp$  57$\hp&\hp$ 3.6$ \\ 
      SN~2002kg &\hp$0.09$      &\hp\hp\hp$0.89$ &\hp\hp\hp$0.07$ &\hp\hp$  26$\hp&\hp$ 3.8$ \\ 
      SN~2003gm &\hp$0.38$      &\hp\hp\hp$0.02$ &\hp\hp$2.1$\hp &$5300$\hp\hp &\hp$ 2.2$ \\ 
       SN~2008S &\hp$0.43$      &\hp\hp\hp$0.12$ &\hp\hp\hp$0.04$ &\hp\hp$ 6.7$\hp &\hp$ 4.8$ \\ 
     NGC~300-OT &\hp$0.24$      &\hp\hp\hp$0.07$ &\hp\hp\hp$0.07$ &\hp$ 140$\hp\hp &\hp$ 2.4$ \\ 
\enddata
\label{tab:objects2}
\tablecomments{The three mass estimates are the upper limit for radiatively
  driven mass loss, $\dot{M}_{max}\Delta t_{1.5}$ (Eqn.~\ref{eqn:mdotmax}) the estimate
  from the photon diffusion time scale, $M_{diff}$ (Eqn.~\ref{eqn:mdiff}) and the
  estimate $M_E$ based on the radiative efficiency $f$ (Eqn.~\ref{eqn:me}) where
  SN have $f \simeq 0.01$, near maximal radiatively driven winds have
  $f \simeq 1$ and normal radiatively driven winds have $f \gg 1$.  
  The optical depth $\tau_V$ is estimated at the epoch of the measured
  SED based on $M_E$ and assuming $\kappa_V =100 \kappa_{100}$~cm$^2$/g.  The scaling
  of $\tau_V$ with $f$ and velocity $v_w$ is indicated.  The 
  expected peak wavelength of the SED is estimated for a luminosity
  of $L_*= 10^6 L_6 L_\odot$.  The mass and optical depth limits for
  SN~1999bw are lower limits because there is only a lower bound
  on $\Delta t_{1.5}$ (see Table.~\ref{tab:expect}). 
  }
\end{deluxetable}

\begin{deluxetable}{llllrrrrrrrrrrr}
\rotate
\scriptsize
\tablecaption{Spitzer Observations}
\tablewidth{0pt}
\tablehead{
  \colhead{SN}
  &\colhead{Date} &\colhead{MJD} &\colhead{PI/Program}
  &\colhead{$F_{3.6}$}
  &\colhead{$F_{4.5}$}
  &\colhead{$F_{5.8}$}
  &\colhead{$F_{8.0}$}
  &\colhead{$F_{24}$} 
  &\colhead{Comment/$F_{70}$}\\
   &
  & & 
  &\colhead{($\mu$Jy)}
  &\colhead{($\mu$Jy)}
  &\colhead{($\mu$Jy)}
  &\colhead{($\mu$Jy)}
  &\colhead{($\mu$Jy)}
  &\colhead{($\mu$Jy)}
  }
\startdata
  SN~1954J  & \multicolumn{3}{c}{merged data (1)}       & $<70$          & $<45$         & $<281$         &$<855$         & $<1570$ \\
  SN~1954J  & \multicolumn{3}{c}{merged data (1)}       & $70\pm21$      & $45\pm13$     & $281\pm65$     & $855\pm198$   & $<1570$ \\
  SN~1997bs & 2004-05-22 & 53148 &   Kennicutt/159      & $<68$          & $<72$         & $336$          & $1372$        \\
  SN~1997bs & 2004-05-22 & 53148 &   Kennicutt/159      & $68\pm20$      & $72\pm16$     & $336\pm68$     & $1372\pm286$  \\
  SN~1999bw & 2004-04-30 & 53126 &   Kennicutt/159      & $20\pm9$       & $40\pm4$      & $144\pm19$     & $354\pm68$    &         \\
            &            &       &                      & $20\pm10$      & $40\pm20$     & $110\pm20$     & $190\pm40$    &         &S04\\
            & 2005-11-29 & 53704 &   Sugerman/20320     & $13\pm6$       & $19\pm4$      & $93\pm22$      & $230\pm62$    & $<1100$ \\
            & 2006-05-03 & 53859 &   Sugerman/20320     & $15\pm8$       & $15\pm4$      & $91\pm25$      & $226\pm67$    & $<940$  \\
            & 2008-05-13 & 54602 &   Meixner/40010      & $15\pm6$       & $12\pm5$      & $58\pm19$      & $224\pm72$    & $<1400$ \\
  SN~2000ch & 2007-12-27 & 54461 &   Kennicutt/40204    & $68\pm6$       & $50\pm4$      & $51\pm4$       & $40\pm10$     \\
            & 2008-05-20 & 54607 &   Fazio/40301        &                &               &                &               & $<89$   \\
  SN~2001ac & 2008-01-05 & 54471 &   Fazio/40349        &                &               &                &               & $<530$  \\
            & 2008-06-20 & 54638 &   Fazio/40349        & $19\pm3$       & $22\pm2$      & $26\pm7$       & $40\pm14$     \\
  SN~2002bu & 2004-04-25 & 53120 &   Fazio/69           &                &               &                &               & $750\pm26$   &$<4200$ \\
            & 2004-05-02 & 53128 &   Fazio/69           & $544\pm7$      & $806\pm9$     & $1098\pm12$    & $1204\pm20$   \\
  SN~2002kg & \multicolumn{3}{c}{merged data (2)}       & $3\pm2$        & $6\pm2$       & $<35$          & $<21$         & $<630$  \\
  SN~2003gm & 2009-08-08 & 55051 &   Sheth/60007        & $44\pm11$      & $35\pm9$ \\
\enddata
\tablecomments{
   Many of the observations assigned dates were observed over a few days to a week.  In these cases, the
   Date is the date of the first observation and the MJD is the mean MJD of the sequences.  The 
   Comment/$F_{70}$ column contains either the $70\mu$m flux or the reference of the alternate
   source of photometry for the epoch, where S04 is \cite{Sugerman2004}.
   (1) The merged data for SN~1954J combined IRAC programs 159 (Kennicutt, 2004-10-08, 2004-10-12), 
   226 (Van Dyk, 2004-10-07, 2004-11-01, 2005-03-24), 20256 (Meikle, 2005-10-20, 2006-03-23), 
   30292 (Meikle, 2006-10-28, 2007-04-02), 30494 (Sugerman, 2006-10-31, 2007-04-2, 2007-11-19), 40010 (Meixner,
   2007-11-23, 2008-04-12) and 40619 (Kotak, 2007-11-24, 2008-04-07) and MIPS programs 226 (Van Dyk, 2004-10-14)
   30494 (Sugerman, 2006-12-01, 2007-04-13, 2007-11-29), 40010 (Meixner, 2007-10-24, 2008-04-14) and 
   40619 (Kotak, 2007-10-24, 2008-04-15).  (2) The merged data for SN~2002kg removes program 226
   and adds IRAC program 61002 (Freedman, 2009-11-14, 2009-12-03, 2009-12-20) from the list for SN~1954J.
   }
\label{tab:log}
\end{deluxetable}

\begin{deluxetable}{llllcccc}
%\rotate
%\tabletypesize{\scriptsize}
%\scriptsize
\tablecaption{Hubble Observations}
\tablewidth{0pt}
\tablehead{
  \colhead{SN}
  &\colhead{Date} &\colhead{MJD} &\colhead{PI/Program}
  &\multicolumn{4}{c}{Filters and Limits (mag) }
  }
\startdata
  SN~1999bw & 2008-04-03 & 54559 &   Meixner/11229      &$\hbox{F110W}$  &$\hbox{F160W}$ &$\hbox{F205W}$  \\
            &            &       &                      &$<23.11$        &$<22.20$       &$<21.18$ \\
  SN~2002bu & 2005-03-20 & 53456 &   Filippenko/10272   &$\hbox{F435W}$  &$\hbox{F555W}$ &$\hbox{F625W}$ &$\hbox{F814W}$ \\
            &            &       &                      &$<25.67$        &$<25.25$       &$<24.83$       &$<24.40$ \\
\enddata
\tablecomments{
   SN~1999W was observed with NICMOS/NIC2 and SN~2002bu was observed with ACS/HRC.  These are $3\sigma$ upper limits derived from
   aperture photometry with standard aperture corrections.
   }
\label{tab:hstphot}
\end{deluxetable}

\begin{deluxetable}{lllrrrr}
%\rotate
\scriptsize
\tablecaption{LBT Observations}
\tablewidth{0pt}
\tablehead{
  \colhead{SN}
  &\colhead{Date} &\colhead{MJD}
  &\colhead{U}
  &\colhead{B}
  &\colhead{V}
  &\colhead{R} \\
  &
  &
  &\colhead{[mag]}
  &\colhead{[mag]}
  &\colhead{[mag]}
  &\colhead{[mag]}
  }
\startdata
   SN~1954J & \multicolumn{2}{c}{reference image} &  $22.69\pm0.06$  &  $23.35\pm0.06$  &  $22.77\pm0.12$  &  $22.36\pm0.12$  \\
  SN~1997bs & \multicolumn{2}{c}{reference image} &  $23.14\pm0.06$  &  $22.54\pm0.06$  &  $22.44\pm0.06$  &  $22.42\pm0.09$  \\
  SN~2002kg & 2008-03-09 & 54534 &  $18.91\pm0.01$  &  $19.62\pm0.02$  &  $19.68\pm0.01$  &  $19.49\pm0.03$  \\
            & 2008-05-05 & 54591 &  $18.78\pm0.01$  &  $19.60\pm0.02$  &  $19.64\pm0.01$  &  $19.49\pm0.03$  \\
            & 2008-11-22 & 54792 &  \nodata         &  $19.54\pm0.02$  &  $19.58\pm0.01$  &  \nodata         \\
            & 2008-11-24 & 54794 &  \nodata         &  $19.59\pm0.02$  &  $19.62\pm0.01$  &  \nodata         \\
            & 2009-01-30 & 54861 &  $18.84\pm0.01$  &  $19.66\pm0.02$  &  $19.73\pm0.01$  &  $19.61\pm0.02$  \\
            & 2009-01-31 & 54862 &  $18.88\pm0.01$  &  $19.68\pm0.02$  &  $19.73\pm0.01$  &  $19.60\pm0.02$  \\
            & 2010-10-01 & 55470 &  $18.97\pm0.01$  &  $19.78\pm0.01$  &  $19.87\pm0.01$  &  $19.77\pm0.02$  \\
            & 2010-12-06 & 55536 &  $19.14\pm0.01$  &  $19.92\pm0.01$  &  $20.02\pm0.01$  &  $19.96\pm0.01$  \\
            & 2010-12-08 & 55538 &  $19.15\pm0.01$  &  $19.92\pm0.01$  &  $20.05\pm0.01$  &  $19.97\pm0.01$  \\
            & 2010-12-10 & 55540 &  $19.19\pm0.01$  &  $19.93\pm0.01$  &  $20.03\pm0.01$  &  $19.98\pm0.01$  \\
            & 2010-12-13 & 55543 &  $19.16\pm0.01$  &  $19.91\pm0.01$  &  $20.03\pm0.01$  &  $19.99\pm0.01$  \\
            & 2011-02-07 & 55599 &  $19.33\pm0.00$  &  $20.05\pm0.01$  &  $20.19\pm0.01$  &  \nodata         \\
            & 2011-11-24 & 55889 &  \nodata         &  $20.04\pm0.01$  &  $20.21\pm0.01$  &  $20.22\pm0.01$  \\
\enddata
\tablecomments{
The light curve uncertainties include both the ISIS errors and the DAOPHOT uncertainties in the reference images.
The magnitude calibration errors for SN~1954J and SN~2002kg are $0.04$, $0.02$, $0.01$ and $0.03$~mag in U, B, V 
and R respectively, while those for SN~1997bs are $0.04$, $0.04$, $0.02$ and $0.01$~mag.
   }
\label{tab:lbtphot}

\end{deluxetable}

\begin{deluxetable}{lllrrrr}
%\rotate
\scriptsize
\tablecaption{ISIS LBT Light Curves}
\tablewidth{0pt}
\tablehead{
  \colhead{SN}
  &\colhead{Date} &\colhead{MJD}
  &\colhead{$F_U$}
  &\colhead{$F_B$}
  &\colhead{$F_V$}
  &\colhead{$F_R$} \\
  &
  &
  &\colhead{[$\mu$Jy]}
  &\colhead{[$\mu$Jy]}
  &\colhead{[$\mu$Jy]}
  &\colhead{[$\mu$Jy]}
  }
\startdata
   SN~1954J & 2008-03-09 & 54534 &  $ 0.45\pm0.12$  &  $ 0.53\pm0.11$  &  $ 0.16\pm0.09$  &  $0.67\pm0.10$   \\
            & 2008-05-05 & 54591 &  $ 0.21\pm0.07$  &  $ 0.19\pm0.05$  &  $ 0.58\pm0.05$  &  $1.17\pm0.06$   \\
            & 2008-11-22 & 54792 &  \nodata         &  $ 0.28\pm0.06$  &  $ 0.16\pm0.05$  &  \nodata         \\
            & 2008-11-24 & 54794 &  \nodata         &  $ 0.11\pm0.04$  &  $ 0.07\pm0.03$  &  \nodata         \\
            & 2009-01-30 & 54861 &  $-0.02\pm0.09$  &  $ 0.24\pm0.08$  &  $ 0.58\pm0.05$  &  $0.86\pm0.08$   \\
            & 2009-01-31 & 54862 &  $ 0.14\pm0.07$  &  $ 0.08\pm0.05$  &  $ 0.22\pm0.04$  &  $0.52\pm0.05$   \\
            & 2010-10-01 & 55470 &  $ 0.02\pm0.11$  &  $ 0.23\pm0.09$  &  $ 1.03\pm0.07$  &  $1.07\pm0.07$   \\
            & 2010-12-06 & 55536 &  $-0.18\pm0.07$  &  $ 0.37\pm0.06$  &  $-0.04\pm0.05$  &  $0.18\pm0.06$   \\
            & 2010-12-08 & 55538 &  $ 0.09\pm0.06$  &  $-0.08\pm0.04$  &  $-0.30\pm0.04$  &  $0.06\pm0.05$   \\
            & 2010-12-10 & 55540 &  $-0.02\pm0.10$  &  $-0.08\pm0.09$  &  $-0.36\pm0.07$  &  $0.02\pm0.08$   \\
            & 2010-12-13 & 55543 &  $-0.10\pm0.06$  &  $-0.00\pm0.04$  &  $-0.20\pm0.04$  &  $0.04\pm0.04$   \\
            & 2011-02-07 & 55599 &  $-0.37\pm0.09$  &  $ 0.21\pm0.09$  &  $-0.15\pm0.07$  &  $1.04\pm0.07$   \\
            & 2011-11-24 & 55889 &  \nodata         &  $ 0.04\pm0.09$  &  $ 0.54\pm0.07$  &  $1.24\pm0.08$   \\
  SN~1997bs & 2008-05-04 & 54590 &  $ 0.11\pm0.04$  &  $ 0.14\pm0.03$  &  $ 0.29\pm0.04$  &  $-0.03\pm0.05$  \\
            & 2009-01-30 & 54861 &  $ 0.10\pm0.04$  &  $-0.02\pm0.04$  &  $ 0.49\pm0.06$  &  $ 0.05\pm0.06$  \\
            & 2009-03-22 & 54912 &  $-0.02\pm0.05$  &  $ 0.10\pm0.05$  &  $-0.22\pm0.08$  &  $-0.10\pm0.06$  \\
            & 2010-01-12 & 55208 &  $-0.04\pm0.06$  &  $-0.27\pm0.06$  &  $ 0.05\pm0.07$  &  \nodata         \\
            & 2010-12-13 & 55543 &  $ 0.14\pm0.04$  &  $-0.09\pm0.04$  &  $ 0.22\pm0.04$  &  $ 0.06\pm0.06$  \\
            & 2011-02-11 & 55603 &  $ 0.03\pm0.05$  &  $-0.14\pm0.05$  &  $ 0.31\pm0.05$  &  \nodata         \\
            & 2012-01-01 & 55927 &  $ 0.02\pm0.03$  &  $ 0.03\pm0.03$  &  \nodata         &  $-0.18\pm0.04$  \\
\enddata
\tablecomments{These light curves are changes in flux relative to the reference image.  Since we lack a clear
   measurement of the source fluxes in the reference images, there is some freedom in selecting the constant
   flux that needs to be added to these values. 
   }
\label{tab:isisphot}
\end{deluxetable}


\begin{thebibliography}{}


\bibitem[Alard \& Lupton(1998)]{Alard1998} Alard, C., \& Lupton, R.~H.\ 1998, \apj, 503, 325
\bibitem[Alard(2000)]{Alard2000} Alard, C.\ 2000, \aaps, 144, 363
\bibitem[Ayani et al.(2002)]{Ayani2002} Ayani, K., Kawabata, T., 
\& Yamaoka, H.\ 2002, \iaucirc, 7864, 4 
\bibitem[Beckmann \& Li(2001)]{Beckmann2001} Beckmann, S., \& Li, W.~D.\ 2001, \iaucirc, 7596, 1 
\bibitem[Berger et al.(2009)]{Berger2009} Berger, E., et al.\ 2009, \apj, 699, 1850
\bibitem[Bond et al.(2009)]{Bond2009} Bond, H.~E., Bedin, L.~R.,
Bonanos, A.~Z., Humphreys, R.~M., Monard, L.~A.~G.~B., Prieto, J.~L.,
\& Walter, F.~M.\ 2009, \apjl, 695, L154
\bibitem[Botticella et al.(2009)]{Botticella2009} Botticella, M.~T., et al.\ 2009, \mnras, 398, 1041
\bibitem[Botticella et al.(2011)]{Botticella2011} Botticella, M.~T., Smartt, S.~J., Kennicutt, R.~C., Jr., et al.\ 2011, arXiv:1111.1692 
\bibitem[Clark et al.(2009)]{Clark2009} Clark, J.~S., Crowther, P.~A., Larionov, V.~M., Steele, I.~A., Ritchie, B.~W., \& Arkharov, A.~A.\ 2009, \aap, 507, 1555 
\bibitem[Chu et al.(2004)]{Chu2004} Chu, Y.-H., Gruendl, R.~A., 
Stockdale, C.~J., Rupen, M.~P., Cowan, J.~J., 
\& Teare, S.~W.\ 2004, \aj, 127, 2850 
\bibitem[Davidson \& Ruiz(1975)]{Davidson1975} Davidson, K., \& Ruiz, M.~T.\ 1975, \apj, 202, 421 
\bibitem[Davidson(1987)]{Davidson1987} Davidson, K.\ 1987, \apj, 317, 760 
\bibitem[Dessart et al.(2010)]{Dessart2010} Dessart, L., Livne, E., \& Waldman, R.\ 2010, \mnras, 405, 2113 
\bibitem[Draine \& Lee(1984)]{Draine1984} Draine, B.~T., \& Lee, H.~M.\ 1984, \apj, 285, 89 
\bibitem[Draine(2011)]{Draine2011} Draine, B.~T.\ 2011, Physics of
the Interstellar and Intergalactic Medium by Bruce T.~Draine.~Princeton
University Press, 2011.~ISBN: 978-0-691-12214-4,
\bibitem[Elitzur \& Ivezi\'c(2001)]{Elitzur2001} Elitzur, M., \& Ivezi\'c, Z.\ 2001, \mnras, 327, 403 
\bibitem[Falk \& Arnett(1977)]{Falk1977} Falk, S.~W., \& Arnett, W.~D.\ 1977, \apjs, 33, 515 
\bibitem[Fern\'andez-Laj\'us et al.(2009)]{Fernandez2009} Fern\'andez-Laj\'us, E., et al.\ 2009, \aap, 493, 1093 
\bibitem[Filippenko et al.(1995)]{Filippenko1995} Filippenko, A.~V., 
Barth, A.~J., Bower, G.~C., Ho, L.~C., Stringfellow, G.~S., Goodrich, 
R.~W., \& Porter, A.~C.\ 1995, \aj, 110, 2261 
\bibitem[Filippenko(1997)]{Filippenko1997} Filippenko, A.~V.\ 1997, \araa, 35, 309 
\bibitem[Filippenko et al.(1999)]{Filippenko1999} Filippenko, A.~V., 
Li, W.~D., \& Modjaz, M.\ 1999, \iaucirc, 7152, 2 
\bibitem[Filippenko(2000)]{Filippenko2000} Filippenko, A.~V.\ 2000, 
\iaucirc, 7421, 3 
\bibitem[Foley et al.(2007)]{Foley2007} Foley, R.~J., Smith, N., Ganeshalingam, M., Li, W., Chornock, R., 
\& Filippenko, A.~V.\ 2007, \apjl, 657, L105 
\bibitem[Freedman et al.(2001)]{Freedman2001} Freedman, W.~L., et al.\ 2001, \apj, 553, 47 
\bibitem[Garcia-Segura et al.(1996)]{Garcia1996} Garcia-Segura, G., Mac Low, M.-M., \& Langer, N.\ 1996, \aap, 305, 229 
\bibitem[Garnavich et al.(1999)]{Garnavich1999} Garnavich, P., Jha, 
S., Kirshner, R., Calkins, M., \& Brown, W.\ 1999, \iaucirc, 7150, 1 
\bibitem[Gieren et al.(2005)]{Gieren2005} Gieren, W., 
Pietrzy\'nski, G., Soszy\'nski, I., Bresolin, F., Kudritzki, R.-P., 
Minniti, D., \& Storm, J.\ 2005, \apj, 628, 695 
\bibitem[Goobar(2008)]{Goobar2008} Goobar, A.\ 2008, \apjl, 686, L103 
\bibitem[Goodrich et al.(1989)]{Goodrich1989} Goodrich, R.~W., 
Stringfellow, G.~S., Penrod, G.~D., 
\& Filippenko, A.~V.\ 1989, \apj, 342, 908 
\bibitem[Horiuchi et al.(2011)]{Horiuchi2010}  Horiuchi, S., Beacom, 
J.~F., Kochanek, C.~S., et al.\ 2011, \apj, 738, 154 
\bibitem[Humphreys \& Davidson(1994)]{Humphreys1994} Humphreys, R.~M., \& Davidson, K.\ 1994, \pasp, 106, 1025 
\bibitem[Humphreys et al.(1999)]{Humphreys1999} Humphreys, R.~M., Davidson, K., \& Smith, N.\ 1999, \pasp, 111, 1124 
\bibitem[Humphreys et al.(2011)]{Humphreys2011} Humphreys, R.~M., Bond, H.~E., Bonanos, A.~Z., et al.\ 2011, arXiv:1109.5131 
\bibitem[Ivezic \& Elitzur(1997)]{Ivezic1997} Ivezic, Z., \& Elitzur, M.\ 1997, \mnras, 287, 799 
\bibitem[Ivezic et al.(1999)]{Ivezic1999} Ivezic, Z., Nenkova, M., \& Elitzur, M.\ 1999, User Manual for DUSTY, 
  University of Kentucky Internal Report http://www.pa.uky.edu/$\sim$moshe/dusty/
\bibitem[Kashi et al.(2010)]{Kashi2010} Kashi, A., Frankowski, A., \& Soker, N.\ 2010, \apjl, 709, L11
\bibitem[Khan et al.(2010)]{Khan2010} Khan, R., Stanek, K.~Z.,
Prieto, J.~L., Kochanek, C.~S., Thompson, T.~A.,
\& Beacom, J.~F.\ 2010, \apj, 715, 1094
\bibitem[Kochanek et al.(2008)]{Kochanek2008} Kochanek, C.~S., 
Beacom, J.~F., Kistler, M.~D., Prieto, J.~L., Stanek, K.~Z., Thompson, 
T.~A., Yuksel, H.\ 2008, \apj, 684, 1336 
\bibitem[Kochanek(2009)]{Kochanek2009} Kochanek, C.~S.\ 2009, \apj, 707, 1578
\bibitem[Kochanek et al.(2011)]{Kochanek2010} Kochanek, C.~S., 
Szczygie{\l}, D.~M., \& Stanek, K.~Z.\ 2011, \apj, 737, 76 
\bibitem[Kochanek(2011a)]{Kochanek2011} Kochanek, C.~S.\ 2011a, \apj, 741, 37 
\bibitem[Kochanek(2011b)]{Kochanek2011b}  Kochanek, C.~S.\ 2011b, \apj, 743, 73 
\bibitem[Kudritzki \& Puls(2000)]{Kudritzki2000} Kudritzki, R.-P., \& Puls, J.\ 2000, \araa, 38, 613 
\bibitem[Leaman et al.(2010)]{Leaman2010} Leaman, J., Li, W., Chornock, R., \& Filippenko, A.~V.\ 2010, arXiv:1006.4611 
\bibitem[Li(1999)]{Li1999} Li, W.~D.\ 1999, \iaucirc, 7149, 1 
\bibitem[Li et al.(2002)]{Li2002} Li, W., Filippenko, A.~V., 
Van Dyk, S.~D., Hu, J., Qiu, Y., Modjaz, M., 
\& Leonard, D.~C.\ 2002, \pasp, 114, 403 
\bibitem[Li et al.(2011)]{Li2011} Li, W., Leaman, J., Chornock, R., et al.\ 2011, \mnras, 412, 1441 
\bibitem[Lien et al.(2010)]{Lien2010} Lien, A., Fields, B.~D., \& Beacom, J.~F.\ 2010, \prd, 81, 083001 
\bibitem[Marigo et al.(2008)]{Marigo2008} Marigo, P., Girardi, L., Bressan, A., Groenewegen, M.~A.~T., Silva, L., \& Granato, G.~L.\ 2008, \aap, 482, 883
\bibitem[Martin et al.(2006)]{Martin2006} Martin, J.~C., Davidson, K., \& Koppelman, M.~D.\ 2006, \aj, 132, 2717
\bibitem[Matheson \& Calkins(2001)]{Matheson2001} Matheson, T., \& Calkins, M.\ 2001, \iaucirc, 7597, 3 
\bibitem[Mathis et al.(1977)]{Mathis1977} Mathis, J.~S., Rumpl, W., \& Nordsieck, K.~H.\ 1977, \apj, 217, 425
S.~J., Kudritzki, R.~P., Podsiadlowski, P., 
\& Gilmore, G.~F.\ 2004, \nat, 427, 129 
\bibitem[Maund et al.(2006)]{Maund2006} Maund, J.~R., et al.\ 2006, \mnras, 369, 390 
\bibitem[Maund et al.(2008)]{Maund2008} Maund, J.~R., et al.\ 2008, \mnras, 387, 1344 
\bibitem[Ohsawa et al.(2010)]{Ohsawa2010} Ohsawa, R., et al.\ 2010, \apj, 718, 1456 
\bibitem[Ostriker et al.(2001)]{Ostriker2001} Ostriker, E.~C., Stone, J.~M., \& Gammie, C.~F.\ 2001, \apj, 546, 980
\bibitem[Owocki et al.(2004)]{Owocki2004} Owocki, S.~P., Gayley, K.~G., \& Shaviv, N.~J.\ 2004, \apj, 616, 525 
\bibitem[Papenkova \& Li(2000)]{Papenkova2000} Papenkova, M., \& Li, W.~D.\ 2000, \iaucirc, 7415, 1 
\bibitem[Pastorello et al.(2010)]{Pastorello2010} Pastorello, A., et 
al.\ 2010, arXiv:1006.0504 
\bibitem[Patat et al.(2003)]{Patat2003} Patat, F., Pastorello, 
A., \& Aceituno, J.\ 2003, \iaucirc, 8167, 3 
\bibitem[Prieto(2008)]{Prieto2008a} Prieto, J.~L.\ 2008, The Astronomer's Telegram, 1550, 1
\bibitem[Prieto et al.(2008)]{Prieto2008} Prieto, J.~L., et al. 2008b, \apjl, 681, L9
\bibitem[Prieto et al.(2009)]{Prieto2009} Prieto, J.~L., Sellgren,
K., Thompson, T.~A., \& Kochanek, C.~S.\ 2009, \apj, 705, 1425
\bibitem[Prieto et al.(2010a)]{Prieto2010a} Prieto, J.~L., et al.\ 
2010, The Astronomer's Telegram, 2406, 1 
\bibitem[Puckett \& Gauthier(2002)]{Puckett2002} Puckett, T., \& Gauthier, S.\ 2002, \iaucirc, 7863, 1 
\bibitem[Puls et al.(2008)]{Puls2008} Puls, J., Vink, J.~S., \& Najarro, F.\ 2008, \aapr, 16, 209
\bibitem[Pumo et al.(2009)]{Pumo2009} Pumo, M.~L., et al.\ 2009, \apjl, 705, L138 
\bibitem[Rest et al.(2011)]{Rest2011} Rest, A., Prieto, J.~L., Walborn, N.~R., et al.\ 2011, arXiv:1112.2210 
\bibitem[Robinson et al.(1973)]{Robinson1973} Robinson, G., Hyland, 
A.~R., \& Thomas, J.~A.\ 1973, \mnras, 161, 281 
\bibitem[Robinson et al.(1987)]{Robinson1987} Robinson, G., Mitchell, R.~M., Aitken, D.~K., Briggs, G.~P., 
\& Roche, P.~F.\ 1987, \mnras, 227, 535 
\bibitem[Sahu et al.(2006)]{Sahu2006} Sahu, D.~K., Anupama, 
G.~C., Srividya, S., \& Muneer, S.\ 2006, \mnras, 372, 1315 
\bibitem[Schlegel et al.(1998)]{Schlegel1998} Schlegel, D.~J., 
Finkbeiner, D.~P., \& Davis, M.\ 1998, \apj, 500, 525 
\bibitem[Schlegel(1990)]{Schlegel1990} Schlegel, E.~M.\ 1990, \mnras, 244, 269
\bibitem[Schwartz et al.(2003a)]{Schwartz2003a} Schwartz, M., Li, W., 
Filippenko, A.~V., \& Chornock, R.\ 2003, \iaucirc, 8051, 1 
\bibitem[Schwartz et al.(2003b)]{Schwartz2003b} Schwartz, M., 
Holvorcem, P., \& Li, W.\ 2003, \iaucirc, 8164, 1 
\bibitem[Sirianni et al.(2005)]{Sirianni2005} Sirianni, M., Jee, M.~J., Benitez, N. et al.\ 2005, \pasp, 117, 1049
\bibitem[Smith et al.(2001)]{Smith2001} Smith, N., Humphreys, R.~M., \& Gehrz, R.~D.\ 2001, \pasp, 113, 692 
\bibitem[Smith et al.(2003)]{Smith2003} Smith, N., Gehrz, R.~D., Hinz, P.~M., et al.\ 2003, \aj, 125, 1458
\bibitem[Smith et al.(2004)]{Smith2004} Smith, N., Vink, J.~S., \& de Koter, A.\ 2004, \apj, 615, 475 
\bibitem[Smith \& Owocki(2006)]{Smith2006} Smith, N., \& Owocki, S.~P.\ 2006, \apjl, 645, L45 
\bibitem[Smith et al.(2009)]{Smith2009a} Smith, N., et al.\ 2009, \apjl, 697, L49
\bibitem[Smith(2009)]{Smith2009b} Smith, N.\ 2009, arXiv:0906.2204 
\bibitem[Smith et al.(2011)]{Smith2011} Smith, N., Li, W., Silverman, J.~M., Ganeshalingam, M., 
\& Filippenko, A.~V.\ 2011, \mnras, 415, 773 
\bibitem[Stetson(1987)]{Stetson1987} Stetson, P.~B.\ 1987, \pasp, 99, 191
\bibitem[Sugerman et al.(2004)]{Sugerman2004} Sugerman, B., Meixner, 
M., Fabbri, J., \& Barlow, M.\ 2004, \iaucirc, 8442, 2 
\bibitem[Szczygie{\l} et al.(2011)]{Szczygiel2011} 
Szczygie{\l}, D.~M., Prieto, J.~L., Kochanek, C.~S., Stanek, K.~Z., Thompson, T.~A., Beacom,
J.~F., Garnavich, P.~M., \& Woodward, C.~E.\ 2011, ApJ, submitted
\bibitem[Tammann \& Sandage(1968)]{Tammann1968} Tammann, G.~A., \& Sandage, A.\ 1968, \apj, 151, 825 
\bibitem[Thompson et al.(2009)]{Thompson2009} Thompson, T.~A.,
Prieto, J.~L., Stanek, K.~Z., Kistler, M.~D., Beacom, J.~F.,
\& Kochanek, C.~S.\ 2009, \apj, 705, 1364
\bibitem[Toal\'a \& Arthur(2011)]{Toala2011}Toal\'a, J.~A., \& Arthur, S.~J.\ 2011, \apj, 737, 100
\bibitem[Treffers et al.(1997)]{Treffers1997} Treffers, R.~R., Peng, 
C.~Y., Filippenko, A.~V., Richmond, M.~W., Barth, A.~J., 
\& Gilbert, A.~M.\ 1997, \iaucirc, 6627, 1 
2006, \aj, 132, 729 
\bibitem[Tully et al.(2008)]{Tully2008} Tully, R.~B., Shaya, 
E.~J., Karachentsev, I.~D., Courtois, H.~M., Kocevski, D.~D., Rizzi, L., 
\& Peel, A.\ 2008, \apj, 676, 184 
\bibitem[Van Dyk et al.(1999)]{Vandyk1999} Van Dyk, S.~D., Peng, 
C.~Y., Barth, A.~J., \& Filippenko, A.~V.\ 1999, \aj, 118, 2331 
\bibitem[Van Dyk et al.(2000)]{Vandyk2000} Van Dyk, S.~D., Peng, 
C.~Y., King, J.~Y., Filippenko, A.~V., Treffers, R.~R., Li, W., 
\& Richmond, M.~W.\ 2000, \pasp, 112, 1532 
\bibitem[Van Dyk et al.(2002)]{Vandyk2002} Van Dyk, S.~D., 
Filippenko, A.~V., \& Li, W.\ 2002, \pasp, 114, 700 
\bibitem[Van Dyk et al.(2005)]{Vandyk2005} Van Dyk, S.~D., Filippenko, A.~V., Chornock, R., Li, W., 
\& Challis, P.~M.\ 2005, \pasp, 117, 553 
\bibitem[Van Dyk \& Matheson(2011)]{Vandyk2011} Van Dyk, S.~D., \& Matheson, T.\ 2011, arXiv:1112.0299 
\bibitem[van Genderen \& The(1984)]{vanGenderen1984} van Genderen, A.~M., \& The, P.~S.\ 1984, \ssr, 39, 317 
\bibitem[Vink(2009)]{Vink2009} Vink, J.~S.\ 2009, in Eta Carinae and the Supernova Impostors, R. Humphreys \& K. Davidson, eds. (Springer) [arXiv:0905.3338]
\bibitem[Wagner et al.(2004)]{Wagner2004} Wagner, R.~M., et al.\ 2004, \pasp, 116, 326 
\bibitem[Walmswell \& Eldridge(2011)]{Walmswell2011} Walmswell, J.~J., \& Eldridge, J.~J.\ 2011, \mnras, 1859 
\bibitem[Weis \& Bomans(2005)]{Weis2005} Weis, K., \& Bomans, D.~J.\ 2005, \aap, 429, L13 
G.~C., Campbell, A., Barlow, M.~J., Sugerman, B.~E.~K., Meixner, M., 
\& Bank, S.~H.~R.\ 2007, \apj, 669, 525 
\bibitem[Wesson et al.(2010)]{Wesson2010} Wesson, R., et al. 2010, \mnras, 403, 474
\bibitem[Whitelock et al.(1994)]{Whitelock1994} Whitelock, P.~A., Feast, M.~W., Koen, C., Roberts, G., \& Carter, B.~S.\ 1994, \mnras, 270, 364 
\end{thebibliography}
\end{document}